%% file: bfkefr.tex
\documentclass{eptcs}
\providecommand{\event}{TARK 2019}
\input defn

\newcommand{\fullv}{\commentout}
\newcommand{\shortv}[1]{#1}
\shortv{\newcommand{\citeyear}{\cite}}

\newcommand{\gap}{\mathit{gap}}

\fullv{\usepackage{chicagor}
\bibliographystyle{chicagor}}

\newcommand{\dbip}{\langle B'_i \rangle}

\newcommand{\strat}{\mathbf{s}}
\newcommand{\intension}[1]{[\![ #1 ]\!]_M}
\newcommand{\intensionc}[1]{[\![ #1 ]\!]_{M^c}}
\newcommand{\intensioncp}[1]{[\![ #1 ]\!]_{M^c}'}
\newcommand{\PR}{\mathcal{PR}}
\renewcommand{\L}{\mathcal{L}}
\newcommand{\TT}{\mathcal{T}}
\newcommand{\pr}{\mathit{pr}}
\renewcommand{\S}{\Sigma}

\usepackage{amssymb}
\usepackage{times}
\renewcommand{\Z}{{\cal Z}}
\newcommand{\undominated}{\mathit{undominated}}
\newenvironment{oldthm}[1]{\par\noindent{\bf Theorem #1:} \em \noindent}{\par}
\newenvironment{oldlem}[1]{\par\noindent{\bf Lemma #1:}
  \em \noindent}{\par}
\newenvironment{oldpro}[1]{\par\noindent{\bf Proposition #1:}
  \em \noindent}{\par}

\newcommand{\othm}[1]{\begin{oldthm}{\ref{#1}}}
\newcommand{\eothm}{\end{oldthm} \medskip}
\newcommand{\olem}[1]{\begin{oldlem}{\ref{#1}}}
\newcommand{\eolem}{\end{oldlem} \medskip}
\newcommand{\opro}[1]{\begin{oldpro}{\ref{#1}}}
\newcommand{\eopro}{\end{oldpro} \medskip}

\newcommand{\init}{\mathit{init}}

\renewcommand{\E}{\mathbf{E}}
\newcommand{\Adv}{\Delta^{\sigma',\sigma}}
\newcommand{\eps}{\epsilon}
\newcommand{\RATzero}{\L^0\mbox{-}\RAT}
\newcommand{\RATzeropi}{\L^0\mbox{-}\RAT^+_i}
\newcommand{\RATzeropj}{\L^0\mbox{-}\RAT^+_j}
\newcommand{\GRATzero}{\mathit{GRAT}}
\newcommand{\GRAT}{\mathit{GRAT}}
\newcommand{\GRATthree}{\L^2\mbox{-}\mathit{GRAT}}
\newcommand{\RATthree}{\L^2\mbox{-}\RAT}
\newcommand{\RATL}{\L\mbox{-}\RAT}
\newcommand{\RATEFR}{\L^0_{\mathit{EF}}\mbox{-}\RAT}
\newcommand{\RATEFRp}{\L^0_{\mathit{EF}}\mbox{-}(\RAT')}
\fullv{\newcommand{\PLAYCONzero}{\mathit{\L^0\mbox{-}PLAYCON}}}
\shortv{\newcommand{\PLAYCONzero}{\mathit{PLAYCON}}}
\newcommand{\PLAYCONthree}{\mathit{\L^2\mbox{-}PLAYCON}}
\newcommand{\PLAYCONefr}{\mathit{\L^0_{\mathit{EF}}\mbox{-}PLAYCON}}
\newcommand{\NSD}{\mathit{NSD}}
\newcommand{\NWD}{\mathit{NWD}}
\newcommand{\NCD}{\mathit{NCD}}
\newcommand{\node}{\mathbf{n}}
\newcommand{\vecmu}{\vec{\mu}}
\renewcommand{\<}{(}
\renewcommand{\>}{)}
\newcommand{\Cond}{\rightarrow}

\fullv{\title{That's All I know:\\
A Logical Characterization of Iterated Admissibility and Extensive Form
Rationalizability
\thanks{A preliminary version of this paper, with the title ``A logical
characterization of iterated admissibility'', appeared in the
Proceedings of Twelfth Conference on Theoretical Aspects of
Rationality and Knowledge, 2009.}
  }}
\shortv{\title{A Conceptually Well-Founded Characterization of
        Iterated Admissibility Using an ``All I Know'' Operator}}

\author{Joseph Y. Halpern and Rafael Pass
  \institute{
  Department of Computer Science\\
Cornell University}
\email{halpern@cs.cornell.edu \quad rafael@cs.cornell.edu}}
\def\titlerunning{A Conceptually Well-Founded Characterization of
    Iterated Admissibility}
\def\authorrunning{J. Y. Halpern \& R. Pass}

\begin{document}
\maketitle

\begin{abstract}
Brandenburger, Friedenberg, and Keisler provide an epistemic
characterization of \emph{iterated admissibility} (IA), also known as
\emph{iterated deletion of 
weakly dominated strategies}, where uncertainty is represented using LPSs
(lexicographic probability sequences).  Their characterization holds in
a rich structure called a \emph{complete} structure, where all types are
possible.
\fullv{A logical characterization of iterated admissibility
is given that holds in all
structures, not just complete structures.  
Roughly speaking, 
our characterization shows that iterated admissibility captures 
the intuition that ``all the agent knows''
is that agents
satisfy the appropriate rationality assumptions.  
We additionally show that essentially the same logical formula
characterizes the notion of extensive-form rationalizability if
uncertainty instead is represented using LPSs.
The analysis brings out the relationship between the
``all I know'' notion, iterated admissibility, and extensive-form
rationalizability. }
\shortv{In earlier work, we gave a characterization of iterated
  admissibility using an ``all I know'' operator, that captures the
  intuition that ``all the agent knows'' is that agents
satisfy the appropriate rationality assumptions.   That
characterization did not need complete structures and used probability
structures, not LPSs.  However, that
characterization did not deal with Samuelson's conceptual concern
regarding IA, namely, that at higher levels, players do not consider
possible strategies 
  that were used to justify their choice of strategy at lower levels.  
In this paper, we give a characterization of IA
using the all I know operator that does deal with Samuelson's
concern.  However, it uses LPSs.  We then show how to modify the
characterization using notions of ``approximate belief'' and
``approximately all I know'' so as to deal with Samuelson's concern
while still working with probability structures.}
 \end{abstract}


\section{Introduction}
A strategy $\sigma_i$ for player $i$ is \emph{admissible}
with respect to a set $\Sigma = \Sigma_1 \times \cdots \times
\Sigma_n$ of strategy profiles if it is a best
response to some belief of player $i$ that puts positive probability
on all the strategy profiles in $\Sigma$.
That is, there is some probability $\mu$ on $\Sigma_{-i}$ such that
no strategy in $\Sigma_i$ gives player $i$ a higher expected utility
than $\sigma_i$ with respect to the beliefs $\mu$.
As Pearce 
\citeyear{Pearce84} has shown, a strategy $\sigma_i$ for
player $i$ is admissible with respect to $\Sigma$ iff it is not weakly
dominated; that is, there 
is no strategy $\sigma_i'$ for player $i$ that gives $i$ at least as high
a payoff as $\sigma_i$ no matter what strategy in $\Sigma_{-i}$ the
other players are using, and sometimes gives $i$ a higher payoff.
It seems natural for a rational player not to play an inadmissible
strategy.  If we delete all strategy profiles from
$\Sigma$ that involve an inadmissible strategy, we get a new set
$\Sigma'$ of strategy profiles.  We can then consider which
strategies are inadmissible with respect to $\Sigma'$, and iterate
this process.
This leads to the solution concept of 
\emph{iterated admissibility} (IA) (also known as \emph{iterated
  deletion of weakly dominated strategies}), one of the most studied
solution concepts in normal-form games.

As Samuelson \citeyear{Sam92} pointed out, 
there is a conceptual problem when it comes to dealing with IA.
As he says, ``the process appears to
initially call for agents to assume that opponents may play any of their
strategies but to subsequently assume that opponents will certainly not play
some strategies.''
%
Brandenburger, Friedenberg, and Keisler \citeyear{BFK04} (BFK from now
on) resolve this paradox
by assuming that strategies are not really eliminated.  Rather, they
assume that strategies that are weakly dominated occur with
infinitesimal (but nonzero) probability.
Formally, they capture this by using what they call a
\emph{full-support} LPS---\emph{lexicographically  
  ordered probability 
  sequence} \cite{BBD1,BBD2}.
Recall that a \emph{lexicograpic probability space} is a tuple
$(\Omega,\F,(\mu_0,\mu_1, 
\ldots, \mu_k))$, where $\F$ is a $\sigma$-algebra over $\Omega$ and $\mu_0,
\ldots, \mu_k$ are probability distributions on $(\Omega,\F)$;
$\vec{\mu} = (\mu_0, \ldots, \mu_k)$ is an LPS.
Intuitively, the first measure in the sequence $\vec{\mu}$, $\mu_0$, is the most
important one, followed by $\mu_1$, $\mu_2$, and so on.
The full-support requirement says that the union of the supports of
$\mu_0, \ldots, \mu_k$ is $\Omega$.
In this paper, for simplicity, we assume that all sets are measurable
and keep $\Omega$ implicit when we speak of an LPS.

BFK define a notion of belief that they call
\emph{assumption},
where an event $E$ is assumed in an LPS $\vec{\mu} = (\mu_0, \ldots,
\mu_k)$ if $E$ is \emph{infinitely more likely} than $\overline{E}$ under
$\vec{\mu}$, and $E$ is infinitely more likely than $F$ for 
events $E$ and $F$ if, for all
$\omega \in E$ and $\omega' \in F$, there is some $i$ such that
$\mu_i(\omega) > 0$ and if there exists $j$ such that $\mu_j(\omega')
> 0$, then there exists $j' < j$ 
such that $\mu_{j'}(\omega) > 0$.%
\footnote{This definition of infinitely more likely is due to Blume,
  Brandenburger, and Dekel \citeyear{BBD1}.
  BFK give a somewhat general definition that applies
  even if not all sets are meaurable.  BFK also require that the
  measures in an LPS have disjoint supports.  While we do not require this,
  our results would continue to hold with essentially  no change in proof if we
  also imposed this requirement.  We remark that the idea of requiring
  strategies that survive $k+1$ rounds of interated deletion to be
  infinitely more likely than strategies that survive only $k$ rounds
  of iterated deletion, used by BFK, goes back to Stahl \citeyear{Stahl95}.}
They then show that 
strategies that survive $k$ rounds of iterated deletion are 
exactly the ones played in states in a \emph{complete} type structure
where there is a $k$th-order assumption of
rationality; that is, everyone assumes that everyone assumes $\ldots$ ($k-1$
times) that everyone is rational.
Complete type structures are particularly rich structures, where all
types are possible.
By considering LPSs with full support, BFK guarantee that strategies
are not really eliminated; that is, no strategies are ever assigned
probability 0.  But full support in complete type structures also forces
agents to ascribe positive probability to many other events; in
particular, they must consider possible all beliefs that other agents could
have about beliefs that other agents could have $\ldots$ about
strategies that an agent is using.  The use of complete type structures
also leads to other technical problems.  For example, although \emph{common
assumption of rationality} (RCAR) ($k$th-order assumption of rationality for
all $k$) is consistent, BFK show that it cannot hold in a complete and
continuous type structure.

There has been a great deal of follow-on work on IA. We briefly discuss some
of the results here.  With regard to
the latter point, Keisler and Lee \citeyear{KL15} show that RCAR is
satisfiable in complete (but not continuous) type structure.  In their
construction, the structure depends on the game; Lee \citeyear{Lee16} provides a
general game-independent construction.   Yang \citeyear{Yang16}
defines a notion of \emph{weak assumption} that, as the name suggests,
is weaker than assumption, and shows that common weak assumption of
rationality is satisfiable in continuous type structures.
Catonini and de Vito \citeyear{CD16} point out that the
full-support condition depends crucially on the topology of the type
space; they replace the full-support condition by what they call
\emph{cautiousness}, which requires only that all strategy profiles
are considered possible, and provide a characterization of IA in
complete type spaces using a notion of common cautious belief in
rationality. 
Finally, Perea \citeyear{Perea12}, using the same notion of
cautiousness as Catonini and de Vito, provides a characterization of
IA using his version of common assumption of rationality. Perea does
 not need to consider complete type spaces; indeed, he shows that his
notion of common assumption of rationality is satisfied even in finite
spaces.  

\fullv{
A related line of work considers \emph{extensive-form
  rationalizability} \cite{Pearce84} (EFR), a standard solution
concept in extensive-form games.  It too involves an 
iterated deletion of strategies that are in a certain sense
dominated \cite{SW98}.  Battigalli and Sinischalchi \citeyear{BS02} 
provide an epistemic characterization of EFR using a notion of
\emph{strong belief}; these are beliefs that are maintained unless
evidence shows that the beliefs are inconsistent.  For example, if player 1
has a strong belief of player 2's rationality, then whatever moves
player 2 makes, player 1 will revise her beliefs and, in particular, her
beliefs about player 2's beliefs, in such a way that she
continues to believe that player 2 is rational (so that she believes that
player 2 is making a best response to his beliefs), unless it is
inconsistent for her to believe that player 2 is rational.  Battigalli
and Sinischalchi characterize EFR in terms of common strong belief of
rationality.  Like that of BFK, the Battigalli-Siniscalchi
characterization holds only in complete type structures.

As Battigalli \citeyear{Bat97} and Shimoji
\citeyear{Shimoji04} show, IA and EFR are closely related.
Moreover, as pointed out by BFK, there are deep connections between
the BFK analysis of iterated admissibility and the Battigalli-Sinischalchi
analysis of EFR and, in particular, between the notions of assumption
and strong  belief.  Yet, on the surface, the characterizations of IA
and EFR seem quite different.  
}

\fullv{
In this paper, we consider a novel ``all I know'' operator, and show
that it can be used to provide quite similar epistemic characterizations of IA
and EFR.}
In earlier work \citeyear{HP08a}, we provided a characterization of IA
using an ``all I know'' operator.  
Roughly speaking, instead of assuming only that agents know (or
assume) that all other agents satisfy appropriate levels of 
rationality, we assume that ``all the agents know'' 
is that the other agents satisfy the appropriate rationality
assumptions.
\fullv{
We are using the phrase ``all agent $i$ knows'' here in
essentially the same 
sense that it is used by Levesque \citeyear{Lev5} and Halpern and
Lakemeyer \citeyear{HalLak94}.}  We
\fullv{formalize}
\shortv{formalized}
this notion by 
requiring that the agent ascribes positive probability to all
formulas of some language $\cal L$ that are consistent with his
rationality assumptions.  (This admittedly fuzzy 
description is made precise in Section~\ref{sec:strong}.) 
We show that the formula $\psi_k^{\cal L}$ that, roughly speaking, says
that ``all that players know with
respect to language $\cal L$ is that all that players know with
respect to $\cal L$  $\ldots$ ($k$ times) is that all players are rational''
characterizes $k$ levels of iterated deletion of weakly dominated
strategies, both in the case $\cal L$ that just describes
the set of possible strategies played and in the case
that $\cal L$ describes, not only players' strategies, but also
players' beliefs about what strategies other players are using
(including higher-order beliefs about other players beliefs).
That is, we show that if the formula $\psi_k^{\cal L}$ (for these two choices of
${\cal L}$) holds at some state in an
arbitrary model, then the strategies used at that state survive $k$
rounds of iterated deletion.  Conversely, if a strategy $\sigma$ survives $k$
rounds of iterated deletion, then there is a state in some model $M$ where
$\psi_k^{\cal L}$  holds and strategy $\sigma$ is played.  
If the language ${\cal L}$ just talks about strategies, then we can take $M$
to be finite; we do not need to work with complete type structures to
characterize IA.  When we consider the language of strategies, ``all I
know'' can be viewed as roughly analogous to Catonini and de Vito's
\citeyear{CD16} notion of cautiousness.
On the other hand, if ${\cal L}$ talks about strategies and beliefs,
then the structures we use are essentially complete type structures.
\fullv{
Interestingly, if ${\cal L}$ is
  the empty language, then $\psi_k^{\cal L}$ characterizes $k$ rounds
  of iterated   deletion of \emph{strongly} dominated strategies.  }

\fullv{We actually consider two (closely related) characterizations
  of IA.  The first uses the standard notion of belief and the corresponding
``all I know'' operator.  In this
construction, we do not have an analogue to the full-support
condition, even for strategies.  In a state where $\psi_k^{\cal L}$
holds, an agent considers possible only strategies that survive $k-1$
rounds of iterated deletion.  The second characterization does force a
full-support condition, at least for strategies.}
\shortv{The problem  with this characterization of IA is that the
  formula $\psi_k^{\cal L}$ says that players ascribe positive probability
to all and only the strategies of other players that survive $k-1$
rounds of iterated 
deletion.  Thus, while it does provide an elegant characterization of IA,
it does not deal with Samuelson's concern
  that, at higher levels, players do not consider possible strategies
  that were used to justify their choice of strategy at lower levels.  
In this paper we give a characterization using an ``all I know''
operator in the spirit of our earlier characterization, but that does
enforce a full-support condition, at least for strategies.}

To do this, we use a generalized belief operator.  Roughly
speaking, $B_i(\phi_1, \ldots, \phi_\ell)$ says that agent $i$ believes that
$\phi_1$ is true, but if it not, then $\phi_2$ is true, and 
if neither $\phi_1$ nor $\phi_2$ is true, then $\phi_3$ is true, and
so on.
Thus, with this generalized belief operator, $i$ describes not only
his beliefs, but his ``plan of retreat'' in case his beliefs turn
out to be false.  There is an ``all I know'' operator that corresponds
to this generalized belief operator in a natural way.
The combination of the generalized belief operator and the
corresponding ``all I know'' operator leads to a characterization of
IA with a full-support requirement on strategies.

\fullv{
For the first characterization, because we use just standard belief,
we can assume that agents represent their uncertainty using standard
probability measures (as is the case for a characterization of IA
introduced by Barelli and Galanis \citeyear{GB13}, although their
approach is quite different in spirit from the one suggested here).
For the second approach,}
\shortv{In our earlier work, we were able to use standard beliefs and
  represent uncertainty using standard probability.  However, to give
  semantics to the generalized belief operator, we need LPSs (we could
  equally well use other approaches that can represent
infinitesimal probability, like conditional probability spaces
or nonstandard probability spaces.}
Just as with our earlier work (and unlike BFK), 
we do not need to 
use complete structures; indeed, it suffices to work with finite
structures to get our characterization.  Nor do we assume \emph{a
  priori} that the LPS is a  
full-support LPS; the formula that characterizes IA forces any LPS
that satisfies it to be a full-support LPS (at least with respect to
strategies). 

\fullv{Although the second approach requires}
\shortv{Although our new approach requires}
LPSs, it does lead to an arguably more elegant epistemic
characterization
that more directly deals with Samuelson's concern (in
much the same way that BFK's approach does).
That said, LPSs require agents to make very fine probability
distinctions.  In Section~\ref{sec:approx}, we show how we can modify
the new approach using notions of ``approximate belief'' and
``approximately all I know'' so as to deal with Samuelson's concern
while still allowing us to work with probability structures, rather
than LPSs.
Roughly speaking, our result says that a strategy for agent $i$
survives $k$ rounds 
of iterated deletion if it is played at a state where  all agent $i$
approximately knows is that all other agents are $k$-level rational, but
if $i$ were to find  out that they are not, then all $i$ approximately
knows is that they are $(k-1)$-level rational, and so on.

\fullv{For both of our approaches,
it is easy to show that 
if there are strategies that survive $k-1$ levels of iterated deletion
but not $k$ levels, then the formulas $\psi^{\cal L}_k$ and
$\psi^{\cal L}_{k-1}$
are inconsistent.  Roughly speaking, according
to the second formula, all strategies that survive $k-2$ rounds of
iterated deletion must get non-infinitesimal probability, while the
first formula says that only formulas that survive $k-1$ rounds of
iterated deletion get non-infinitesimal probability.  Thus, in general, the
infinite conjunction $\land_{k = 0}^\infty \psi^{\cal L}_k$, which can
be thought of as the analogue of RCAR, is inconsistent.  However,
for every finite game $\Gamma$, there exists a $k^*$ such that the
infinite conjunction $\land_{k = k^*}^\infty \psi^{\cal L}_k$ is
satisfiable.  Indeed, if ${\cal L}$ just talks about strategies, it is
satisfiable in a finite model, so continuity trivially holds.
}

\commentout{
More precisely, let $\RATL_i$ be true iff player $i$ is playing a
best response to his belief
and he considers possible all formulas in $\L$. 
Define $\RATL_i^{k+1}$ to be true iff 
$\RATL_i$ holds, 
player $i$ is playing a strategy compatible with $\RATL_i^{k}$, and
for all $j\neq i$, $i$ knows that $\RATL_j^k$ holds, and that is
``all that agent $i$ knows'' about players $j \neq i$. 
That is, $\RATL_i^{k+1}$ holds (i.e., player $i$ is $(k+1)$-level
rational with respect to $\L$) 
iff player $i$ is playing  
a best response to his beliefs, 
using a strategy consistent with $k$-level rationality,
and the only thing he knows about the other players is that they are
$k$-level rational with respect to $\L$, so that he considers possible
all formulas in $\L$ compatible with $k$-level rationality for the other
players.   
As we show, for natural choices of languages $\L$,
a strategy $\sigma$ for player $i$ survives $k$ levels of iterated
deletion of weakly dominated strategies 
iff there is a structure and a state where $\sigma$ is played by player
$i$ and the formula $\RATL^k_i$ holds.
}

\commentout{
However, they prove only that their
characterization of iterated admissibility holds in particularly rich
structures called \emph{complete} structures,
where all types are  possible.
%
Here, we provide an alternate logical characterization of IA.
The characterization has the advantage that it
holds in all structures, not just complete structures, and assumes  
that agents represent their uncertainty using standard probability
measures, rather than LPSs or nonstandard probability measures (as is
done in a characterization of Rajan \citeyear{Rajan98}).  Moreover, 
while complete structures must be uncountable, we show that our
formula characterizing IA is satisfiable in a
structure with finitely many 
states.


For this result to hold, $\L$ must be reasonably expressive; in
particular, it must be possible to express in $\L$ a player's beliefs
about the strategies that other players are playing.
Interestingly, for less expressive languages $\L$, $\RATL^k_i$ instead 
characterizes iterated deletion of \emph{strongly} dominated strategies.
For instance, if $\L$ is empty, then ``all
agent $i$ knows is $\phi$'' is equivalent to ``agent $i$ knows $\phi$'';  
it then easily follows (given standard assumptions about knowledge) that 
$\RATL_i^{k+1}$ is equivalent to the statement that player $i$ is
rational 
in the standard sense (i.e., player $i$  is making a best response to his
beliefs, with no constraints on the beliefs)
and knows that everyone is rational and knows that everyone
knows that everyone knows \ldots ($k-1$ times) that everyone is rational.
Tan and Werlang \citeyear{TW88} and Brandenburger and Dekel
\citeyear{BD87a} show that
this formula characterizes rationalizability, and hence, by results
of Pearce \citeyear{Pearce84}, that it also characterizes iterated
deletion of strongly dominated strategies.
Thus, essentially the same logical formula characterizes both iterated
deletion of strongly and weakly dominated strategies; in a sense, the
only difference between these notions is the expressiveness of the
language used by the players to reason about each other. 
\commentout{
For all choices of $k$ and all languages $\L$ that we consider, we can
find a state in a countable 
structure where $\RATL^k_i$ holds for all agents.
Indeed, for some choices of $\L$ that characterize
iterated admissibility, we can take the structure to be finite.
(Interestingly, depending on the choice of $\L$, we cannot necessarily
find a state where $\RATL^k$ holds for all $k$.  It is not the case that 
$\RATL^{k+1}$ implies $\RATL^{k}$.)
There are computational and conceptual advantages to having these
formulas satisfied in finite structures; these are much easier for an
agent to work with.
However, in earlier work on ``all I know'' (e.g., \cite{HalLak94}) there
has been a great deal of emphasis on working with what is called 
the \emph{canonical}
structure, which has a state corresponding to every satisfiable collection of
formulas.  Of course, this requirement means that the canonical has an
uncountable number of states.   
The complete structures used by BFK are also quite rich.  
We can show that the canonical structure is complete in the sense of
BFK. Moreover, under a technical assumption, every complete structure
is essentially canonical (i.e., it has a state corresponding to every
satisfiable collection of formulas).  This sequence of results allows us
to connect iterated admissibility, complete structures, canonical
structures, and the notion of ``all I know''.  
}
}


\fullv{
We then turn our attention to EFR, and show that we can get an
epistemic characterization of EFR using slight variants of
the formulas use to characterize IA.  
Our characterization of IA and EFR using all I know brings out the
relationship between them more clearly.}

\fullv{The rest of the paper is organized as follows.  
In Section
  \ref{sec:charL0} and \ref{sec:strong} contain our characterizations
of iterated admissibility using ``all I know''.  In Section
\ref{sec:charL0} we define ``all I know'' using a simple language;
 Section \ref{sec:strong} considers the effect of using more expressive
languages. 
We provide our characterization of EFR using ``all I know'' in
Section~\ref{sec:efr}, 
and conclude with a discussion in Section
\ref{sec:discussion}. }

\section{Probability Structures, Rationalizability, and
Admissibility}\label{sec:admissibility}
\shortv{The material in this section is taken almost verbatim from our
  earlier paper \cite{HP08a}.}

We consider normal-form games with $n$ players.  
Given a (normal-form) $n$-player game $\Gamma$, let
$\Sigma_i^{\Gamma}$ 
denote the strategies of player $i$ in $\Gamma$, and let
$u_i^{\Gamma}$ denote the utility function of player $i$ in $\Gamma$.
We omit the superscript $\Gamma$ when it is clear from context or
irrelevant.  Let 
$\vec{\Sigma} = \Sigma_1 \times \cdots \times \Sigma_n$.
We restrict to finite games, so we assume that $\vec{\Sigma}$ is finite.
We further assume, without loss of generality (since the game is
finite), that for each player 
$i$, the range of $u_i$ is $[0,1]$. 
Let $\L^1$ be the language where we start with $\true$ and the special
primitive 
proposition $\RAT_i$ and close off under modal operators $B_i$ and
$\dbi$, for $i = 1, \ldots, n$, conjunction, and negation.   
We think of $B_i \phi$ as saying that, according to player $i$, $\phi$
holds with probability 1, and 
$\dbi \phi$ as saying, accoding to $i$, that $\phi$ holds with
positive probability. 
As we shall see, $\dbi$ is definable as $\neg B_i \neg$ if we make the
appropriate measurability assumptions.

To reason about the game $\Gamma$,
we consider a class of probability structures corresponding to
$\Gamma$.
A \emph{probability structure $M$ appropriate for $\Gamma$} 
is a tuple 
$(\Omega,\strat,\F, \PR_1, \ldots, \PR_n)$, where
$\Omega$ is a set of  
states; $\strat$ associates with each state $\omega \in \Omega$ a
pure strategy profile $\strat(\omega)$ in the game $\Gamma$; 
$\F$ is a $\sigma$-algebra over $\Omega$; and, for each
player $i$, $\PR_i$ associates with each state $\omega$ a probability
distribution 
$\PR_i(\omega)$ on $(\Omega,\F)$.
Intuitively, $\strat(\omega)$ is the strategy profile used at state
$\omega$ and $\PR_i(\omega)$ is player $i$'s probability distribution at
state $\omega$.  As is standard, we require that each player knows his
strategy and his beliefs.  Formally, we require that 
\begin{enumerate}
\item
for each strategy $\sigma_i$ for player $i$,
$\intension{\sigma_i} = \{\omega: \strat_i(\omega) = \sigma_i\} \in \F$, 
where $\strat_i(\omega)$ denotes player $i$'s strategy in the strategy
profile $\strat(\omega)$; 
\item
$\PR_i(\omega)(\intension{\strat_i(\omega)}) 
= 1$;
\item for each probability
measure $\pi$ on $(\Omega, \F)$ and player $i$, $\intension{\pi,i}
=\{\omega : \PR_i(\omega) = \pi\} \in \F$; and 
\item
$\PR_i(\omega)(\intension{\PR_i(\omega),i}) = 1$.
\end{enumerate}


The semantics is given as follows:  
\begin{itemize}
\item $(M,\omega) \sat \true$ (so $\true$ is vacuously true).

\item $(M,\omega) \sat \RAT_i$ if $\strat_i(\omega)$ is a best response,
given player $i$'s beliefs on the strategies of other players induced by
$\PR_i(\omega)$.
That is, $i$'s expected utility with $\strat_i(\omega)$ is at least
as high as with any other strategy in $\Sigma_i$, given $i$'s beliefs.
(Because we restrict to appropriate structures, a
player's expected utility at a state $\omega$ is well defined, so we can
talk about best responses.)
\item $(M,\omega) \sat \neg \phi$ if $(M,\omega) \not\sat
\phi$.  
\item $(M,\omega) \sat \phi \land \phi'$ iff $(M,\omega) \sat \phi$ and 
$(M,\omega) \sat \phi'$.
 \item $(M,\omega) \sat B_i \phi$ if there exists a set $F \in \F$ such
that $F \subseteq \intension{\phi}$ and $\PR_i(\omega)(F) = 1$, where
$\intension{\phi} = \{\omega: (M,\omega) \sat \phi\}$.
\item $(M,\omega) \sat \dbi \phi$ if there exists a set $F \in \F$ such
that $F \subseteq \intension{\phi}$ and $\PR_i(\omega)(F) > 0$.
\end{itemize}
We say that $\phi$ is \emph{valid (for game $\Gamma$)} if $(M,\omega)
\sat \phi$ for all structures $M$ appropriate for game $\Gamma$ and all states
$\omega$ in $M$.  We say that $\phi$ is \emph{satisfiable (for game
$\Gamma$)} if $(M,\omega) \sat \phi$ for some state $\omega$ in some 
structure $M$ appropriate for $\Gamma$.

Note that here we do not assume that $\intension{\phi}$ is
measurable.  Thus, we cannot take $B_i \phi$ to mean that agent $i$
ascribes probability 1 to $\intension{\phi}$.  Rather, we take it to
mean that there is a set of probability 1 contained in
$\intension{\phi}$.  Put another way, we are requiring that the
\emph{inner measure} of $\intension{\phi}$ is 1.  Similarly, $\dbi
\phi$ does not quite say that $i$ ascribes $\intension{\phi}$ positive
probability; rather, it says that the inner measure of
$\intension{\phi}$ is positive.  
Given a language (set of formulas) $\L$, $M$   is
\emph{$\L$-measurable} if $M$ is appropriate (for some game $\Gamma$) and 
$\intension{\phi} \in \F$ for all formulas $\phi \in \L$.  It is easy to
check that in an $\L^1$-measurable structure, $B_i \phi$ means that
$i$ ascribes probability 1 to $\intension{\phi}$, $\dbi \phi$ means
that $i$ ascribes positive probability to $\intension{\phi}$, and 
$\dbi \phi$ is equivalent to $\neg B_i \neg \phi$.

\fullv{
To put our results on iterated admissibility into context, we first
consider rationalizability \cite{Ber84,Pearce84}.
We first recall the definition of rationalizability given by Osborne
and Rubinstein \citeyear{OR94}:%
\footnote{This definition is not quite equivalent to the original
definition of rationalizability due to Bernheim \citeyear{Ber84} and
Pearce \citeyear{Pearce84}, since it allows for opponents' strategies
to be correlated, where as Bernheim and Pearce require them to be
independent.} 
\dfn\label{rat1} A strategy
$\sigma$ for player~$i$ in game~$\Gamma$  
is \emph{rationalizable} if, for each player $j$, there is a set $\Z_j   
\subseteq \Sigma_j(\Gamma)$ and, for each strategy $\sigma' \in \Z_j$,  a   
probability measure $\mu_{\sigma'}$ on $\Sigma_{-j}(\Gamma)$ whose
support is a subset of
$\Z_{-j}$ such that     
\begin{itemize}   
\item $\sigma \in \Z_i$; and
\fullv{\item} for each player $j$ and strategy $\sigma' \in \Z_j$, 
strategy $\sigma'$ is a best response to (the beliefs) $\mu_{\sigma'}$.   
\end{itemize}
\edfn

Pearce \citeyear{Pearce84} also considered a characterization of
rationalizability in terms of 
iterated deletion.
\dfn\label{rat2} A strategy
$\sigma$ for player~$i$ in game~$\Gamma$  
is \emph{rationalizable$'$} if
and only if,
for each player $j$, there exists a
sequence $X_j^0, X_j^1, X_j^2, \ldots$ of sets of strategies for player
$j$ such that $X_j^0 = \Sigma_j$ and, for each strategy $\sigma' \in
X_j^k$, $k \ge 1$, a probability measure $\mu_{\sigma',k}$ whose support
is a subset of 
$\vec{X}_{-j}^{k-1}$ such that 
\fullv{\begin{itemize}   
\item $\sigma \in \inter_{j=0}^\infty X_i$; and}
\shortv{ $\sigma \in \inter_{j=0}^\infty X_i$ and,}
\fullv{\item} for each player $j$, each strategy $\sigma' \in X_j^k$ is
a best 
response to the beliefs $\mu_{\sigma',k}$. 
\fullv{\end{itemize}}
\edfn
Intuitively, $X_j^1$ consists of strategies that are best
responses to some belief of player $j$, and $X_j^{h+1}$ consists of
strategies 
in $X_j^h$ that are best responses to some belief of player $j$ with
support $X_{-j}^{h}$; that is, beliefs that assume that everyone else is
best responding to some beliefs assuming that everyone else is responding
to some  beliefs assuming \ldots ($h$ times).
\pro\label{pro:rat} {\rm \cite{Pearce84}} A strategy is rationalizable
iff it is 
rationalizable$'$.  
\epro

We now state the epistemic characterization of rationalizability due to
Tan and Werlang \citeyear{TW88} 
and Brandenburger and Dekel \citeyear{BD87a}
in our language; it just says that a 
strategy is rationalizable iff it can be played in a state where
rationality is common knowledge.

Let $\RAT$ be an abbreviation of $\RAT_1 \land \ldots \land \RAT_n$;
let $E \phi$ be an abbreviation of $B_1 \phi \land \ldots \land B_n
\phi$; and define $E^{k}
\phi$ for all $k$ inductively by taking $E^0 \phi$ to be $\phi$ and
$E^{k+1} \phi$ to be $E(E^k \phi)$.  
Common knowledge of $\phi$ holds iff
$E^k \phi$ holds for all $k \ge 0$.
For convenience, we take $E^{-1} \phi$ to be the formula $\true$.

\thm\label{thm:charrat} The following are equivalent:
\begin{itemize}
\item[(a)] $\sigma$ is a rationalizable strategy for
$i$ in game $\Gamma$; 
\item[(b)] there exists an $\L^1$-measurable structure  $M$
that is appropriate for $\Gamma$ and a state $\omega$ such that
$\strat_i(\omega) = \sigma$ and $(M,\omega) \sat B_i E^{k} \RAT$ for all
$k \ge 0$;  
\item[(c)] there exists a structure  $M$
that is appropriate for $\Gamma$ and a state $\omega$ such that
$\strat_i(\omega) = \sigma$ and $(M,\omega) \sat B_i E^{k} \RAT$ for all
$k \ge 0$.  
\end{itemize}
\ethm

\fullv{
\prf 
The proof is similar in spirit to that of Tan and Werlang
\citeyear{TW88}; we include it here for completeness.
Suppose that $\sigma$ is rationalizable.  Choose 
$\Z_j \subseteq \Sigma_j(\Gamma)$ and measures $\mu_{\sigma'}$ for each
strategy $\sigma' \in \Z_j$ guaranteed to exist by
Definition~\ref{rat1}.  Define an appropriate structure 
$M = (\Omega,\strat,\F, \PR_1, \ldots, \PR_n)$, where
\begin{itemize}
\item $\Omega = \Z_1 \times \cdots \times \Z_n$;
\item $\strat_i(\vec{\sigma}) = \sigma_i$;
\item $\F$ consist of all subsets of $\Omega$;
\item $\PR_i(\vec{\sigma})(\vec{\sigma}')$ is $0$ if $\sigma'_i \ne
\sigma_i$ and is $\mu_{\sigma_i}(\sigma'_{-i})$ otherwise.
\end{itemize}
Since each player is best responding to his beliefs at every state, it
is easy to see that $(M,\vec{\sigma}) \sat \RAT$ for all states
$\vec{\sigma}$.  It easily follows (formally, by 
induction on $k$) that $(M,\vec{\sigma}) \sat E^k \RAT$.
Since $(M,\vec{\sigma}) \sat E^{k+1} \RAT$, it must be the case that 
$(M,\vec{\sigma}) \sat B_i E^{k} \RAT$.
Clearly $M$ is $\L_1$-measurable.
This shows that (a) implies (b).

The fact that (b) implies (c) is immediate.

Finally, to see that (c) implies (a), suppose that 
$M$ is a structure appropriate for $\Gamma$ and $\omega$ is a state in
$M$ such that $\strat_i(\omega) = \sigma$ and $(M,\omega) \sat B_i
E^{k} \RAT$ for all $k \ge 0$.   
Let $X^0_j = \Sigma_j$ and let
$X^{k+1}_j = \{\strat_j(\omega'): (M,\omega') \sat B_j E^k \RAT\}$ for
$k \ge 0$.  We prove by induction that the sets $X^k_j$ satisfy the
requirements of Definition~\ref{rat2}.  This is true by definition for
$k=0$, since $X^0_j = \Sigma_j$.    
If $\sigma' \in X^k_j$ for $k \ge 1$, 
choose some state $\omega'$ such that $(M,\omega') \sat B_j E^{k-1}\RAT$ 
and $\strat_j(\omega') = \sigma'$.   Let $\mu = \PR_j(\omega')$.
It is easy to check that $B_j E^{k-1} \phi \rimp
B_j B_{j'} E^{k-2} \phi$ is valid for all $k$. 
Thus, at all states $\omega''$ in the support of $\mu$, we have
$(M,\omega'') \sat B_{j'} E^{k-2} \phi$.  
(Recall that we have defined $E^{-1} \phi$ to be $\true$.)
It follows that $\strat_{j'}(\omega'') \in X^{k-1}_{j'}$.  In the case
that $k > 1$, this follows from the definition; in the case that
$k=1$, it follows since $X^0_{j'} = \Sigma_{j'}$.  
Let $\mu_{\sigma',k}$ be the projection of $\mu$ onto 
$\Sigma_{-j}$.  The observations above show that the support of
$\mu_{\sigma',k}$ is contained in $X^{k-1}_{-j}$.  
Since players know their own strategies and beliefs, and
$(M,\omega') \sat B_j \RAT$, it easily follows 
that $(M,\omega') \sat \RAT_j$, so $\sigma'_j$ must be a best response
to the beliefs $\mu_{\sigma',k}$, as desired.  
Thus, by Definition~\ref{rat2}, $\sigma$ is rationalizable$'$ and, by
Proposition~\ref{pro:rat}, $\sigma$ is rationalizable.
\eprf
}

We now consider iterated deletion of strongly dominated
(resp., weakly dominated) strategies.  
}

\dfn
\fullv{
Strategy $\sigma$ for player is $i$ \emph{strongly dominated by 
  a mixed strategy
  $\sigma'$ with respect to $\Sigma'_{-i} \subseteq \Sigma_{-i}$} if 
$u_i(\sigma', \tau_{-i}) > u_i(\sigma, \tau_{-i})$ for all $\tau_{-i} \in
\Sigma'_{-i}$.  
}
Strategy $\sigma$ for player is $i$ \emph{weakly dominated by 
  mixed strategy
  $\sigma'$ with respect to $\Sigma'_{-i} \subseteq \Sigma_{-i}$} if 
$u_i(\sigma', \tau_{-i}) \ge u_i(\sigma, \tau_{-i})$ for all $\tau_{-i} \in
\Sigma'_{-i}$ and   $u_i(\sigma', \tau'_{-i}) > u_i(\sigma, \tau'_{-i})$
for some $\tau'_{-i} \in \Sigma'_{-i}$.

\fullv{
Strategy $\sigma$ for player $i$ survives $k$ rounds of iterated
deletion of strongly dominated strategies if, 
for each player $j$, there exists a
sequence $\NSD_j^0, \NSD_j^1, \NSD_j^2, \ldots, \NSD_j^k$ of sets of
strategies for 
player 
$j$ such that $\NSD_j^0 = \Sigma_j$ and, if $h < k$, then $\NSD_j^{h+1}$
consists of the strategies in $\NSD_j^h$ not strongly 
dominated by any mixed strategy with respect to $\NSD_{-j}^h$, and $\sigma \in
\NSD_i^k$.  Strategy $\sigma$ for player $j$ survives iterated deletion
of strongly 
dominated strategies if it survives $k$ rounds
of iterated deletion of strongly dominated strategie for all $k$, that is, if 
$\sigma \in \NSD^\infty_j = \inter_k \NSD^k_j$.  
We can similarly define the sets $\NWD^k_j$ for all $k$, the strategies
for player $j$ 
that survive $k$ rounds of iterated deletion of weakly dominated
strategies and $\NWD^\infty_j$, the strategies for player $j$ that
survive iterated deletion of weakly dominated strategies, by replacing
``strong'' by ``weak'' everywhere above.
}
\shortv{Strategy $\sigma$ for player $i$ survives $k$ rounds of iterated
deletion of weakly dominated strategies if, 
for each player $j$, there exists a
sequence $\NWD_j^0, \NWD_j^1, \NWD_j^2, \ldots, \NWD_j^k$ of sets of
strategies for 
player 
$j$ such that $\NWD_j^0 = \Sigma_j$ and, if $h < k$, then $\NWD_j^{h+1}$
consists of the strategies in $\NWD_j^h$ not weakly
dominated by any mixed strategy with respect to $\NWD_{-j}^h$, and $\sigma \in
\NWD_i^k$.  Strategy $\sigma$ for player $j$ survives iterated deletion
of weakly
dominated strategies if it survives $k$ rounds
of iterated deletion of weakly dominated strategies for all $k$, that is, if 
$\sigma \in \NWD^\infty_j = \inter_k \NWD^k_j$.}  
\edfn

The following well-known result 
\fullv{connects strong and weak dominance to best responses.}
\shortv{connects weak dominance to best responses.}
\pro\label{pro:Pearce} {\rm \cite{Pearce84}}
\fullv{
\begin{itemize}
\item A strategy $\sigma$ for player $i$ is not strongly dominated by
    any mixed strategy with
respect to $\Sigma'_{-i}$ iff there is a belief
$\mu_\sigma$ of player $i$ whose support is a subset of $\Sigma'_{-i}$
such that $\sigma$ is a best response with respect to $\mu_\sigma$.  
\item}
  A strategy $\sigma$ for player $i$ is not weakly dominated by any
  mixed
  strategy with respect to $\Sigma'_{-i}$ iff there 
is a belief $\mu_\sigma$ of player $i$ whose support is all of
$\Sigma'_{-i}$ such that $\sigma$ is a best response with respect to
$\mu_\sigma$.   
\fullv{\end{itemize}}
\epro

\fullv{
It immediately follows from 
Proposition~\ref{pro:Pearce} (and is well known)
that a strategy is 
rationalizable iff it survives 
iterated deletion of strongly dominated strategies.  Thus, the
characterization of rationalizability in Theorem~\ref{thm:charrat} is
also a characterization of strategies that survive iterated deletion of
strongly dominated strategies.  
We now give a slightly different characterization that allows us to relate
iterated deletion of strongly and weakly dominated strategies.
For each player $i$, define the formulas $\RAT_i^k$ inductively by taking
$\RAT^0_i$ to be $\true$ and $\RAT^{k+1}_i$ 
to be an abbreviation of 
$$\RAT_i \land B_i (\RAT^k_{-i}),$$
where $\RAT^k_{-i}$ is an abbreviation of $\land_{j \ne i} \RAT^k_j$.%
\footnote{We use similar abbreviations in the sequel without comment.}
That is, $\RAT^{k+1}_i$ holds (i.e., player $i$ is $(k+1)$-level rational)
iff player $i$ is playing a best response to his beliefs, and he knows
that all players are $k$-level rational.
Note that $\RAT_i^1$ is equivalent to $\RAT_i$.

Two formulas $\phi$ and $\psi$ are \emph{logically
equivalent} if $\phi \Leftrightarrow \psi$ is valid.

\lem\label{Ckchar}
The following formulas are logically equivalent for all $k \ge 0$.
\begin{itemize}
\item[(a)] $\RAT^{k+1}_i$;
  \item[(b)] $B_i \RAT^{k+1}_i$;
\item[(c)] $\RAT_i \land B_i(\land_{j\ne i} \RAT^k_j)$.
\end{itemize}
Moreover, $\RAT^{k+1}_i \rimp \RAT^k_i$ is valid for all $k \ge 0$.
\elem

\prf A straightforward induction on $k$. 
 \eprf

The following alternative characterization of iterated
deletion of strongly dominated strategies 
can be proved in much as the same way as Theorem \ref{thm:charrat}; we
omit the details here.
\thm\label{thm:charsd} The following are equivalent:
\begin{itemize}
\item[(a)] the strategy $\sigma$ for player $i$ survives $k$ rounds of
iterated  
deletion of strongly dominated strategies
in game $\Gamma$; 
\item[(b)] there is an $\L^1$-measurable structure $M^{k}$
appropriate for $\Gamma$ and a state $\omega^{k}$ in $M^{k}$ such that
$\strat_i(\omega^{k}) = \sigma$ and $(M^{k},\omega^{k}) \sat \RAT^{k}_i$;
\item[(c)] there is a structure $M^{k}$
appropriate for $\Gamma$ and a state $\omega^{k}$ in $M^{k}$ such that
$\strat_i(\omega^{k}) = \sigma$ and $(M^{k},\omega^{k}) \sat \RAT^{k}_i$.
\end{itemize}
\ethm

The following corollary now follows from Theorem \ref{thm:charsd}, Lemma
\ref{Ckchar}, and 
the fact that the deletion procedure converges after a finite number of steps.
\cor\label{thm:charsd2} The following are equivalent:
\begin{itemize}
\item[(a)] The strategy $\sigma$ for player $i$ 
survives iterated deletion of strongly dominated strategies 
in game $\Gamma$; 
\item[(b)] there exists an $\L^1$-measurable structure  $M$
that is appropriate for $\Gamma$ and a state $\omega$ such that
$\strat_i(\omega) = \sigma$ and $(M,\omega) \sat \RAT_i^{k} $ for all
$k \ge 0$;  
\item[(c)] there exists a structure $M$
that is appropriate for $\Gamma$ and a state $\omega$ such that
$\strat_i(\omega) = \sigma$ and $(M,\omega) \sat \RAT_i^{k} $ for all
$k \ge 0$.  
\end{itemize}
\ecor

We next turn to characterizing iterated deletion of weakly dominated
strategies.
}

\fullv{\section{Characterizing Iterated Admissibility}}
\shortv{\section{The Earlier Characterization of IA}\label{sec:strong}}

In this section, we review our earlier characterization of iterated
admissibility, to set the stage for the new results.  Again, the
exposition is taken almost verbatim from our earlier paper.

For each player $i$, define the formulas $\RAT_i^k$ inductively by taking
$\RAT^0_i$ to be $\true$ and $\RAT^{k+1}_i$ 
to be an abbreviation of 
$$\RAT_i \land B_i (\RAT^k_{-i}),$$
where $\RAT^k_{-i}$ is an abbreviation of $\land_{j\ne
  i}\RAT^k_{j}$.%
\footnote{We use similar abbreviations in the sequel without comment.}
That is, $\RAT^{k+1}_i$ holds (i.e., player $i$ is $(k+1)$-level rational)
iff player $i$ is playing a best response to his beliefs, and he knows
that all players are $k$-level rational.
\fullv{In the standard 
treatment (which is essentially the one considered in
Section~\ref{sec:admissibility},}
\shortv{Thus, with these definitions,}
player $i$ is taken to be $(k+1)$-level
rational iff player $i$ is rational (i.e., playing a best response to
his beliefs), and knows that all other player are $k$-level rational.%
\footnote{We should perhaps say ``believes'' here rather than ``knows'',
since a player can be mistaken.  We are deliberately blurring the
subtle distinctions between ``knowledge'' and ``belief'' here.}
But what else do players know?  

We want to consider a situation where, intuitively,
\emph{all} an agent knows about the other agents is that they satisfy
the appropriate rationality assumptions.  
More precisely, we modify the formula $\RAT^{k+1}_i$ to require
that not only does player $i$ know that 
the 
players are $k$-level
rational, but this is the \emph{only}  
thing that he knows about the other players. 
That is, we say that agent $i$ is $(k+1)$-level rational if player $i$ is
rational,  
he knows that the players are $k$-level rational, and this is all
player $i$ knows about the other players. 
We here use the phrase ``all agent $i$ knows'' in
essentially
the same 
sense that it is used by Levesque \citeyear{Lev5} and Halpern and
Lakemeyer \citeyear{HalLak94}, but formalize it a bit differently.
Roughly speaking, we interpret ``all agent $i$ knows is $\phi$'' as
meaning that agent $i$ believes $\phi$, and considers possible every
\emph{formula} about the other players that is consistent with $\phi$.
Thus, what 
``all I know'' means is very sensitive to the choice of the language.
To stress this point, we talk about ``all I knows \emph{with respect to
  language} $\L$''.
%
\fullv{
We actually provide two (closely related) epistemic characterizations of IA
using the notion of ``all I know''.  The first is somewhat simpler,
but is arguably not completely epistemic.  The second uses a
generalized belief operator (which may be of independent interest) and
a corresponding ``all I know'' operator, and is more purely epistemic.

\subsection{The first characterization of IA}}

To define the ``all I know'' operator, we use a modal operator $\Diamond$
that characterizes consistency, which is defined as follows: 
\begin{itemize}
\item $(M,\omega) \sat \Diamond \phi$ iff there is some structure $M'$
appropriate for $\Gamma$ and state $\omega'$ such that $(M',\omega')
\sat \phi$.
\end{itemize}
Intuitively, $\Diamond \phi$ is true if there is some state and
structure where $\phi$ is true; 
that is, if $\phi$ is satisfiable.  Note that if $\Diamond \phi$ is
true at some state, then it is true at all states in all structures.
Define $O^{\L}_i\phi$ (read ``all agent $i$ knows with
respect to the language $\L$'') to be
an abbreviation of
$$B_i \phi \land (\land_{\psi \in \L} (\Diamond
(\phi \land \psi) \rimp \dbi \psi)).$$

\fullv{
Since $O^{\emptyset}_i$ (where $\emptyset$ denotes the empty language) 
is equivalent to $B_i$ (under the standard identification of the empty
conjunction with the formula $\true$), it follows that $\RAT^{k+1}_i$ is
just $\RAT_i \land O^{\emptyset}_i 
(\RAT^k_{-i})$.
Thus, we can get a characterization of iterated deletion of strongly
dominated strategies using $O^{\emptyset}_i$.

We next consider a slightly richer language, whose formulas can talk
about strategies (but not beliefs)}

\shortv{In this paper, we focus on just one of the languages
  considered in our earlier paper,  whose formulas can talk
about strategies (but not beliefs) 
of the players.}
\fullv{
  (In Section \ref{sec:strong}, we
consider languages in which we can talk about 
both the strategies and beliefs of the players.)}
Define the primitive proposition $\play_i(\sigma)$ as follows:
\begin{itemize}
\item
$(M,\omega) \sat \play_i(\sigma)$ iff $\omega \in
\intension{\sigma}$.
\end{itemize}
Let $\play (\vec{\sigma})$ be an abbreviation of
$\land_{j=1\,}^n \play_j(\sigma_j)$, and 
let $\play_{-i}(\sigma_{-i})$ be an abbreviation of $\land_{j\ne i}
\play_j(\sigma_j)$.  Intuitively, $(M,\omega) \sat \play(\vec{\sigma})$
iff $\strat(\omega) = \sigma$, and $(M,\omega) \sat
\play_{-i}(\sigma_{-i})$ if, at $\omega$,  
the players other than $i$ are playing strategy profile $\sigma_{-i}$.
%
Let $\L^0(\Gamma)$ be the language whose only formulas are (Boolean
combinations 
of) formulas of the form $\play_i(\sigma)$, $i = 1, \ldots, n$, $\sigma
\in \Sigma_i$. Let $\L^0_i(\Gamma)$ consist of just the formulas of the form
$\play_i(\sigma)$, and let $\L^0_{-i}(\Gamma) = \union_{j \ne i}
\L^0_j(\Gamma)$.  
Again, we omit the parenthetical $\Gamma$ when it is clear from context
or irrelevant.

\fullv{
We would now like to define $\RATzero_i^{k+1}$ by replacing $B_i$ by
$O_i^{\L^0_{-i}}$ in the definition of $\RAT_i^{k+1}$.  
That is, rather than just saying that agent $i$ believes that all agents
are rational up to level $k$, we want to say that this is \emph{all}
agent $i$ believes.
This is, in fact, what we do with regard to $i$'s beliefs about the
other players; all player $i$ believes about player $j \ne i$ is that
$j$ is rational up to level $k$, that is, that $\RATzero_j^{k}$ holds.
But clearly that is not all $i$ believes about himself, since $i$
knows what strategy he is using.  Indeed, $i$ cannot even believe that
he himself is $k$-level rational, since $i$ believing $\RATzero_i^{k}$
entails that all $i$ knows about $j \ne i$ is that $j$ 
is $(k-1)$-level rational, and that is inconsistent with all $i$
knowing about $j$ is that $j$ is $k$-level rational! For example, if
all $i$ knows about $j$ is that $j$ is 2-level rational, then $i$ 
must consider possible (i.e., assign positive probability to) all
strategies for $j$ consistent with 1-rationality, while if all $i$
knows about $j$ is that $j$ is 1-level rational, then  $i$ must
consider possible all strategies for $j$.  We do want to require that
$i$ believe that his strategy is consistent with $k$-level
rationality.  As we now show, this turns out to be enough to capture IA.

In more detail, the first step in getting an
appropriate analogue to $\RAT_i^k$ is to remove the conjunct
$\RATzero_i^k$ from 
the scope of $O_i^\L$.
\fullv{As Lemma~\ref{Ckchar}(c) shows, this change would have}
\shortv{It is not hard to show that this change has} 
no impact on the definition of $\RAT^{k+1}_i$.
We then add a conjunct saying that $i$ believes that his strategy is
consistent with $k$-level rationality.
Thus,  for each player $i$, we
define the formulas $\RATzero_i^k$ inductively by}
\shortv{The sense in which a player is rational when playing a
  strategy that survives iterated deletion is captured by the formulas
  $\RATzero_i^k$, which are defined inductively by}
taking $\RATzero^0_i$ to be $\true$ and  
$\RATzero^{k+1}_i$ to be an abbreviation of 
$$\RAT_i \land 
B_i(\PLAYCONzero^k_i) \land
O^{\L^0_{-i}}_i (\RATzero^k_{-i})
,$$
where
$\PLAYCONzero^k_i$ (read ``player $i$ plays a strategy consistent with
$k$-level rationality'') is an abbreviation of $\land_{\sigma' \in
\Sigma_i(\Gamma)} (\play_i(\sigma') \rimp
\Diamond(\play_i(\sigma') \land \RATzero^k_i))$. 
That is, $\RATzero^{k+1}_i$ holds (i.e., player $i$ is $(k+1)$-level rational)
iff player $i$ is rational,  
believes that he is playing a strategy that is consistent with
$k$-level rationality, 
knows that other players are $k$-level rational, and that is all
player $i$ knows about the \emph{strategies} of the other players.%
\fullv{
\footnote{$\RATzero^k_i$ is essentially the formula $\RAT^k_i$ from the
introduction.  However, since we consider a number of variants of
$\RAT^k_i$, we find it useful to distinguish them.}}

By expanding the modal operator $O$, it easily follows that $\RATzero^{k+1}_i$
implies $\RAT_i \land B_i (\RATzero^k_{-i})$. 
an easy induction on $k$ then shows that $\RATzero^{k+1}_i$ implies
$\RAT^{k+1}_i$. 
But $\RATzero^{k+1}_i$ requires more;
it requires  player $i$ to assign positive
probability to each strategy profile for the other players that is
compatible with $\RATzero^{k}_{-i}$ (i.e., with level-$k$ rationality).
\fullv{
  As we now show,}
\shortv{As shown in our earlier paper \cite{HP08a},}
the formula $\RATzero^{k}_i$ characterizes strategies
that survive  
iterated deletion of weakly dominated strategies.

\thm\label{thm:charwd} The following are equivalent:
\begin{itemize}
\item[(a)] the strategy $\sigma$ for player $i$ survives $k$ rounds of
iterated deletion of weakly dominated strategies in game $\Gamma$;
\item[(b)]  there exists an $\L^0$-measurable structure  $M^{k}$
appropriate for $\Gamma$ and a state $\omega^{k}$ in $M^{k}$ such that
$\strat_i(\omega^{k}) = \sigma$ and $(M^{k},\omega^{k}) \sat
\RATzero^{k}_i$;   
\item[(c)]   there exists a structure  $M^{k}$
appropriate for $\Gamma$ and a state $\omega^{k}$ in $M^{k}$ such that
$\strat_i(\omega^{k}) = \sigma$ and $(M^{k},\omega^{k}) \sat
\RATzero^{k}_i$.
\end{itemize}
In addition,
if $\vec{\sigma} \in \NWD^k$, then
there is a finite structure $\bar{M}^k =
(\Omega^k,\strat^k,\F^k, \PR_1^k, \ldots, \PR_n^k)$ such that 
$\Omega^k = \{(k',i,\vec{\sigma}): 0 \le k' \le k, 1 \le i \le n, \vec{\sigma}
\in \NWD^{k'}\}$,
$\strat^k(k',i,\vec{\sigma}) = 
\vec{\sigma}$, $\F^k = 2^{\Omega^k}$, and
for all states 
$(k', i, \vec{\sigma}) \in \Omega^k$,  
$(\bar{M}^k,(k',i, \vec{\sigma})) \sat \RATzero_{-i}^{k'}$.
\ethm

\fullv{
\prf 
We proceed by induction on $k$, proving both the equivalence of (a),
(b), and (c) 
and the existence of a structure $\bar{M}^k$ with the required properties.

The result clearly holds if $k=0$.  
Suppose that the result holds for $k$; we prove it for $k+1$.
We first show that (c) implies (a).  Suppose that 
$(M^{k+1},\omega^{k+1}) \sat \RATzero^{k+1}_i$ and $\strat_i^k(\omega^{k+1}) =
\sigma$. 
It follows that $\sigma$ is a best response to the belief $\mu_{\sigma}$ on the strategies
of other players induced by $\PR^{k+1}_i(\omega)$.  Since 
$(M^{k+1},\omega^{k+1}) \sat B_i(\RATzero^{k}_{-i})$, it
follows from the 
induction hypothesis that 
the support of $\mu_{\sigma_i}$ is contained in $\NWD^k_{-i}$.  Since
$(M^{k+1},\omega^{k+1}) \sat \land_{\sigma_{-i} \in \S_{-i}}
(\Diamond(\play_{-i}(\sigma_{-i}) \land (\RATzero^k_{-i}) \rimp 
\dbj (\play_{-i}(\sigma_{-i})))$, it follows from the
induction hypothesis that the support of $\mu_{\sigma}$ is all of 
$\NWD^k_{-i}$.  Since $(M^{k+1},\omega^{k+1}) \sat \PLAYCONzero^{k}_i$,
it follows from the 
induction hypothesis that $\sigma_i \in \NWD^k_i$.  Thus, 
since $(M^{k+1,i},\omega^{k+1,i}) \sat \RAT_i$, it follows by Proposition~\ref{pro:Pearce} that
$\sigma_i \in \NWD^{k+1}_i$.  

We next construct the structure $\bar{M}^{k+1} =
(\Omega^{k+1},\strat^{k+1},\F^{k+1}, \PR_1^{k+1}, \ldots, \PR_n^{k+1})$.  
As required, we define
$\Omega^{k+1} = \{(k',i,\vec{\sigma}): k' \le k+1, 1 \le i \le n,
\vec{\sigma} \in
\NWD^{k'}$, $\strat^{k+1}(k', i, \vec{\sigma}) = \vec{\sigma}$, $\F^{k+1} =
2^{\Omega^{k+1}}$.  
For a state $\omega$ of the form $(0, i,\vec{\sigma})$, let 
$\PR_j^{k+1}(\omega)$ be the uniform distribution over states (we could
actually use an arbitrary distribution here); for 
a state $\omega$ of the form $(k', i,\vec{\sigma})$, 
where $k'\geq 1$,
since $\sigma_j \in \NWD^{k'}_j$,
by Proposition \ref{pro:Pearce},
there exists a distribution
$\mu_{k',\sigma_j}$
on strategies
whose support is all of $\NWD^{k'-1}_{-j}$ such that
$\sigma_j$ is a best response to $\mu_{\sigma_j}$.  
Extend
$\mu_{k',\sigma_j}$ to a distribution $\mu_{k',i,\sigma_j}'$ on $\Omega^{k+1}$
as follows:
\begin{itemize}
\item if $i\neq j$, then $\mu_{k',i,\sigma_j}'(k'',i',\vec{\tau}) = 
\mu_{k',\sigma_j}(\vec{\tau}_{-j})$ if $i'=j, k'' = k'-1$, and $\tau_j =
\sigma_j$, and 0 otherwise;   
\item$\mu_{k',j,\sigma_j}'(k'',i',\vec{\tau}) = 
\mu_{k',\sigma_j}(\vec{\tau}_{-j})$ if $i'=j, k'' = k'$, and $\tau_j = \sigma_j$, and 0 otherwise.  
\end{itemize}
Let $\PR_j(k',i,\vec{\sigma}) = \mu_{k',i,\sigma_j}$.
We note for future reference that if $k' \ge 1$, then the support of
$\PR_j(k',j,\vec{\sigma})$ consists of all states of the form
$(k',j,\vec{\tau})$ such that $\tau_j = \sigma_j$ and
$\vec{\tau}_{-j} \in \NWD^{k'-1}$, while if $i' \ne
j$, then the support of
$\PR_j(k',i',\vec{\sigma})$ consists of all states of the form
$(k'-1,j,\vec{\tau})$ such that $\tau_j = \sigma_j$ and
$\vec{\tau}_{-j} \in \NWD^{k'-1}$.
We leave it to the reader to check that this
structure is appropriate.  
We prove by induction on $k'$ that 
$(\bar{M}^{k+1},(k',i,\vec{\sigma})) \sat  \RATzero^{k'}_{-i}$.
The result is trivial if $k' = 0$.  Suppose that the result holds for $k'$;
we prove it for $k'+1$.  Fix $j \ne i$.
Recall that 
the support of $\PR_j(k'+1,i,\vec{\sigma})$ consists of all states of the
form $(k',j,\vec{\tau})$ such that $\tau_j = \sigma_j$ and
$\vec{\tau}_{-j} \in \NWD^{k'}$.  Moreover, the construction of
$\PR_j(k'+1,i,\vec{\sigma})$ guarantees that
$(\bar{M}^{k+1},(k'+1,i,\vec{\sigma})) \sat \RAT_j$ for all players
$j$.  Since $\sigma_j \in \NWD^{k'+1}$, $\sigma_j \in \NWD^{k'}$, so
by the induction hypothesis, $\sigma_j$ is consistent with
$\RATzero^{k'}_j$.  Thus, $(\bar{M}^{k+1},(k',j,\vec{\tau}))
\sat \PLAYCONzero^{k'}_j$ for all states $(k',j,\vec{\tau})$ in the
support of $\PR_j(k'+1,i,\vec{\sigma})$.  It follows that
$(\bar{M}^{k+1},(k'+1,i,\vec{\sigma})) \sat B_j(\PLAYCONzero^{k'}_j)$.  
Finally, we must show that
$(\bar{M}^{k+1},(k'+1,i,\vec{\sigma})) \sat O^{\L^0_{-i}}_j (
\RATzero^{k'}_{-j}) $.  It immediately follows from the
induction hypothesis that for all states of the form
$(k',j,\vec{\tau})$ in the support of $\PR_j(k',i,\vec{\sigma})$, we
have that $(\bar{M}^{k+1},(k',j,\vec{\tau})) \sat 
\RATzero^{k'}_{-j} $.  Thus,
$(\bar{M}^{k+1},(k'+1,i,\vec{\sigma})) \sat B_j (
\RATzero^{k'}_{-j}) $.  Moreover, since all strategies
$\vec{\tau}_{-j} \in \NWD^{k'}_{-j}$ are given positive probability
by $\PR_j(k'+1,i,\vec{\sigma})$, we get that 
$(\bar{M}^{k+1},(k'+1,i,\vec{\sigma})) \sat O^{\L^0_{-i}}_j (
\RATzero^{k'}_{-j}) $, as desired.

To see that (a) implies (b), suppose that $\sigma_j \in \NWD^{k+1}_j$.
Choose a state $\omega$ in 
$\bar{M}^{k+1}$ of the form $(k+1,i,\vec{\sigma})$, where $i \neq j$.  
As we just showed,
$(\bar{M}^{k+1}, \omega) \sat \RATzero^{k+1}_j$ and
$\strat_j(\omega) = \sigma_j$.  Moreover, $\bar{M}^{k+1}$ is

Clearly (b) implies (c).
\eprf
}

\fullv{
Note that there is no analogue of Corollary~\ref{thm:charsd2} here.
This is because there is no state where $\RATzero^k_i$ holds for all $k \ge 0$;
it cannot be the case that $i$ places positive probability on all
strategies (as required by $\RATzero^k_1$) and that $i$ places positive
probability only on strategies that survive one round of iterated
deletion (as required by $\RATzero^k_2$), unless all strategies survive one
round on iterated deletion.  We can say something slightly weaker
though.  There is some $k^*$ such that the process of iterated deletion
converges after $k^*$ steps; that is, $\NWD^{k^*}_j = \NWD^{k^*+1}_j$ for all $j$ (and hence
$\NWD^{k^*}_j = \NWD^{k'}_j$ for all $k' \ge k^*$).
That means that there is a
state where $\RATzero^{k'}_i$ holds for all $k' > k^*$.  Thus, we can show that 
a strategy $\sigma$ for player $i$ survives iterated deletion of weakly
dominated strategies iff there exists a $k^*$ and a state $\omega$ such
that $\strat_i(\omega) = \sigma$ and $(M,\omega) \sat \RATzero^{k'}_i$ for all
$k' > k^*$.  Since $\RAT^{k+1}_i$ implies $\RAT^k_i$, an analogous result holds
for iterated deletion of strongly dominated strategies, with $\RATzero^{k'}_i$
replaced by $\RAT^{k'}_i$.}


\fullv{\subsection{The second characterization of IA}\label{sec:second}
In our first characterization of IA, }
\shortv{\section{The new characterization of IA}\label{sec:second}
In our earlier characterization of IA,}
in a state where $\RATzero^k$
holds, player $i$ does \emph{not} consider all strategies possible,
but only the ones consistent with the appropriate level of
rationality.  
That is, because of the $B_i(\PLAYCONzero^{k-1}_i)$ conjunct in
$\RATzero_i^{k}$, player $i$ ascribes positive probability only to
strategies consistent with $(k-1)$-level rationality.
This means that the characterization of the earlier paper does
not address Samuelson's concern.  More 
specifically, it does not provide an epistemic explanation for
\emph{why}, at higher
levels, players do not consider possible strategies that were used to
justify their choice of strategy at lower levels; it just assumes that
they do.  We deal with this
\fullv{problem in our second characterization of IA.  The second}
\shortv{problem in our new characterization of IA.  The new}
characterization forces the agent to ascribe positive probability to
all strategies,
and thus can be viewed as forcing a full-support requirement, at the
level of strategies.  

As a first step to getting this characterization, we
introduce a notion of \emph{generalized belief}, which may be of
independent interest.  Specifically, we
consider formulas of the form $B_i(\phi_1, \ldots, \phi_\ell)$.  As we said in
the introduction, this formula can be read ``agent $i$ believes that
$\phi_1$ is true, but if it not, then $\phi_2$ is true, and 
if neither $\phi_1$ nor $\phi_2$ is true, then $\phi_3$ is true,
\ldots, and if none of $\phi_1, \ldots, \phi_{\ell-1}$ is true, then $\phi_\ell$
is true.  We give semantics to such formulas in an LPS $\vecmu = \<\mu_0, \ldots,
\mu_k\>$.%
\footnote{There is nothing special about the use of LPSs here.
  Battigalli and Sinischalchi use \emph{conditional probability
    systems} to define their notion of strong belief, and we could
  equally well use conditional probability systems here.  We could
  also easily use \emph{nonstandard probability measures}.  Readers
  familiar with these representations of uncertainty (see \cite{Hal26}
  for an overview) should have no difficulty giving analogues of our
  semantic definitions using these alternative approaches.}

To give semantics to generalized belief, we use 
\emph{LPS structures}, that is, structures of the form $M =
(\Omega,\strat,\F, \PR_1, \ldots, \PR_n)$, where now $\PR_i$
associates with each state an LPS.  
To define the semantics of the generalized belief operator, we need to
recall the definition of conditioning in LPSs \cite{BBD1}.  
For simplicity, we restrict our attention to structures $M$
where $\Omega$ is finite and $\F$ consists of all the subsets of
$\Omega$; that is, every set is measurable;
we refer to such structures as \emph{fully measurable}.
Given a measurable set $U$ and $\vecmu = \<\mu_0, \ldots, \mu_k\>$, define
$$\vecmu|U = \<\mu_{k_0}(\cdot \mid U), \mu_{k_1}(\cdot \mid U), \ldots
\>,$$ where $(k_0, k_1, \ldots)$ is the subsequence of all indices for which the
probability of $U$ is positive.  Formally, $k_0 = \min\{k: \mu_k(U) > 0\}$
and, if $\mu_{k_h}$ has been defined and there exists  an index $h'$
such that $k_h <h' \le k$ and $\mu_{h'}(U) > 0$, then $k_{h+1} = \min\{h':
\mu_{h'}(U) > 0, \, k_h < h' \le k\}$.  Note that $\vecmu|U$ is
undefined if $\vecmu(U) = \vec{0}$ (i.e., $\mu_j(U) = 0$ for $j=0,
\ldots, k$) and that the length of  the
sequence $\vecmu|U$ depends on $U$.   If $(\vecmu|U) = \<\mu_{k_0}, \ldots\>$, 
then we write $\vecmu(V \mid U)_0$ to denote $\mu_{k_0}(V \mid U)$,
the conditional probability according to the first
probability measure in the LPS $\vecmu|U$.  

If $M = (\Omega,\strat,\F, \PR_1, \ldots, \PR_n)$ and $\PR_i(\omega) =
\vec{\mu} = (\mu_0, \ldots, \mu_k)$, then
$$
\begin{array}{ll}
(M,\omega) \sat B_i(\phi_1, \ldots, \phi_\ell) \mbox{ if}\!\!\!\!
&\vec{\mu}(\neg \phi_1 \land \ldots \land \neg \phi_{\ell-1}) \ne
  \vec{0}, \   \mu_0(\intension{\phi_1}) = 1,\
    (\vecmu(\intension{\phi_2} \mid \intension{\neg\phi_1})_0 = 1,
  \\  &
  \ldots,\
    (\vecmu(\intension{\phi_\ell} \mid \intension{\neg\phi_1 \land
    \ldots \land \neg \phi_{\ell-1}})_0 = 1.
  \end{array}
  $$
That is, $\phi_1$ gets probability 1 at the top level, $\phi_2$ get
probability 1 at the top level conditional on $\phi_1$ being false, and so on.
(The first requirement, that $\vec{\mu}(\neg \phi_1 \land \ldots \land
\neg \phi_{\ell-1}) \ne \vec{0}$, ensures that all the conditional
probabilities are well defined.)

There is also a corresponding ``all I know'' operator,
$O^{\L}_i(\phi_1,\ldots, \phi_{\ell})$, which again is taken with
respect to a language  $\L$, defined as follows:
$$\begin{array}{ll}
  (M,\omega) \sat O^{\L}_i(\phi_1,\ldots, \phi_{\ell}) 
 \mbox{ if}\!\!\!\! &(M,\omega) \sat B_i(\phi_1,\ldots, \phi_{\ell}) \mbox{
  and,}\\ &\mbox{for all $\psi \in \L$, if } (M,\omega) \sat \Diamond(\phi_1
   \land \psi) \mbox{ then } 
\mu_0(\intension{\psi}) \ne  \vec{0} \mbox{ and,}\\ &\mbox{for all $h$ with $2 \le h
   \le \ell$, if } (M,\omega) \sat \Diamond(\neg \phi_1 \land \ldots
 \land \neg \phi_{h-1} \land \phi_h \land \psi)\\ &\mbox{\ \ \ \ \ \  then }
 (\vecmu(\intension{\psi } \mid \intension{\neg \phi_1
 \land \ldots \land \neg 
  \phi_{h-1}})_0 \ne \vec{0}.\end{array}$$
It is easy to see that the new definition $O^{\L}_i(\phi_1)$ is
identical to the earlier 
definition.  The generalized version $O^{\L}_i(\phi_1, \ldots,
\phi_k)$ requires all formulas $\psi$ consistent with $\phi$ to have
positive probability at the top level and, in addition, for $h \ge 2$,
all formulas $\psi$ consistent with $\neg \phi_1 \land \ldots
\land \neg \phi_{h-1} \land \phi_h$  must have positive probability at
the top level conditional on $\neg \phi_1 \land \ldots
 \land \neg \phi_{h-1}$.

Before going on, we briefly review how best response is defined in LPS
structures.  Since  
 player $i$'s beliefs at a state $\omega$ are defined by an LPS
 $(\mu_0, \ldots, \mu_k)$, we take the expected utility associated
 with $i$'s strategy $\strat_i(\omega)$ at $\omega$ to be a tuple
 $(u_0,\ldots,u_k)$, where $u_j$ is the expected utility of
 $\strat_i(\omega)$ with respect to probability $\mu_j$.  We can then
 compare two expected utilities lexicographically: $(u_0, \ldots, u_k)
 > (u_0', \ldots, u_k')$ if there exists a $j \le k$ such that $u_0 =
 u_0'$, \ldots, $u_{j-1} = u_{j-1}'$, and $u_j > u_j'$.  With this
 definition, we can still take $\RAT_i$ to hold at $\omega$ if
 $\strat_i(\omega)$ is a best response, given $i$'s about the
 strategies of other players at $\omega$.
 
  We can now define the formulas $\GRATzero^k_i$ (the $G$ stands for
``generalized'') inductively by taking $\GRATzero^0_i$ to be $\true$ and  
$\GRATzero^{k+1}_i$ to be an abbreviation of 
  $$\RAT_i \land O^{\L^0_{-i}}_i (\GRATzero^k_{-i},
\ldots, \GRATzero^0_{-i}).$$
That is, all agent $i$ knows is that the other agents are $k$-level
rational, but if they are not, then are $(k-1)$-level rational, and
if they are not, they are $(k-2)$-level rational, and so on.

\fullv{
  
\commentout{
Moreover, since $\mu_0(\intension{\GRATzero^k_{-i}}) = 1$
and for all $h$ with $0 \le h < k$,  
$(\vecmu|\intension{\land_{h=h'+1}^k \neg
  \GRATzero^h_{-i}})_0( \GRATzero^{h'}_{-i}) = 1$
it follows that $\vecmu$  
gives ``infinitely higher'' probability to strategy profiles all of
whose strategies are compatible with level $h$ rationality than to
strategy profiles  where at least one strategy is not compatible with
level $h$ rationality.  
That is, if for some $h \le
k$ we have that (a) there exists a state $\omega'$ such that
$\tau_{-i} = \strat_{-i}(\omega')$ and $(M,\omega') \sat
  \GRATzero^{h}$,
  (b) there is no state $\omega''$ and $h''$ with $h \le h' \le k$ such that
  $\strat_{-i}(\omega'') = \tau'_{-i}$ and $(M,\omega'') \sat
  \GRATzero^{h'}$, and (c) for all $k'' \le m$, if 
  $\mu_{k'}(\intension{\tau'_{-i}}) > 0$, then there
  then exists $k'' < k'$ such that $\mu_{k''}(\tau_{-i}) > 0$.
}
}
\thm\label{thm:charwdg} The following are equivalent:
\begin{itemize}
\item[(a)] the strategy $\sigma$ for player $i$ survives $k$ rounds of
iterated deletion of weakly dominated strategies in $\Gamma$;
\item[(b)] there exists
  a fully measurable
LPS structure $M^{k}$
appropriate for $\Gamma$ and a state $\omega^{k}$ in $M^{k}$ such that
$\strat_i(\omega^{k}) = \sigma$ and $(M^{k},\omega^{k}) \sat
\GRATzero^{k}_i$.%
\footnote{It follows from the proof that we can take all the LPSs in $M^k$
  to have length $k+1$.}
\end{itemize}
In addition,
if $\vec{\sigma} \in \NWD^k$, then
there is a 
fully measurable
LPS structure $\bar{M}^k =
(\Omega^k,\strat,\F, \PR_1^k, \ldots, \PR_n^k)$ such that 
$\Omega^k = \{(k',i,\vec{\sigma}): 0 \le k' \le k, 1 \le i \le n, \vec{\sigma}
\in \NWD^{k'}\}$,
$\strat^k(k',i,\vec{\sigma}) = 
\vec{\sigma}$, $\F^k = 2^{\Omega^k}$, and
for all states 
$(k', i, \vec{\sigma}) \in \Omega^k$,  
$(\bar{M}^k,(k',i, \vec{\sigma})) \sat \GRATzero_{-i}^{k'}$.
\ethm

The proof of this and other results can be found in the full paper.
However, we mention here one of the key propositions used in proving
the theorem, since it also gives some intuition for the $\GRATzero_j$
operator and will allow us to compare our results to those of others.
Suppose that $M$ is a model appropriate for a game $\Gamma$, 
$(M,\omega) \sat \GRATzero_i^{k+1}$, and $\PR_i(\omega) = \vec{\mu}$. 
Part (a) of the proposition says that player $i$ satisfies
cautiousness under $\vec{\mu}$ in the sense of Catonini and de Vito
\citeyear{CD16} and Perea \citeyear{Perea12}: for all strategy
profiles $\vec{\tau}_{-i} \in \Sigma_{-i}$, we have
$\vec{\mu}(\intension{\play(\tau_{-i})}) \ne \vec{0}$.
Part (b) says
that, if there are at least two players not all
of whose strategies survive iterated deletion, then the formulas
$\GRAT_i^h$ for $h = 1, 2, 3, \ldots$, are mutually exclusive.  Part
(c) says that for all $h \le k$, strategy profiles compatible with
$\GRATzero_{-i}^{h}$ are infinitely more likely those not compatible
with $\GRATzero_{-i}^h \lor \ldots \lor \GRATzero_{-i}^k$ under $\vec{\mu}$.  But our
sense of ``infinitely more likely than'' is weaker than that of
Blume, Brandenburger, and Dekel \citeyear{BBD1}, and closer in spirit
to that of Lo \citeyear{Lo99}.  Formally, we use the notion of domination,
where event $E$ \emph{$\mu$-dominates} $F$, 
written $E \gg_{\vec{\mu}} F$, if $\min \{\ell: \mu_\ell(E) > 0\} < 
\min \{\ell: \mu_\ell(F) > 0\}$ (where we take $\min(\emptyset) = \infty$).

  \pro\label{lem:GRAT}
  Suppose that $M$ is an appropriate model for game $\Gamma$,
  $(M,\omega) \sat \GRATzero_i^{k+1}$, and $\PR_i(\omega) = \vec{\mu}
  = (\mu_0, \ldots, \mu_m)$.
\begin{itemize}
  \item[(a)] For all strategy profiles $\vec{\tau}_{-i}
    \in\Sigma_{-i}$, we have 
    $\vec{\mu}(\intension{\play(\tau_{-i})}) \ne \vec{0}$. 
  \item[(b)] If $\NWD^1_j \ne \NWD^0_j$ for at least two players $j$,
    then $(M,\omega) \sat \neg \GRATzero^1_i \land
      \ldots \land \neg  \GRATzero^{k}_i$.
    \item[(c)] If $h < h' \le k$, then $\intension{H^{h'}_{-i}}
      \gg_{\mu} \intension{H^{h}_{-i}}$, 
      where $H^{k'}_{-i}$ is an abbreviation of the formula
$\GRAT_{-i}^{k'} \land \neg
      \GRAT_{-i}^{k'+1} \land \ldots \land \neg \GRAT_{-i}^{k}$ (so
      $H^k_{-i}$ is $\GRAT_{-i}^k$).
      Moreover, 
      for all $h \le k$ and strategy profiles
      $\vec{\tau}_{-i}$ and $\vec{\tau}'_{-i}$, if
      $\intension{\GRAT_{-i}^h \land \play(\vec{\tau}_{-i})} \ne
  \emptyset$, and     $\intension{(\GRAT_{-i}^h \lor \ldots \lor
    \GRAT_{-i}^k) \land 
  \play(\vec{\tau}_{-i}')} =  \emptyset$, then
$\intension{\play(\vec{\tau}_{-i})} 
      \gg_{\vec{\mu}} \intension{\play(\vec{\tau}'_{-i})}$. 
\end{itemize}
\epro

It follows from part (b) that if some strategies of at least two players
are weakly dominated, then the analogue of common assumption of rationality
cannot hold.  There is no state where $\GRAT^k_i$ holds for all $k$;
indeed, there is not even a state where $\GRAT^k_i$ holds for all
sufficiently large $k$.  (The same comment applies to the $\RATzero^k_i$
operators used in Theorem~\ref{thm:charwd}.)
By  way of contrast, Catonini and de Vito
\citeyear{CD16} and Perea \cite{Perea12} show that their variants of
common assumption do hold, while for BFK, $k$-level assumption for all $k$
larger that some $k^*$ holds (but which $k^*$ it is depends on the game
$\Gamma$).  Unlike Catonini and DeVito and BFK, but like Perea, we are
able to characterize IA using only finite structures.
Perhaps the biggest difference between Perea's characterization and
ours is that we have different notions of caution.  For Perea's notion of
\emph{$(k+1)$-fold assumption of rationality} to hold for player $i$ at a state
  $\omega$, each strategy profile
  $\vec{\tau}_{-i}$ compatible with a $k$-fold assumption 
  of rationality must get positive probability (i.e., if
  $\Pr_i(\omega) = \vec{\mu} = (\mu_0,\ldots, \mu_m)$, then
  $\mu_h(\vec{\tau}_{-i}) > 
  \vec{0}$ for some $h$).  On the other hand, if $\GRAT_i^{k+1}$
  holds at $\omega$, then  
  for each strategy profile $\vec{\tau}_{-i}$ compatible with
  $\GRAT_{-i}^{k}$ we have $\mu_0(\vec{\tau}_{-i}) > 0$.  The fact that we
  require  $\mu_0(\vec{\tau}_{-i}) > 0$ rather than just
  $\mu_h(\vec{\tau}_{-i}) > 0$ for some $h$ will play an important role in
  the characterization of IA given in the next section that uses only
  standard probability.  Perea's approach does not lead to an obvious
  analogue of that result.

\section{Using approximate belief and probability structures}\label{sec:approx}
While the approach described in Section~\ref{sec:second}
deals with Samuelson's concern, it does so by assuming that
the agents' beliefs are characterized by LPSs.  Our earlier approach
characterized IA using (standard) probability structures, but did not
deal with Samuelson's concerns.  We now show that we can characterize
IA using standard probability structures, while still dealing with
Samuelson's concern, by considering approximate belief in an appropriate sense.

\commentout{
We start by adding an operator enabling us to reason about probabilistic
beliefs.
Let $\L^2(\Gamma)$ be the language that extends $\L^0(\Gamma)$ by allowing
formulas of the form $\pr_i(\phi \mid \psi) \ge \alpha$ where $\alpha$ is a rational number in $[0,1]$
(and then closing off under conjunction and negation);
$\pr_i(\phi \mid \psi) \ge \alpha$ can be read as ``the probability of
$\phi$ conditioned on $\psi$
according to $i$ is at least $\alpha$''. 
Given a probability structure $(\Omega,\strat,\F, \PR_1, \ldots,
\PR_n)$, we define semantics of $\pr_i$ as follows:
\begin{itemize}
\item $(M,\omega) \sat \pr_i(\phi \mid \psi)\ge \alpha$ iff 
there exists some  measurable $F_1,F_2 \in \F$ such that $F_1 \subseteq
\intension{\phi \land \psi}$, $\intension{\psi} \subseteq F_2$,
  $\PR_i(\omega)(F_2) \ge 0$ and
  $$\frac{\PR_i(\omega)(F_1)}{\PR_i(\omega)(F_1)}  \geq \alpha.$$
\end{itemize}
We analogously define $\pr_i(\phi \mid \psi) > \alpha$ and let 
$\pr_i(\phi)$ be an abbreviation for $\pr_i(\phi \mid \true)$.
We next introduce a more quantitative version of the ``all I
know'' operator.
Define $O^{\L,\nu,\epsilon}_i\phi$ (read ``all agent $i$ approximately
knows with respect to $\L$ is $\phi$'')
to be
an abbreviation for 
$$(\pr_i (\phi) \geq 1-\epsilon) \land (\land_{\psi \in \L} (\Diamond
(\phi \land \psi) \rimp \pr(\psi \mid \phi) \geq \nu)).$$
Note that this is different from our earlier definition of ``all i
know'' in two aspects: a) first, $i$ only needs to assign probability
$1-\epsilon$ to $\phi$---that is, it is ``almost certain'' that $\phi$
holds; b) we lower bound the probability that that $\psi$ conditioned on $\phi$ holds
by $\nu$ (as opposed to simply requiring it to be positive).
$O^{\L,\nu,\epsilon}_i\phi$ means that $i$ is almost certain that $\phi$
holds, and if $i$ were to find out that $\phi$ indeed does hold, then
this is all that $i$ knows with respect to $\L$.

For our characterization, we will also need a
conditional notion of ``all I know''.
Define $O^{\L,\delta,\eps}_i(\phi  \mid \theta)$ (read ``if agent $i$ were to
find out that $\theta$ holds, then all agent $i$ approximately knows
with respect to $\L$ 
is $\phi$'') to be an abbreviation for
$$(\pr_i (\theta) > 0) \land (\pr_i (\phi \mid \theta) \geq
1-\epsilon) \land (\land_{\psi \in \L} (\Diamond 
(\phi \land \psi \land \theta) \rimp \pr(\psi \mid \phi \land \theta) \geq \nu)).$$

Finally, let $O_i^{\L,\delta,\eps}(\phi_1,\ldots, \phi_{\ell})$ be an
abbreviation for
$$O_i^{\L,\delta,\eps}(\phi_1) \land O_i^{\L,\delta,\eps}(\phi_2 \mid \neg
\phi_1),\ldots, 
O_i^{\L,\delta,\eps}(\phi_\ell \mid \neg
\phi_1 \land \ldots\land \neg \phi_{\ell-1})$$

We now define the formulas $\GRAT_i^k(\delta,\epsilon)$ in exactly the
same way as 
$\GRATzero^k$ except that we replace the LPS-based $O_i^{\L}$$ operator with 
$O_i^{\L}(\delta,\epsilon)$
$\GRAT^{k}_{-i}(\epsilon,\nu)$ to be an abbreviation of 
$\land_{j \ne i}\GRAT^k_{j}(\epsilon,\nu)$
and $\GRAT^{k+1}_i(\epsilon,\nu)$ to be an abbreviation of 
$$\RAT_i \land O^{\L^0_{-i}, \epsilon,\nu}_i (\GRAT^k_{-i}(\epsilon,\nu),
\ldots, \GRAT^0_{-i}(\epsilon,\nu)).$$
That is, all agent $i$ approximately knows is that all other
agents are $k$-level rational, but if $i$ were to find out that they
are not, then all $i$ approximately knows is that they are
$(k-1)$-level rational and so on.
We omit the parenthetical $(\epsilon,\nu)$ whenever they are clear from the context.
}
We start with a quantitative analogues of the 
belief operators $B_i, \dbi$ and also define a 
conditional belief operators. Just as we did in the previous section, for simplicity, we restrict
our attention to fully measurable structures.
If $M = (\Omega,\strat,\F, \PR_1, \ldots, \PR_n)$ is a fully
measurable probability
structure, then
\begin{itemize}
\item $(M,\omega) \sat B_i^\delta  \phi$ 
if $\PR_i(\omega)(\intension{\phi}) \geq 1-\delta$, and $(M,\omega)
\sat \dbi^\delta  \phi$ if  
$\PR_i(\omega)(\intension{\phi}) \geq \delta$,
\item $(M,\omega) \sat B_i^\delta  (\phi \mid \theta)$ if 
    $\PR_i(\omega)(\intension{\theta} ) > 0$ and
$\PR_i(\omega)(\intension{\phi} \mid \intension{\theta} ) \geq
1-\delta$, and analogously for $\dbi^\delta(\phi \mid \theta)$.
\end{itemize}
That is, $B_i^\delta  \phi$ means that player $i$ is ``almost certain''
that $\phi$ holds---$i$ assigns probability at least $1-\delta$ to
$\phi$ holding---and $B_i^\delta  (\phi \mid \theta)$ means that if $i$
learns that $\theta$ holds, then $i$ is almost certain that
$\phi$ holds. 

The analogous ``all I approximately know'' operator,
$O^{(\L,\delta,\epsilon)}_i$, takes two
parameters, $\delta$ and $\epsilon$.  As with the approximate belief
operator $B_i^{\delta}$, the $\delta$ tells
us how close to 1 agent $i$'s 
beliefs have to be.  The $\epsilon$ gives us a lower bound on how
likely each formula in $\L$ consistent with what is believed must be.
Again, we also consider a conditional version of the operator.
Define $O^{\L,\nu,\epsilon}_i\phi$ (read ``all agent $i$ approximately
knows with respect to $\L$ is $\phi$'')
to be
an abbreviation for 
$$B^\delta_i (\phi) \land (\land_{\psi \in \L} (\Diamond
(\phi \land \psi) \rimp \dbi^{\epsilon}(\psi))).$$
and define $O^{\L,\delta,\eps}_i(\phi  \mid \theta)$ (read ``if agent $i$ were to
find out that $\theta$ holds, then all agent $i$ approximately knows
with respect to $\L$ 
is $\phi$'') to be an abbreviation for
$$B^{\delta}_i (\phi \mid \theta) \land (\land_{\psi \in \L} (\Diamond
(\phi \land \psi \land \theta) \rimp \dbi^{\epsilon}(\psi \mid \theta))).$$
Finally, let $O_i^{\L,\nu,\eps}(\phi_1,\ldots, \phi_{\ell})$ be an
abbreviation for
$$O_i^{\L,\delta,\eps}(\phi_1) \land O_i^{\L,\delta,\eps}(\phi_2 \mid \neg
\phi_1) \land \ldots \land
O_i^{\L,\delta,\eps}(\phi_\ell \mid \neg
\phi_1 \land \ldots\land \neg \phi_{\ell-1}).$$
To relate this definition to the definition in LPS structures,
let $B_i^\delta(\phi_1, \ldots, \phi_\ell)$ be an abbreviation for
$$B_i^{\delta}(\phi_1) \land B_i^{\delta}(\phi_2 \mid \neg
\phi_1) \land \ldots \land
B_i^{\delta}(\phi_\ell \mid \neg
\phi_1 \land \ldots\land \neg \phi_{\ell-1}).$$
Note that
$$\begin{array}{ll}
  (M,\omega) \sat O^{\L,\delta,\epsilon}_i(\phi_1,\ldots, \phi_{\ell}) 
 \mbox{ iff}\!\!\!\! &(M,\omega) \sat B_i^\delta(\phi_1,\ldots,
                      \phi_{\ell}) \mbox{
   and,}\\ &\mbox{for all $\psi \in \L$, if } (M,\omega) \sat \Diamond(\phi_1
 \land \psi) \mbox{ then } 
  \mu(\intension{\psi\mid \phi_1}) > \epsilon \mbox{ and,}\\ &\mbox{for all $h$ with $2 \le h
   \le \ell$, if } (M,\omega) \sat \Diamond(\neg \phi_1 \land \ldots
 \land \neg \phi_{h-1} \land \phi_h \land \psi)\\ &\mbox{\ \ \ \ \ \  then }
 (\mu(\intension{\psi} \mid \intension{\neg \phi_1 \land \ldots \land \neg
 \phi_{h-1}})_0 > \epsilon.\end{array}$$
Thus, $O^{\L,\delta,\epsilon}_i(\phi_1,\ldots, \phi_{\ell})$ really is
the ``approximate'' analogue  of $O^{\L}_i(\phi_1,\ldots, \phi_{\ell})$.
We now define the formulas $\GRAT_i^{k,\delta,\epsilon}$ in exactly the
same way as 
$\GRATzero^k$ except that we replace the LPS-based $O_i^{\L}$ operator with 
$O_i^{\L,\delta,\epsilon}$.
In more detail, define $\GRAT^{0,\delta,\epsilon}_i$ to be $\true$, and
$\GRAT^{k,\delta,\epsilon}_{i}$ to be an abbreviation of
$$\RAT_i \land O^{\L^0_{-i},\delta,\epsilon}_i (\GRAT^{k,\delta,\epsilon}_{-i},
\ldots, \GRAT^{0,\delta,\epsilon}_{-i}).$$
That is, all agent $i$ approximately knows is that all other
agents are $k$-level rational, but if $i$ were to find out that they
are not, then all $i$ approximately knows is that they are
$(k-1)$-level rational and so on.

\thm\label{thm:charwdg1} For all finite games $\Gamma$ and all
sufficiently small $\epsilon > 0$, there 
   exists some $\delta > 0$ such that 
the following are equivalent: 
\begin{itemize}
\item[(a)] the strategy $\sigma$ for player $i$ survives $k$ rounds of
iterated deletion of weakly dominated strategies in $\Gamma$;
\item[(b)] there exists a 
fully measurable structure  $M^{k}$
appropriate for $\Gamma$ and a state $\omega^{k}$ in $M^{k}$ such that
$\strat_i(\omega^{k}) = \sigma$ and $(M^{k},\omega^{k}) \sat
\GRAT^{k,\delta,\epsilon}_i$.
\end{itemize}
In addition,
if $\vec{\sigma} \in \NWD^k$, then
for all sufficiently small $\epsilon > 0$, there exists some $\delta >
0$ and 
a fully measurable structure $\bar{M}^{k,\delta} =
(\Omega^{k,\delta},\strat^{k,\delta},\F^{k,\delta}, \PR_1^{k,\delta},
\ldots, \PR_n^{k,\delta})$ such that  
$\Omega^{k,\delta} = \{(k',i,\vec{\sigma}): 0 \le k' \le k, 1 \le i \le n, \vec{\sigma}
\in \NWD^{k'}\}$,
$\strat^{k,\delta}(k',i,\vec{\sigma}) = 
\vec{\sigma}$, $\F^{k,\delta} = 2^{\Omega^k}$, 
and
for all states 
$(k', i, \vec{\sigma}) \in \Omega^{k,\delta}$,   we have 
$(\bar{M}^{k,\delta},(k',i, \vec{\sigma})) \sat
\GRAT_{-i}^{k',\delta,\epsilon}$.%
\footnote{How small $\epsilon$ has to be depends only on the game.  
Although the choice of
  $\delta$ depends on the choice of $\epsilon$, $\epsilon$ plays no role in the
construction of $\bar{M}^{k,\delta}$ (which is why we did not write
$\bar{M}^{k,\delta,\epsilon}$).}
\ethm


\commentout{
The idea is that given
a probability structure $M = (\Omega,\strat,\F, \PR_1, \ldots,
\PR_n)$, 
as in the first approach, we construct an LPS structure
$M^+ = (\Omega,\strat, \F,\PR_1^+, \ldots, \PR_n^+)$, where
$\Pr_i^+(\omega)$ associates with state $\omega$ an LPS that can be
viewed as a ``canonical extension'' of $\Pr_i(\omega)$,   
and then essentially apply the second approach to $M^+$.

Suppose that $\Pr_i(\omega) = \mu^{\omega,i}$.  Then
$\Pr_i^+(\omega)=  (\mu^{\omega,i})^+ =
((\mu^{\omega,i}_0)^+,(\mu^{\omega,i}_1)^+, \ldots)$. 
We define $(\mu_h^{\omega,i})^+$ inductively.  We take
$(\mu_0^{\omega,i})^+ = \mu^{\omega,i}$, 
so that agent $i$'s top-level beliefs with $\Pr_i^+(\omega)$ are given by
$\Pr_i(\omega)$.  The intuition behind $(\mu_1^{\omega,i})^+$ is that, since $i$
views each other agent $j$ as rational, $i$ does not want to discount
$j$'s beliefs completely.   Player $i$ takes into account other
players' beliefs by setting  $(\mu_1^{\omega,i})^+(\omega')$ to be
the average belief 
that the other agents ascribe to $\omega'$; that is,
$(\mu_1^{\omega,i})^+(\omega') = \sum_{j \ne i} \mu^{\omega,j}(\omega')/(n-1)$.
To define $(\mu_2^{\omega,i})^+$, we want to consider $i$'s beliefs
about other agents' beliefs about other agents' beliefs.
That is, for each state $\omega'$ and each player $j \ne i$, player
$i$ wants to take into account $j$'s beliefs about others' beliefs at
$\omega'$.  But $(\mu_1^{\omega',j})^+$ already takes into account
$j$'s beliefs about others at $\omega'$.  Thus, we take
$$(\mu_2^{\omega,i})^+ = \sum_{j \ne i} (\sum_{\omega' \in \Omega}
\mu^{\omega,i}(\omega')(\mu_1^{\omega',j})^+)/(n-1),$$
and, more generally,
\begin{equation}\label{eq:cautious}
(\mu_{h+1}^{\omega,i})^+ = \sum_{j \ne i} (\sum_{\omega' \in \Omega}
  \mu^{\omega,i}(\omega')(\mu_h^{\omega',j})^+)/(n-1).
  \end{equation}

We can now prove the following analogue of Theorem~\ref{thm:charwdg}:
\thm\label{thm:charwdg1} The following are equivalent:
\begin{itemize}
  \item[(a)] the strategy $\sigma$ for player $i$ survives $k$ rounds of
iterated deletion of weakly dominated strategies in game $\Gamma$;
\item[(b)] there exists an $\L^0$-measurable probability structure  $M^{k}$
appropriate for $\Gamma$ and a state $\omega^{k}$ in $M^{k}$ such that
$\strat_i(\omega^{k}) = \sigma$ and $((M^{k})^+,\omega^{k}) \sat
\GRATzero^{k}_i$.   
\end{itemize}
In addition,
if $\vec{\sigma} \in \NWD^k$, then
there is a finite probability structure $\bar{M}^k =
(\Omega^k,\strat,\F, \PR_1^k, \ldots, \PR_n^k)$ such that 
$\Omega^k = \{(k',i,\vec{\sigma}): 0 \le k' \le k, 1 \le i \le n, \vec{\sigma}
\in \NWD^{k'}\}$,
$\strat(k',i,\vec{\sigma}) = 
\vec{\sigma}$, $\F = 2^{\Omega^k}$, and
for all states 
$(k', i, \vec{\sigma}) \in \Omega^k$,  
$((\bar{M}^k)^+,(k',i, \vec{\sigma})) \sat \land_{j \ne i} \GRATzero_j^{k'}$.
\ethm

\prf The fact that (b) implies (a) follows
immediately from Theorem~\ref{thm:charwdg}.  Now suppose that 
$\vec{\sigma} \in \NWD^k$.  By
Theorem~\ref{thm:charwd}, there exists
there is a finite structure $M^k =
(\Omega^k,\strat,\F, \PR_1^k, \ldots, \PR_n^k)$ such that 
$\Omega^k = \{(k',\vec{\sigma}): 0 \le k' \le k, \vec{\sigma} \in X^{k'}_1
\times \cdots \times X^{k'}_n\}$, $\strat(k',\vec{\sigma}) =
\vec{\sigma}$, $\F = 2^{\Omega^k}$, where $X^{k'}_j$ consists of all
strategies for player $j$ that survive $k$ rounds of iterated deletion
of weakly dominated strategies, 
and, for all states $(k', \vec{\sigma}) \in \Omega^k$,  
$(M^k,(k',\vec{\sigma})) \sat \RATzero_1^{k'} \land \ldots \land
\RATzero_n^{k'}$.  It clearly suffices to show that
$((\bar{M}^k)^+,(k',i, \vec{\sigma})) \sat \land_{j \ne i} \GRATzero_j^{k'}$.
To do this, we first show that
if $\PR_j^+((k',i, \vec{\sigma}) = (\mu_0^{k',i,\vec{\sigma},j},
\mu_1^{k',i,\vec{\sigma},j}, \ldots$, and if $1 \le h \le k'$, then 
$\mu_h^{k',i,\vec{\sigma},j}$ has support consisting of all states of
the form $(k'',i',\vec{\tau}) \in \Omega^k$ such that $k'' - h \le k''
\le k'$.  We proceed by induction on $h$.  The case that $h=0$ follows
from the definition of $\PR_j^+$, and the general case is immediate
from the induction hypothesis, the observation made in the course of
proving Theorem~\ref{thm:charwd}.
that the support of
$\PR_j(k',j,\vec{\sigma})$ consists of all states of the form
$(k'-1,j,\vec{\tau})$ such that $\tau_j = \sigma_j$, while if $i' \ne
j$, then the support of
$\PR_j(k',i',\vec{\sigma})$ consists of all states of the form
$(k',j,\vec{\tau})$ such that $\tau_j = \sigma_j$.  
From the definition of $M^k$, we have that if 
$(k',i, \vec{\sigma}) \in \Omega^k$ then $\vec{\sigma} \in \NWD^{k'}$,
and for every profile $\vec{\sigma} \in \NWD^{k'}$, every state of the
  form $(k,i,\vec{\sigma}) \in \Omega^{k}$.  A straightforward
  induction on $k'$ now shows that 
$((\bar{M}^k)^+,(k',i, \vec{\sigma})) \sat \land_{j \ne i} \GRATzero_j^{k'}$.
\eprf

We can also characterize admissible strategies more directly using
probability structures.  For this, we need some definitions.
Define the
formula $\RAT_i^+$ by taking $\RAT_i^+$ to be true
at a state $\omega$ in a probability structure $M$ if $\RAT_i$ is true
at $\omega$ in the 
corresponding LPS structure.  More precisely:
$$(M,\omega) \sat \RAT_i^+ \mbox{ iff } (M^+,\omega) \sat \RAT_i.$$
Now define the formulas $(\RATzeropi)^k$ 
inductively much as we defined $\RATzero_i^k$, except that (a) we replace
$\RAT+i$ by $\RAT_i^+$ and (b) we do not need the $\PLAYCONzero$
conjunct.  That is, 
$(\RATzeropi)^0$ is $\true$ and  
$(\RATzeropi)^{k+1}$ is an abbreviation of 
$$\RAT_i^+ \land 
O^{\L^0_{-i}}_i (\land_{j \ne i}(\RATzeropj)^k).$$

\thm\label{thm:charwd2} The following are equivalent:
\begin{itemize}
\item[(a)] the strategy $\sigma$ for player $i$ survives $k$ rounds of
iterated deletion of weakly dominated strategies in game $\Gamma$;
\item[(b)] there is a an $\L^0$-measurable probability structure  $M^{k}$
appropriate for $\Gamma$ and a state $\omega^{k}$ in $M^{k}$ such that
$\strat_i(\omega^{k}) = \sigma$ and $(M^{k},\omega^{k}) \sat (\RATzeropi)^k$.  
\end{itemize}
In addition,
if $\vec{\sigma} \in \NWD^k$, then
there is a finite structure $\bar{M}^k =
(\Omega^k,\strat,\F, \PR_1^k, \ldots, \PR_n^k)$ such that 
$\Omega^k = \{(k',i,\vec{\sigma}): 0 \le k' \le k, 1 \le i \le n, \vec{\sigma}
\in \NWD^{k'}\}$,
$\strat(k',i,\vec{\sigma}) = 
\vec{\sigma}$, $\F = 2^{\Omega^k}$, and
for all states 
$(k', i, \vec{\sigma}) \in \Omega^k$,  
$(\bar{M}^k,(k',i, \vec{\sigma})) \sat \land_{j \ne i} (\RATzeropj)^{k'}$.
\ethm

\prf  As usual, we proceed by induction on $k$, proving the equivalence
of (a) and (b) and the existence of the required structure $\bar{M}^k$.
Again, the result is trivial for $k=0$.  Suppose that the result holds
for $k$; we prove it for $k + 1$.

We first show that (a) implies (b).  So suppose that $\vec{\sigma} \in
\NWD^k$.  Let  $\bar{M}^{k+1}$ to be the structure 
constructed in the proof of Theorem~\ref{thm:charwd}.  We show by
induction on $k'$ that for all states $(k',i,\vec{\sigma}') \in
\Omega^{k+1}$, we have that $(M,(k',i,\vec{\sigma}') \sat \land_{j \ne
  i} (\RATzeropj)^k$.  The result is trivially true if $k' = 0$.
Suppose it is true for $k'$; we prove it for $k'+1$.

in $M$, we have 
 $(M,\omega) \sat (\RATzeropi)^k$ iff  $(M^+,\omega) \sat \GRATzero^{k}_i$.  
We prove the result by induction on $k$.  It is trivially true if
$k=0$ and immediate from the definition of $\RAT^+$ if $k-0$.
\eprf

}

\fullv{
\section{Richer Languages}\label{sec:strong}
The formula $\RATzero^{k+1}_i$ was defined with respect to the language 
$\L^0_{-i}$, and thus required that player $i$ assign positive
probability to all and only
strategies consistent with $\RATzero^{k}_{-i}$. 
This is in the spirit of Pearce's definition of weak dominance.
By considering ``all I know'' with respect to $\L^0_{-i}$, in the
second approach, we get beliefs on the space of strategies.

What happens if we consider different languages?  As we saw, if we
consider ``all I know'' with respect to the empty language, we recover
the standard characterization if iterated deletion of
rationalizability.  Our goal in this section is to consider a stronger
language, one that leads us to a result more in the spirit of BFK, so
that we end up with full-support beliefs on spaces that talk about
players' beliefs as well as strategies.

Let $\L^2(\Gamma)$ be the language that extends $\L^0(\Gamma)$ by allowing
formulas of the form $\pr_i(\phi) \ge \alpha$ and $\pr_i(\phi) >
\alpha$, where $\alpha$ is a rational number in $[0,1]$
(and then closing off under conjunction and negation);
$\pr_i(\phi) \ge \alpha$ can be read as ``the probability of $\phi$
according to $i$ is at least $\alpha$'', and similarly for $\pr_i(\phi) >
\alpha$.  We allow
nesting here, so that we can have a formula of the form
$\pr_j(\play_i(\sigma) \land \pr_k(\play_i(\sigma')) > 1/3) \ge 1/4$.  
As we would expect,
\begin{itemize}
\item $(M,\omega) \sat \pr_i(\phi)\ge \alpha$ iff
  $\PR_i(\omega)(\intension{\phi}) \ge \alpha$; and
\item $(M,\omega) \sat \pr_i(\phi) > \alpha$ iff
  $\PR_i(\omega)(\intension{\phi}) > \alpha$. 
\end{itemize}
The restriction to $\alpha$ being rational allows the language to be
countable 
(see the end of this section for further discussion of this point).
As before, we omit the parenthetical $\Gamma$ when the game $\Gamma$ is clear from
context or irrelevant.
Note that $\L^2$ is more expressive that $\L^1$; we can view $B_i\phi$
as an abbreviation for $\pr_1(\phi) \ge 1$.

\commentout{
However, as we show shortly, it is not too serious a 
restriction.  

Let $\L^4(\Gamma)$ be the language that extends $\L^2(\Gamma)$ by closing off 
under (1) countable conjunctions, so that if $\phi_1, \phi_2, \ldots$ are
formulas, then so is $\land_{m=1}^\infty \phi_m$, and (2) formulas of the form
$\pr_i(\phi) > \alpha$, where $\alpha$ is a real number in $[0,1]$.
(We can express $\pr_i(\phi) \ge \alpha$ as the countable 
conjunction $\land_{\beta < \alpha, \beta \in Q \inter [0,1]}
\pr_i(\phi) > \beta$, where $Q$ is the set of rational numbers, so there
is no need to include formulas of the form $\pr_i(\phi) \ge \alpha$
explicitly in $\L^4(\Gamma)$.)   We again omit the parenthetical $\Gamma$ in
$\L^3(\Gamma)$ and $\L^4(\Gamma)$ when the game $\Gamma$ is clear from
context
or irrelevant.

A subset $\Phi$ of $\L^3$ is
$\L^3$-\emph{realizable} if there 
exists an appropriate structure $M$ for $\Gamma$ and state $\omega$ in
$M$ such that, for all formulas $\phi \in \L^3$,  $(M,\omega) \sat 
\phi$ iff $\phi \in \Phi$.%
\footnote{For readers familiar with standard completeness proofs in
modal logic, if we had axiomatized the logic we are implicitly using
here, the $\L^3$-realizable sets would just be the maximal consistent
sets in the logic.}
We can similarly define what it means for a subset of $\L^4$ to be
$\L^4$-realizable.

\lem\label{lem:realizable} Every $\L^3$-realizable set can be uniquely
extended to an $\L^4$-realizable set. \elem

\fullv{
\prf It is easy to see that every $\L^3$-realizable set can be extended 
to an $\L^4$-realizable set.  For suppose that $\Phi$ is
$\L^3$-realizable.  Then there is some state $\omega$ and structure $M$
such that, for every formula $\phi \in \L^3$, we have that $(M,\omega)
\sat \phi$ iff $\phi \in \Phi$.   Let $\Phi'$ consist of the $\L^4$
formulas true at $\omega$.  Then clearly $\Phi'$ is an $\L^4$-realizable
set that extends $\Phi$.

To show that the extension is unique, suppose that there are two
$\L^4$-realizable sets, say $\Phi_1$ and $\Phi_2$, that extend $\Phi$.
We want to show that $\Phi_1 = \Phi_2$.  To do this, we consider two
language, $\L^5$ and $\L^6$, intermediate between $\L^3$ and $\L^4$.  

Let $\L^5 $ be the language that extends $\L^2$ by 
closing off 
under countable conjunctions and formulas of the form 
$\pr_i(\phi) > \alpha$, where $\alpha$ is a rational number in $[0,1]$.
Thus, in $\L^5$, we have countable conjunctions and disjunctions, but
can talk explicitly only about rational probabilities.  Nevertheless,
it is easy to see that for every formula $\phi \in \L^4$, there
is an formula equivalent formula $\phi' \in \L^5$, since if
$\alpha$ is a real number, then $\pr_i(\phi) > \alpha$ is equivalent   
to $\lor_{\beta > \alpha, \, \beta \in [0,1]\inter Q\,} \pr_i(\phi) > \beta$
(an infinite disjunction $\lor_{i=1}^\infty \phi_i$ can be viewed as an
abbreviation of $\neg  \land_{i=1}^\infty \neg \phi_i$).

Next, let $\L^6$ be the result of closing off formulas in
$\L^3$ under countable conjunction and disjunction.  Thus, in
$\L^6$, we can apply countable conjunction and disjunction only at the
outermost level, not inside the scope of $\pr_i$.  
We claim that for every formula $\phi \in \L^5$, there is an
equivalent formula in $\L^6$.  More precisely, for every formula
$\phi \in \L^5$, there exist
formulas $\phi_{ij} \in \L^3$, $1 \le i,j  < \infty$ such that 
$\phi$ is equivalent to $\land_{m=1}^\infty \lor_{n=1}^\infty
\phi_{mn}$.   We prove this by induction on the structure of $\phi$.
If $\phi$ is $\RAT_i$, $\play_i(\sigma)$, or $\true$, then the statement
is clearly true.  The result is immediate from the induction hypothesis
if $\phi$ is a countable conjunction.  If $\phi$ has the form $\neg
\phi'$, we apply the induction hypothesis, and observe 
that $\neg (\land_{m=1}^\infty \lor_{n=1}^\infty \phi_{mn})$ is
equivalent to $\lor_{m=1}^\infty \land_{n=1}^\infty \neg \phi_{mn}$.  
We can convert this to a conjunction of disjunctions by distributing the
disjunctions over the conjunctions in the standard way (just as 
$(E_1 \inter E_2) \union (E_3 \inter E_4)$ is equivalent to
$(E_1 \union E_3) \inter (E_1 \union E_4) \inter (E_2 \union E_3) \inter
(E_2 \union E_4)$).  Finally, if $\phi$ has the form $\pr_i(\phi') > 
\alpha$, we apply the induction hypothesis, and observe that 
$\pr_i(\land_{m=1}^\infty \lor_{n=1}^\infty \phi_{mn}) > \alpha$ is
equivalent to $$\lor_{\alpha' > \alpha, \alpha' \in Q \inter [0,1]}
\land_{M=1}^\infty \lor_{N=1}^\infty \pr_i(\land_{m=1}^M \lor_{n=1}^N
\phi_{mn}) > \alpha'.$$

The desired result follows, since if two states agree on all formulas in
$\L^3$, they must agree on all formulas in $\L^6$, and hence on all
formulas in $\L^5$ and $\L^4$. \eprf

The choice of language turns out to be significant for a
number of our results; we return to this issue at various points below.
}

\lem\label{L3meas} If $M$ is an $\L^3$-measurable structure, and $\phi$
is a formula in $\L^6$, then $\intension{\phi}$ is measurable.
\elem

\prf An easy induction on the structure of formulas shows that
$\intension{\phi}$ is measurable for all formulas $\phi \in \L^6$.  For
example $\intension{\pr_i(\psi) > \alpha} = \union_{r \in {\bf Q}: r >
\alpha} \intension{\pr_i(\psi) > r}$, and so is the union of measurable
sets, and thus must be measurable. 
\eprf

It follows from Lemma~\ref{L3meas} that the notions of $\L^3$-measurability,
$\L^4$-measurability, $\L^5$-measurability, and $\L^6$-measurability
all coincide.

With this background, 
}
Let $\L^2_i$ consist of all formulas in
$\L^2$ of the form $\pr_i(\phi) \ge \alpha$ and $\pr_i(\phi) >
\alpha$ ($\phi$ can mention $\pr_j$, $j\neq i$; it is only the outermost
modal operator that must be $i$).  Intuitively, $\L^2_i$
consists of the formulas describing $i$'s beliefs.  Let 
$\L^2_{i+}$ consist of $\L^2_i$ together with formulas of the form
$\true$, $\RAT_i$, and $\play_i(\sigma)$, for $\sigma \in\S_i$.  
Let $\L^2_{(-i)+}$ be an abbreviation of $\union_{j \ne i} \L^2_{j+}$.

Note that if $\phi \in \L^2_{(-i)+}$, then $O_i^{L^2_{(-i)+}} \phi$
is an abbreviation of the
formula  
$$B_i \phi \land (\land_{\psi \in
\L^2_{(-i)+}}  \Diamond(\phi \land \psi) \rimp \dbj \psi).$$  
Thus, $O^{\L^2_{(-i)+}}_i \phi$ holds if agent $i$ believes $\phi$ but does not know
anything beyond that; he ascribes positive probability to all formulas
in $\L^2_{(-i)+}$ consistent with $\phi$.
\commentout{
Of course, we could go further and define a notion of ``all $i$ knows''
for the language $\L^4$.  
Doing this would give a definition that is 
even closer to that of Halpern and Lakemeyer.
Unfortunately, we cannot require than 
agent $i$  ascribe positive probability to all the formulas  in
$\L^4_{(-i)+}$ consistent with $\phi$; in general, there will be
an uncountable number of distinct and mutually exclusive formulas
consistent with $\phi$, so they
cannot all be assigned positive probability.  This problem does not
arise with $\L^2$, since it is a countable language.  
Halpern and Lakemeyer could allow an agent to consider an uncountable
set of worlds possible, since they were not dealing with probabilistic
systems.  
In the sequel, we thus focus on the language $\L^2$ and let 
$O_i$ denote $O^{\L^2_{(-i)+}}_i$.%
\footnote{Note that the modal operator $\Diamond$ is not in the language
$\L^3$ or $\L^4$.  None of our results would be affected if we had
considered a language that also included $\Diamond$; for ease of
exposition, we have decided not to include $\Diamond$ here.}
}

We again have two approaches for characterizations of IA.  The first
assumes that uncertainty is represented by a standard probability measure,
uses a conjunct $\PLAYCONthree^k_i$, which plays the same role as
$\PLAYCONzero^k_i$.  
Define the formulas $\RATthree^k_i$ inductively
just as we defined $\RATzero^k_i$, except that now we use the language
$\L^2$ rather than $\L^0$.  In more detail, 
we take $\RATthree^0_i$ 
to be the formula $\true$,  and $\RATthree^{k+1}_i$ to be
an abbreviation of 
$$\RAT_i \land 
B_i(\PLAYCONthree^k_i) \land
O_i (\land_{j \ne i}\RATthree^k_{j}),$$
where $\PLAYCONthree^k_i$ is an abbreviation of
$\land_{\sigma' \in \Sigma_i(\Gamma)} (\play_i(\sigma') \rimp
\Diamond(\play_i(\sigma') \land \RATthree^k_i))$. 
Thus, $\RATthree^{k+1}_i$ says that $i$ is rational%
, believes that he plays a $k$-level rational strategy,
knows that all the other
players satisfy level-$k$ rationality (i.e., $\RATthree^k_{j}$), and that is all
that $i$ knows.  
It
is easy to see that $\RATthree^{k+1}_j$ implies
$\RATzero^{k+1}_j$.   
The difference is that instead of requiring just that $j$
assign positive probability to all strategy profiles compatible with
$\RATthree^{k}_{-j}$, it requires that $j$ assign positive probability to all
formulas in $\L^2_{(-i)+}$ compatible with $\RATthree^{k}_{-j}$.  
(Since there are only countably many such formulas, it is possible to
do so.)

The next result shows that 
$\RATthree^k_i$ characterizes iterated admissibility, just as $\RATzero^k_i$ does.

\thm\label{thm:charwd1} The following are equivalent:
\begin{itemize}
\item[(a)] the strategy $\sigma$ for player $i$ survives $k$ rounds of
iterated deletion of weakly dominated strategies in game $\Gamma$;
\item[(b)] there exists an $\L^2$-measurable structure  $M^{k}$
appropriate for $\Gamma$ and a state $\omega^{k}$ in $M^{k}$ such that
$\strat_i(\omega^{k}) = \sigma$ and $(M^{k},\omega^{k}) \sat \RATthree^{k}_i$;
\item[(c)] there is a structure  $M^{k}$
appropriate for $\Gamma$ and a state $\omega^{k}$ in $M^{k}$ such that
$\strat_i(\omega^{k}) = \sigma$ and $(M^{k},\omega^{k}) \sat \RATthree^{k}_i$.
\end{itemize}
\ethm

\shortv{The proof of Theorem \ref{thm:charwd1} is similar in spirit to
the proof of Theorem \ref{thm:charwd}, but is more complicated. Due to
lack of space, we defer it to the full version.} 
\fullv{
\prf The proof is similar in
spirit to the proof of Theorem~\ref{thm:charwd}.  We again proceed by
induction on $k$.  The result clearly holds for $k=0$.  If $k=1$, the
proof that (c) implies (a) is essentially identical to that of
Theorem~\ref{thm:charwd}; we do not repeat it here.  
The fact that (b) implies (c) is immediate.

To prove that (a) implies (b), we need 
three lemmas.  The first shows that a formula is always
satisfied in a state that has probability 0; the 
second shows that  that we can get a
new structure with a world where agent $i$ ascribes positive probability
to each of a countable collection of satisfiable formulas in
$\L_{-i}^2$; and the third shows that formulas in 
$\L^2_{i+}$ for different players $i$ are independent (i.e., if 
$\phi_i \in \L^2_{i+}$ is satisfiable, then so is 
$\phi_1 \land \ldots \land \phi_n$).  

\lem\label{lem:paste0} If $\phi \in \L^2$ is satisfiable in 
an $\L^2$-measurable structure%
, then there exists an $\L^2$-measurable structure
 $M$ and state $\omega$ such that
$(M,\omega)\sat \phi$, $\{\omega\}$ is measurable,
and
$\PR_j(\omega)(\{\omega\}) = 0$ for $j = 1,\ldots,n$.  \elem

\prf Suppose that
$(M',\omega') \sat\phi$, where $M' = (\Omega',\strat',\F', \PR_1', \ldots,
\PR_n')$.
Let $\Omega = \Omega' \union \{\omega\}$, where 
where $\omega$ is a fresh state; let $\F$ be the smallest
$\sigma$-algebra that contains $\F'$ and $\{\omega\}$;
let $\strat$ and $\PR_j$ agree with
$\strat'$ and $\PR_j'$ when restricted to states in $\Omega'$; more
precisely, if $\omega'' \in \Omega'$, then 
$\PR_j(\omega'')(A) = \PR_j'(\omega'')(A \inter \Omega')$ for $j =
1,\ldots, n$.    
Finally, define $\strat_i(\omega)= \strat_i(\omega')$, and take
$\PR_j(\omega)(A) = \PR_j'(\omega')(A \inter \Omega')$ for $j =
1,\ldots,n$.   Clearly $\{\omega\}$ is measurable, and
$\PR_j(\omega)(\{\omega\}) = 0$ for $j = 1,\ldots,n$.  An easy induction
on structure shows that for all formulas $\psi$,
(a) $(M,\omega) \sat \psi$ iff  $(M,\omega') \sat \psi$, and 
(b) for all states $\omega'' \in \Omega'$, we have that 
$(M,\omega'') \sat \psi$ iff $(M',\omega'') \sat \psi$.
It follows that $(M,\omega) \sat \phi$ and that $M$ is $\L^2$-measurable.
\eprf

\lem\label{lem:paste} Suppose that $\vec{\sigma} \in \vec{\Sigma}$,
$\Phi'$ is a countable collection of formulas in $\L^2_{-i}$, $\phi
\in \L^2_{-i}$,  
and $\Sigma'_{-i}$ is a set of strategy profiles in $\Sigma_{-i}$ such
that 
for each formula $\phi' \in \Phi'$, there exists some profile
$\sigma_{-i} \in \Sigma'_{-i}$ such that 
$\phi \land \phi' \land \play_{-i}(\sigma_{-i})$ is satisfied in 
an $\L^2$-measurable structure.
Then, if $\mu_i$ is a probability measure with support 
$\Sigma'_{-i}$, there exists an $\L^2$-measurable structure $M$ and state 
$\omega$ such that $\strat(\omega) = \vec{\sigma}$, 
and
$(M,\omega) \sat
\pr_{i}(\play_{-i}(\sigma_{-i})) \ge \alpha$ 
iff $\mu_i(\sigma_{-i}) \ge \alpha$ (i.e., $\mu_{i}$ agrees with
$\PR_i(\omega)$ when marginalized to strategy profiles in
$\Sigma'_{-i}$), and $(M,\omega) \sat B_i \phi \land \dbi \phi'$  for
all $\phi' \in \Phi'$.
\elem

\prf 
Let $\Phi'$, $\Sigma'_{-i}$, and $\mu_i$ be as in the
statement of the lemma.  Suppose that $\Phi' = \{\phi_1, \phi_2, \ldots,
\ldots\}$.  By assumption, for each formula $\phi_k \in \Phi'$, there
exists some strategy profile $\sigma'_{-i} \in \Sigma'_{-i}$, 
an $\L^2$-measurable structure
$M^k = (\Omega^k, \strat^k, \F^k, \PR_1^k, \ldots, \PR_n^k)$, and
a state
$\omega^k \in \Omega^k$ such that 
$(M^k,\omega^k) \sat \phi \land \phi_k \land \play_{-i}(\sigma_{-i}')$,
for $k = 1, 2, \ldots$.
By Lemma~\ref{lem:paste0}, we can assume without
loss of generality that $\{\omega^k\} \in \F^k$ and
$\PR_j^k(\omega^k)(\{\omega^k\})  = 
0$ for $j = 1, \ldots, n$.  Define $M^\infty = (\Omega^\infty, \strat^\infty, \F^\infty,
\PR_1^\infty, \ldots, \PR_n^\infty)$ as follows:
\begin{itemize}
\item $\Omega^\infty = \union_{k=0}^\infty \Omega^k \union
\{\omega\}$, where $\omega$ is a fresh state; 
\item $\F^\infty$ is the
smallest  $\sigma$-algebra that contains $\{\omega\} \union \F_1 \union
\F_2 \union \ldots$; 
\item $\strat^\infty(\omega) = \vec{\sigma}$ and 
$\strat^\infty$ agrees with $\strat^k$ when restricted to states in
$\Omega^k$, except that $\strat^\infty_i(\omega^k) = \sigma_i$.
\item $\PR_j^\infty$ agrees with
$\PR_j^k$ when restricted to states in $\Omega^k$; more
precisely, if $\omega' \in \Omega^k$, then 
$\PR_j^\infty(\omega')(A) = \PR_j^k(\omega')(A \inter \Omega^k)$, except
that  
$\PR_i^\infty(\omega) = \PR_i^\infty(\omega^1) = \PR_i^\infty(\omega^2) =
\cdots$ is defined to be a distribution with support $\{\omega^1,
\omega^2, \ldots\}$ (so that all these states are given positive
probability) such that $\PR_i^\infty(\omega)$ agrees with $\mu$ when
marginalized to profiles in $\Sigma_{-i}$, 
and $\PR_j^\infty(\omega)(\{\omega\}) = 1$ for $j \ne i$.
It is easy to see that our
assumptions guarantee that this can be done.  
\end{itemize}

We can now prove by a straightforward induction on the structure of
$\psi$ that (a) for 
all formulas $\psi$, $k = 1, 2, 3, \ldots$, and 
states $\omega' \in \Omega^k - \{\omega^k\}$, we have that 
$(M^k,\omega') \sat \psi$ iff $(M^\infty,\omega') \sat \psi$; and 
(b)  for all formulas $\psi \in \L^2_{(-i)^+}$, $k = 1, 2, 3, \ldots$, 
$(M^k,\omega^k) \sat \psi$ iff $(M^\infty,\omega^k) \sat \psi$. (Here it
is important that $\PR_j^\infty(\omega^k)(\{\omega^k\}) = 
\PR_j^k(\omega^k)(\{\omega^k\}) = 0$ for $j
\ne i$; this ensures that $j$'s beliefs about $i$'s strategies and beliefs
is
unaffected by the fact that  $\strat_i^k(\omega^k) \ne
\strat_i^\infty(\omega^k)$ and $\PR_i^k(\omega^k) \ne
\PR_i^\infty(\omega^k)$.)  It easily follows that 
$(M^\infty,\omega) \sat B_i \phi \land \dbi \phi'$ for all $\phi'
\in \Phi'$. \eprf

\lem\label{lem:paste1} If $\phi_i \in \L^2_{i+}$ is
satisfiable 
in an $\L^2$-measurable structure
for $i = 1, \ldots, n$, 
then $\phi_1 \land \ldots \land \phi_n$ is satisfiable
in an $\L^2$-measurable structure.
\elem

\prf Suppose that $(M^i,\omega^i) \sat \phi_i$, where 
$M^i = (\Omega^i,\strat^i,\F^i,\PR_1^i,\ldots, \PR_n^i)$ and 
$\phi_i \in 
\L^2_{i+}$.  By Lemma~\ref{lem:paste0}, we again assume without loss of
generality that $\{\omega^i\} \in \F^i$ and
$\PR_j(\omega^i)(\{\omega^i\}) = 0$.  Let $M^* =
(\Omega^*,\strat^*,\F^*, \PR_1^*, \ldots, \PR_n^*)$, where 
\begin{itemize}
\item $\Omega^* = \union_{i=1}^n \Omega^i$;
\item $\F^*$ is the smallest $\sigma$-algebra containing
$\F^1 \union \ldots \union \F^n$;
\item $\strat^*$ agrees with $\strat^j$ on states in $\Omega^j$ except
that $\strat^*_i(\omega^j) = \strat^i_i(\omega^i)$ (so that
$\strat^*(\omega^1) = \cdots = \strat^*(\omega^n)$);
\item $\PR^*_i$ agrees with $\PR^j_i$ on states in $\Omega^j$ except
that $\PR^*_i(\omega^j) = \PR^i_i(\omega^i)$ (so that
$\PR_i^*(\omega^1) = \cdots = \PR_i^*(\omega^n) = \PR_i^i(\omega^i)$).
\end{itemize}

We can now prove by induction on the structure of $\psi$ that (a) for
all formulas 
$\psi \in \L^2_{i+}$, 
$i = 1, \ldots, n$, and 
states $\omega' \in \Omega^i$, we have that 
$(M^i,\omega') \sat \psi$ iff $(M^*,\omega') \sat \psi$; 
(b)  for all formulas $\psi \in \L^2_{i+}$, $1 \le i, j \le n$,
$(M^i,\omega^i) \sat \psi$ iff $(M^*,\omega^j) \sat \psi$ (again,
here it is important that $\PR_i^*(\omega^j)(\{\omega^j\}) = 0$ for $j = 1,\ldots,
n$).  Note that part (b) implies that the states $\omega^1, \ldots,
\omega^n$ satisfy the same formulas in $M^*$.  It easily follows that 
$(M^*,\omega^i) \sat \phi_1 \land \ldots \land \phi_n$ for $i =
1,\ldots, n$. \eprf

We can now prove the theorem.  
To see that (a) implies (b), suppose that 
$\sigma_i \in \NWD^{k+1}_i$.  By Proposition~\ref{pro:Pearce}, there exists
a distribution $\mu_i$ whose support is $\NWD^k_{-i}$ such that $\sigma_i$
is a best response to $\mu_i$.  By the induction hypothesis, for each
strategy profile $\tau_{-i} \in \NWD^k_{-i}$, and all $j \ne i$, the
formula $\play_{j}(\tau_{j}) \land \RATthree^k_{j}$ is satisfied in 
an $\L^2$-measurable 
structure.  By
Lemma~\ref{lem:paste1}, $\play_{-j}(\tau_{-j}) \land (\land_{j
\ne i} \RATthree^k_{j})$ is satisfied in an $\L^2$-measurable structure.
Taking 
$\phi$ to be  $\land_{j \ne i} \RATthree^k_{j}$, by Lemma~\ref{lem:paste},
there exists an $\L^2$-measurable structure $M$ and state $\omega$ in
$M$ such 
that the marginal  of
$\PR_i(\omega)$ on $\NWD^k_{-i}$ is $\mu_i$, $\strat_i(\omega)$ is
$\sigma_i$, and $(M,\omega) \sat B_i (\land_{j \ne i}\RATthree^k_{j})
\land 
(\land_{\psi \in \L^2_{(-j)+}}  \Diamond(\psi \land (\land_{j \ne i} 
\RATthree^k_{j})) \rimp \dbj \psi)$.  It follows that $(M,\omega)
\sat \RAT_i \land O_i (\land_{j \ne i}\RATthree^k_{j})$.
Moreover, since $\sigma_i \in \NWD^{k+1}_i$, $\sigma_i$ is also in
$\NWD^k_i$.  By the induction hypothesis, $\play(\sigma_i) \land
\RATthree^k_i$ is 
satisfiable.  Thus, $(M,\omega)\sat \PLAYCONthree^k_i$.  
Hence, $(M,\omega) \sat \RATthree^{k+1}_i$, as desired.
\eprf 
}

While this is closer in spirit to the BFK result, it still does not
result in full-support beliefs.  We can get full-support beliefs (on
the richer language) by using LPSs, by considering the analogue of the
approach defined in Section~\ref{sec:second}.  Again, we now work in
LPS structures, rather than probability structures.  We use the same formula as
before, just replacing $\L^0$ by $\L^2$.  Thus,
$\GRATthree^0_i$ is $\true$ and  
$\GRATthree^{k+1}_i$ is an abbreviation of 
$$\RAT_i \land O^{\L^0_{-i}}_i (\land_{j \ne i}\GRATthree^k_{j},
\ldots, \land_{j \ne i}\GRATthree^0_{j}).$$
Now, if $(M,\omega) \sat O^{\L^2_{-i}}_i (\land_{j \ne i}\GRATzero^k_{j},
\ldots, \land_{j \ne i}\GRATzero^0_{j})$ for some $k > 0$ for some $k
\ge 0$, then the LPS 
$\PR_i(\omega)$ must give positive probability (at some level) to all
consistent formulas in $\L^3$.  This makes such a structure very close
in spirit to a full-support LPS in a complete  type structure.

We now have the obvious analogue of Theorem~\ref{thm:charwdg}.
\thm\label{thm:charwdg2} The following are equivalent:
\begin{itemize}
\item[(a)] $\vec{\sigma} \in \NWD^k$;
iterated deletion of weakly dominated strategies;
\item[(b)] there exists an $\L^0$-measurable LPS structure $M^{k}$
appropriate for $\Gamma$ and a state $\omega^{k}$ in $M^{k}$ such that
$\strat_i(\omega^{k}) = \sigma$ and $(M^{k},\omega^{k}) \sat \GRATzero^{k}_i$.  
\end{itemize}
\ethm

\prf The argument that (b) implies (a) is identical to that in the
proof of Theorem~\ref{thm:charwdg}.  The argument that (a) implies (b)
is much like that of the analogous argument in the proof of
Theorem~\ref{thm:charwd1}, 
using straightforward analogues of Lemmas~\ref{lem:paste0},
\ref{lem:paste}, and \ref{lem:paste1} for LPS structures to construct
the required LPS structure.
\eprf

We might consider trying to define an ``all I know'' operator for an
even richer language that allows probabilities to range over the reals.
Unfortunately, we cannot require than 
agent $i$  ascribe positive probability to all the formulas  in
such a language consistent with a formula $\phi$; in general, there will be
an uncountable number of distinct and mutually exclusive formulas
consistent with $\phi$, so they
cannot all be assigned positive probability.  (For example, the formulas
could say that another agent ascribes probability exactly $\alpha$ to
some $\psi$, for all $\alpha \in [0,1]$.) 
This problem does not
arise with $\L^2$, since it is a countable language.  
In their approach, Halpern and Lakemeyer \citeyear{HalLak94} 
could allow an agent to consider an uncountable
set of worlds possible, since 
they
did not deal with probabilistic
systems.

\commentout{
\section{Complete and Canonical Structures}\label{sec:complete}
In this section we relate the canonical structures used in modal logic
to the complete structures used in the game theory.  This section can be
skipped without loss of continuity.

\subsection{Canonical Structures}

The intuition behind ``all $i$ knows is $\phi$'' goes back to Levesque
\citeyear{Lev5}.  The idea is that all $i$ knows is $\phi$ if (1) $i$
knows $\phi$
(so that $\phi$ is true in all the worlds that $i$ considers possible)
and (2) $i$
considers possible all worlds consistent with $\phi$ (so that $\phi$ is
false in all worlds that $i$ does not consider possible).  In the
single-agent case (which is what Levesque considered) it is relatively
easy to make precise the set of worlds that $i$ does not consider
possible, since a world can be identified with a truth assignment.
This is much more complicated in the multi-agent setting considered by
Halpern and Lakemeyer \citeyear{HalLak94}.  They made it precise by
working in the canonical structure.  The advantage of the canonical
structure is that, in a sense, it has all possible worlds, so it is clear
what worlds an agent does not consider possible.   Although our
definition of ``all $i$ knows'' is more language-dependent, our
intuitions for the notion are still grounded in the canonical structure.
Thus, in this section, we consider our definitions in the context of
canonical structures.  The reason we say ``structures'' here rather than
``structure'' is that the notion of canonical structure is also language
dependent. 

Define the \emph{canonical structure $M^c = (\Omega^c, \strat^c,
\F^c, \PR_1^c, 
\ldots, \PR^c_n)$ for $\L^4$} as follows:
\begin{itemize}
\item $\Omega^c = \{\omega_\Phi: \Phi$ is a realizable subset of
$\L^4(\Gamma)\}$; 
\item $\strat^c(\omega_\Phi) = \vec{\sigma}$ iff $\play(\sigma) \in
\Phi$;
\item $\F^c = \{F_\phi: \phi \in \L^4\}$, where $F_\phi =
\{\omega_\Phi: \phi \in \Phi\}$;
\item $\PR_i^c(\omega_\Phi)(F_\phi) = \sup\{\alpha: \pr_i(\phi) > 
\in \Phi\}$.  
\end{itemize}

\lem $M^c$ is an appropriate $\L^2$-measurable structure for $\Gamma$. \elem

\fullv{
\prf 
It is easy to see that $\F^c$ is a $\sigma$-algebra, since the complement
of $F_\phi$ is $F_{\neg \phi}$ and $\inter_{m=1}^\infty F_{\phi_i} =
F_{\land_{m=1}^\infty \phi_m}$.  
Given a strategy $\sigma$ for player $i$, $\intensionc{\sigma} = 
F_{\play_i(\sigma)} \in \F$.
Moreover, each realizable set $\Phi$ that includes $\play_i(\sigma)$
must also include $\pr_i(\play_i(\sigma)) = 1$, so that 
$\PR_i(\omega_\Phi)(\strat_i(\omega_\Phi)) = 
\PR_i(\omega_\Phi)(F_{\play_i(\strat_i(\omega_\Phi))}) = 1$.
Similarly, suppose that $\PR_i(\omega_\Phi) = \pi$.  Then 
$\{\omega \in \Omega^c: \PR_i(\omega) = \pi\} = \inter_{\phi \in \L^3} \inter_{\{
\alpha \in Q \inter [0,1]: \pi(\intensionc{\phi}) \ge \alpha\}} F_{\phi
\ge \alpha} \in \F^c$.  Moreover, if $\alpha \in Q \inter [0,1]$, then
$\pi(\intensionc{\phi}) \ge \alpha$ iff $\pr_i(\phi) \ge \alpha \in
\Phi$.  But if $\pr_i(\phi) \ge \alpha \in \Phi$, then
$\pr_i(\pr_i(\phi) \ge \alpha) = 1 \in \Phi$.  It easily follows that  
$\PR_i(\omega_\Phi)(\{\omega: \PR_i(\omega) = \pi\}) = 1$. 
Finally, the definition
of $\F^c$ guarantees that every set $\intensionc{\phi}$ is measurable
and that $\PR_i(\omega_\Phi)$ is indeed a probability distribution on
$(\Omega^c,\F^c)$.   
\eprf
}

The following result is the analogue of the standard ``truth lemma'' in
completeness proofs in modal logic.

\pro\label{pro:truth} For $\psi \in \L^4$, 
$(M^c,\omega_\Phi) \sat \psi$ iff $\psi \in \Phi$. \epro

\fullv{
\prf A straightforward induction on the structure of $\psi$.
\eprf
}

We have constructed a canonical structure for $\L^4$.  It follows easily
from Lemma~\ref{lem:realizable} that the canonical structure for $\L^3$
(where the states are realizable $\L^3$ sets) is isomorphic to $M^c$.
(In this case, the set $\F^c$ of measurable sets would be
the smallest $\sigma$-algebra containing $\intension{\phi}$ for $\phi
\in \L^3$.)  
Thus, the choice of $\L^3$ vs.~$\L^4$ does not play an important role
when constructing a canonical structure.

\commentout{
A strategy $\sigma_i$ for player $i$ survives iterated
deletion of weakly dominated strategies iff the
formula 
$$\undominated^k(\sigma_i) = \play_i(\sigma_i) \land 
\lor_{k^*=}^\infty (\land_{k=k^*}^\infty F^k_i)$$ 
 is satisfied at 
some state in the canonical structure.  But there are other structures
in which $\undominated(\sigma_i)$ is satisfied.   One way to get such a
structure is by essentially ``duplicating'' states in the canonical
structure.  The canonical structure can be
\emph{embedded} in a structure $M$ if, for all $\L^3$-realizable sets
$\Phi$, there is a state $\omega_\Phi$ in $M$ such that
$(M,\omega_\Phi)\sat \phi$ iff $\phi \in \Phi$.  Clearly
$\undominated(\sigma_i)$ is satisfied in any structure in which the
canonical structure can be embedded.

A structure in which the canonical structure can be embedded is in a
sense larger than the canonical structure.  But $\undominated(\sigma_i)$
can be satisfied in structures smaller than the canonical structure.
\fullv{(Indeed, with some effort, we can show that it is satisfiable in a
structure with countably many states.)}
There are two reasons for this.
The first is that to satisfy $\undominated(\sigma_i)$, there is no need
to consider a structure with states where all the players are
irrational.  It suffices to restrict to states where at least one player
is using a strategy that survives at least one round of iterated
deletion.  This is because players know their strategy; thus, in a state
where a strategy $\sigma_i$ for player $i$ is admissible, player $i$
must ascribe positive probability to all other strategies; however, in
those states, player $i$ still plays $\sigma_i$.  

A perhaps more interesting reason that we do not need the canonical
structure is our use of the language $\L_3$.   
The formulas $F^k_i$  guarantee that player $i$ ascribes positive
probability to all formulas $\phi$ consistent with the appropriate level
of rationality.  
Since a finite conjunction
of formulas in $\L^3$ is also a formula in $\L^3$, player $i$ will
ascribe positive probability to all finite conjunctions of formulas
consistent with rationality.  But a 
state is characterized by a \emph{countable} conjunction of formulas.
Since $\L^3$ is not closed under countable conjunctions, a structure
that satisfies $\undominated(\sigma_i)$ may not have states
corresponding to all $\L^3$-realizable sets of formulas.  If we had used
$\L^4$ instead of $\L^3$ in the definition of $F^k_i$
(ignoring the issues raised earlier with using $\L^4$), then there would
be a state corresponding to every $\L^4$-realizable (equivalently,
$\L^3$-realizable) set of formulas.  Alternatively, if we consider
appropriate structures that are compact in a topology where all sets
definable by formulas (i.e., sets of the form $\intension{\phi}$, for
$\phi \in \L^3$) are closed (in which case they are also open, since
$\intension{\neg \phi}$ is the complement of $\intension{\phi}$), then
all states where at least one player is using a strategy that survives at
least one round of iterated deletion will be in the structure.

Although, as this discussion makes clear, the formula $\RATthree_i^k$ that
characterizes iterated admissibility can be satisfied in structures
quite different from 
the canonical structure, the canonical structure does seem to be the
most appropriate setting for reasoning about statements involving ``all
agent $i$ knows''.
Moreover, as we now show, canonical structures allow us to relate our
approach to that of BFK.
}

\subsection{Complete Structures}
BFK worked with complete structures.   We now want to show that $M^c$ is
complete, in the sense of BFK.  To make this precise, we need to recall
some notions from BFK (with some minor changes to be more consistent
with our notation).   

BFK considered what they called \emph{interactive probability
structures}. These can be viewed as a special case of probability
structures.  A \emph{BFK-like structure} (for a game $\Gamma$) is 
a probability structure $M = (\Omega, \strat, \F, \PR_1,\ldots, \PR_n)$
such that there exist spaces $T_1, \ldots, T_n$ (where $T_i$ can be
thought of as the \emph{type space} for player $i$) such that
$\Omega$ is isomorphic to $\vec{\Sigma} \times \vec{T}$ via some
isomorphism $h$, where $h(\omega) =
(\vec{\sigma}^\omega,\vec{t}^\omega)$ and 
\fullv{\begin{itemize}}
\shortv{\begin{list}{\labelitemii}{\leftmargin=.5\leftmarginii}}
\item $\strat(\omega) = \vec{\sigma}^\omega$; 
\item the support of $\PR_i(\omega)$ is
contained in $\{\omega': \strat_i(\omega') = \sigma_i^\omega,\, t_i^{\omega'} =
t_i^\omega\}$, so that $\PR_i(\omega)$ induces a probability on
$\Sigma_{-i} \times T_{-i}$;
\item $\PR_i(\omega)$ depends only on $t_i^\omega$, in the sense that
if $t_i^\omega = t_i^{\omega'}$, then $\PR_i(\omega)$ and
$\PR_i(\omega')$ induce the same probability distribution on
$\Sigma_{-i} \times T_{-i}$.
\fullv{\end{itemize}}
\shortv{\end{list}}

A BFK-like structure $M$ whose state space is isomorphic to $\vec{\Sigma}
\times \vec{T}$ is \emph{complete} if, for every 
distribution
$\mu_i$ over $\Sigma_{-i} \times T_{-i}$,
(where the measurable sets are the ones induced by the isomorphism $h$ and the
measurable sets $\F$ on $\Omega$),
there is a state $\omega$ in $M$ such that the
probability distribution on $\Sigma_{-i} \times T_{-i}$
induced by $\PR_i(\omega)$ is $\mu_i$.  

\pro\label{pro:canonicaliscomplete} $M^c$ is a complete BFK-like
structure. \epro 

\fullv{
\prf A set $\Phi \subseteq
\L^3_i$ is \emph{$\L^3_i$-realizable} if there 
exists an appropriate structure $M$ for $\Gamma$ and state $\omega$ in
$M$ such that, for all formulas $\phi \in \L^3_i$,  $(M,\omega) \sat 
\phi$ iff $\phi \in \Phi$.  Take the type space $T_i$ to consist of all
$\L^3_i$-realizable sets of formulas.  There is an isomorphism  $h$
between $\Omega^c$ and $\vec{\Sigma} \times \vec{T}$, 
such that $h(\omega) = (\vec{\sigma}^\omega,\vec{t}^\omega)$,
$\vec{\sigma}^\omega = \strat^c(\omega)$, and $t_i^\omega$ consists of 
all formulas of the form $\pr_i(\phi) \ge \alpha$ that are true at
$(M^c,\omega)$. It follows easily from
Lemma~\ref{lem:paste0} that $h$ is a surjection. 
We can identify $\Omega^c$, the state space in the canonical structure,
with $\vec{\S} \times \vec{T}$.  

To prove that $M^c$ is complete, given a probability $\mu$ on
$\Sigma_{-i} \times  T_{-i}$, we must show that there is some state
$\omega$ in $M^c$ such that the probability induced by $\PR_i(\omega)$ on
$\Sigma_{-i} \times  T_{-i}$ is $\mu$.  
Let $M^{\mu} = (\Omega^{\sigma,\mu}, \F^{\mu},
\strat^{\mu}, \PR_1^{\mu}, 
\ldots, \PR_n^{\mu})$, where 
\begin{itemize}
\item $\Omega^{\mu} = \Omega^c
\union \Sigma \times \{\mu\} \times
T_{-i}$; 
\item $\F^{\mu}$ is the smallest $\sigma$-algebra that contains
$\F^c$ and all sets of the form $\vec{\sigma} \times \{\mu\} \times
\intensioncp{\phi}$, where  $\intensioncp{\phi}$ consists of the all type
profiles $t_{-i}$ such that, for some state $\omega$ in $M^c$,
$(M^c,\omega) \sat \phi$ and $t_{-i}^\omega = t_{-i}$;
\item $\strat^\mu(\omega) = \strat^c(\omega)$ for $\omega \in \Omega^c$,
and $\strat^\mu(\vec{\sigma} \times \{\mu\} \times \vec{t}) = \vec{\sigma}$;
\item $\PR_j^\mu(\omega) = \PR_j^c(\omega)$ for $\omega \in \Omega^c$,
$j = 1, \ldots, n$; for $j \ne i$, $\PR_j^\mu(\vec{\sigma} \times
\mu\times t_{-i}) =  \PR_j(\omega)$, where $\strat_j(\omega) = \sigma_j$
and $t_j^\omega = t_j$ (this is well defined, since if
$\strat_j(\omega') = \sigma_j$ and $t_j^{\omega'} = t_j$, then
$\PR_j(\omega) = \PR_j(\omega')$; finally, 
$\PR_i^\mu(\vec{\sigma} \times \mu\times t_{-i})$ is a
distribution whose support is contained in $\{\sigma_i\} \times \Sigma_{-i}
\times \{\mu\} \times T_{-i}$, and $\PR_i^\mu(\vec{\sigma} \times
\mu\times t_{-i})(\vec{\sigma} \times \mu\times \intensioncp{\phi}) =
\mu(\intensioncp{\phi})$.  
\end{itemize}
Choose an arbitrary state $\omega \in \vec{\Sigma} \times \{\mu\} \times
T_{-i}$.  The construction of $M^\mu$ guarantees that for $\phi \in
\L^4_{(-i)+}$, $(M^\mu, \omega) \sat \pr_i(\phi) > \alpha$ iff
$\mu(\intensioncp{\phi}) > \alpha$.  
By the construction of $M^c$, there exists a state $\omega' \in 
\Omega^c$ such that $(M^c,\omega') \sat \psi$ iff $(M^\mu,\omega) \sat
\psi$.  Thus, the distribution on $\Sigma_{-i} \times T_{-i}$ induced by
$\PR_i(\omega)$ is $\mu$, as desired.  This shows that $M^c$ is
complete. 
\eprf
}

We would now like to prove the converse to
Proposition~\ref{pro:canonicaliscomplete}, and
show that every $\L^3$-measurable complete BFK-like
structure is the canonical structure.  This is not quite true.  The set of
formulas true at two different states in the canonical structure must
be different.  This is not necessarily true in a complete structure.
Thus, we prove a slightly weaker property: we show that the canonical
structure can be \emph{embedded} in every 
$\L^3$-measurable complete structure, in the sense that, given an
$\L^3$-measurable complete structure $M$, for every
$\L^3$-realizable set $\Phi$ of formulas, there is a state $\omega_\Phi$ in
$M$ such that $(M,\omega_\Phi)\sat \phi$ iff $\phi \in \Phi$.  This
shows that complete structures are rich in the sense that any
(possibly infinite) collection of formulas that can be satisfied is
satisfied somewhere in a complete structure.

To prove this result, we seem to need to make a nontrivial technical
restriction: we restrict to 
\emph{strongly measurable} complete structures, where a structure is
strongly measurable if it is $\L^3$ measurable and the only measurable sets are
those defined by $\L_4$ formulas (or, equivalently, the set of
measurable sets is the smallest set that contains the sets defined by
$\L_3$ formulas).  
\fullv{
We explain where strong measurability is needed at
the end of the proof of the following theorem.}
\shortv{(We explain the need for strong measurability in the full paper.)}

\thm\label{thm:strong} If $M$ is a strongly measurable complete
BFK-like structure,  
then the canonical structure can be embedded in $M$.    
\ethm

\fullv{
\prf Suppose that $M$ is a strongly measurable complete BFK-like structure.
We can assume without loss of generality that the state space of $M$ has
the form $\vec{\Sigma} \times \vec{T}$, so that a state $\omega$ in $M$
has the form $(\vec{\sigma}^\omega,\vec{t}^\omega)$.
To prove the result, we need
the following lemmas.

\lem\label{lem:depends} If $M$ is BFK-like, the truth of a formula $\phi
\in \L^4_i$ at a state $\omega$ in $M$ depends only on $i$'s type; 
that is,  if $T_i(\omega) = T_i(\omega')$, then $(M,\omega) \sat \phi$
iff $(M,\omega') \sat \phi$.  Similarly, the truth of a formula in
$\L_{i+}$ in $\omega$ depends only on $\vec{\sigma}^\omega$ and
$t_i^\omega$, and the truth of a formula in
$\L^4_{i+}$  in $\omega$  depends only on $t_{-i}^\omega$.
\elem

\prf  A straightforward induction on structure. \eprf

Define a \emph{basic formula} to be one of the form $\psi_1\land \ldots
\land \psi_n$, where $\psi_i \in \L^3_{i+}$ for $i = 1,\ldots,n$.

\lem\label{lem:basic} Every formula in $\L^3$ is equivalent to a finite
disjunction of basic formulas.  
\elem

\prf A straightforward induction on structure. \eprf

\lem\label{lem:basic1} Every formula in $\L^3_{i+}$ is equivalent to a 
disjunction of formulas of the form 
$$
\play_i(\sigma) \land (\neg) \RAT_i
\land (\neg) \pr_i(\phi_1) > \alpha_1 \land \ldots  \land
(\neg)\pr_i(\phi_m) > \alpha_m \land (\neg) \pr_i(\psi_1) \ge \beta_1
\land \ldots  \land (\neg)\pr_i(\psi_{m'}) \ge \beta_{m'},
$$
where $\phi_1, \ldots, \phi_m, \psi_1, \ldots, \psi_{m'} \in
 \L^3_{(-i)+}$ and the
``($\neg$)'' indicates that the presence of negation is optional.
\elem

\prf A straightforward induction on the structure of formulas, using the
 observation that $\neg\play_i(\sigma)$ is equivalent to $\lor_{\{\sigma'
\in \Sigma_i: \sigma' \ne \sigma\}} \play_i(\sigma')$. \eprf

\lem\label{lem:satisfiable} If $\phi \in \L^3$ is satisfiable, then
$\intension{\phi} \ne \emptyset$.  \elem

\prf By Lemma~\ref{lem:basic}, it suffices to prove the result for the
case that $\phi$ is a basic formula.  By Lemma~\ref{lem:basic1}, it
suffices to assume that the ``$i$-component'' of the basic formula
is a conjunction.  We now prove the result by induction on the depth 
of nesting of the modal operator $\pr_i$ in $\phi$.  (Formally, define
$D(\psi)$, the depth of nesting of $\pr_i$ in $\psi$, by induction on
the structure of $\psi$.  If $\psi$ has the form $\play_j(\sigma)$,
$\RAT_j$, or $\true$, then $D(\psi) = 0$; $D(\neg \psi) = D(\psi)$; 
$D(\psi_1 \land \psi_2) = \max(D(\psi_1),D(\psi_2))$; and $D(\pr_i(\psi)
> \alpha) = D(\pr_i(\psi) \ge \alpha) = 1 + D(\psi)$.)  
Because the state space $\Omega$ of $M$ is essentially a product space,
by Lemma~\ref{lem:depends}, it suffices to prove the result for formulas
in $\L^3_{(i)+}$.   It is clear that $\phi$ possibly puts constraints on
what strategy $i$ is using, the probability of strategy profiles in
$\Sigma_{-i}$, and the probability of formulas that appear in the scope
of $\pr_i$ in $\phi$.  If $M' = (\Omega',\strat',\F',\PR_1',\ldots,
\PR_n')$ is a structure and $\omega' \in \Omega'$, 
then $(M',\omega') \sat \phi$ iff $\strat'_i(\omega')$ and
$\PR'_i(\omega')$ satisfy these constraints.  (We leave it to the reader
to formalize this somewhat informal claim.)  
By the induction hypothesis, each formula in the scope
of $\pr_i$ in $\phi$ that is assigned positive probability by
$\PR_i(\omega')$ is satisfied in $M$.  Since $M$ is complete and
$\L^3$-measurable, there is a
state $\omega$ in $M$ such that $\strat_i(\omega) = \strat'_i(\omega')$
and $\PR_i(\omega)$ places the same constraints on formulas that appear
in $\phi$ as $\PR_i$.  We must have $(M,\omega) \sat \phi$.   
\eprf  

Returning to the proof of the theorem, suppose that $M =
(\Omega,\strat,\F, \PR_1, \ldots, \PR_n)$.  
Given a state $\omega \in \Omega^c$, we claim that there must be a state
$\omega'$ in $M$ such that $\strat(\omega') = \strat^c(\omega)$ and, for
all $i = 1, \ldots, n$ and all formulas $\psi \in \L^3$, we have 
$\PR_i^c(\omega)(\intensionc{\psi}) = 
\PR_i(\omega')(\intension{\psi})$.  To show this, because $\Omega$ is a
product space, and $\PR_i(\omega')$ depends only on $T_i(\omega')$, it
suffices to show that, for each $i$, there exists a state $\omega_i$ in
$M$ such that $\PR_i^c(\omega)(\intensionc{\psi}) =
\PR_i(\omega_i)(\intension{\psi})$.  By Lemma~\ref{lem:satisfiable},
if $\intensionc{\psi} \ne \emptyset$, then $\intension{\psi} \ne
\emptyset$.  Thus, the existence of $\omega_i$ follows from
the assumption that $M$ is complete and strongly measurable.

Roughly speaking, 
to understand the need for strong measurability here, note that even
without strong measurability, the argument above tells us that 
there exists an appropriate measure defined on 
sets of the form $\intension{\phi}$ for $\phi$ in $\L^3_{(-i)+}$.  We can
easily extend $\mu$ to a measure $\mu'$ on sets of the form
$\intension{\phi}$ for $\phi$ in $\L^4_{(-i)+}$.  However, if the set
$\F$ of measurable sets in $M$ is much richer than the sets definable by
$\L^4$ formulas, it is not 
clear that we can extend $\mu'$ to a measure on all of $\F$.  In
general, a countably additive measure defined on a subalgebra of  a set
$\F$ of measurable sets cannot be extended to $\F$.  For example, it is
known that, under the continuum hypothesis, 
Lebesgue measure defined on the Borel sets cannot be extended to all
subsets of $[0,1]$ \cite{Ulam30}; (see \cite{KT64} for further
discussion).  Strong measurability allows us to avoid this problem.
\eprf
}
}

\section{Extensive-Form Rationalizability}\label{sec:efr}
Up to now, we have considered games in normal form.  Pearce
\citeyear{Pearce84} defined a notion of rationalizability for games in
extensive form, described by a game tree, that he called
\emph{extensive-form rationalizability (EFR)}.  Pearce's definition
was simplified by 
Shimoji and Watson \citeyear{SW98}, using a notion they call
\emph{conditional dominance}.  We review their definition here.  
%
The idea is that,
like the definition of rationalizability, EFR proceeds by iterated
deletion of strategies, where at each stage, we delete dominated
strategies.  But now the notion of dominance takes into account the game
tree.    

We assume that the reader is familiar with game trees and the notion of
perfect recall.  A game $\Gamma$ in extensive form is
characterized by a game tree.  
Suppose that we are given a game tree
$\Gamma$ where agents have perfect recall.  
As before, we let $\Sigma_i(\Gamma)$ denote the pure strategies for player
$i$ in $\Gamma$, and let $\Sigma(\Gamma)$ denote all the pure strategy
profiles in $\gamma$.  
%
Given a player $i$, an information set $I$ for player $i$, 
a set $\Sigma' = \Sigma'_i \times \Sigma'_{-i}$, a 
(pure)
strategy $\sigma
\in \Sigma'_i$, and a mixed strategy $\sigma'$ over $\Sigma'_i$, let
$\Sigma'_{\sigma,\sigma',I,-i}$ 
denote the strategy profiles $\tau_{-i} \in \Sigma'_{-i}$ such that 
$(\sigma,\tau_{-i})$ reaches $I$ and 
$(\sigma'',\tau_{-i})$ reaches $I$ for every strategy 
$\sigma''$ in the support of $\sigma'$.
Since we restrict to games of perfect recall, the game tree provides a
natural ordering on information sets, and the notion of an information
set $I'$ for player $i$ being below an information set $I$ for player
$i$ is well defined.

\dfn\label{dfn.cd} Given a set $\Sigma' = \Sigma'_{i} \times
\Sigma'_{-i} \subseteq \Sigma(\Gamma)$, a strategy 
$\sigma \in \Sigma'_i$ is \emph{conditionally dominated} on $\Sigma'$ if
there exists an 
information set $I$ for player $i$ and a mixed strategy $\sigma'$ for
player $i$ such that 
(a) $\Sigma'_{\sigma,\sigma',I,-i} \neq \emptyset$, and
(b) $u_i(\sigma',\tau_{-i}) > u_i(\sigma,\tau_{-i})$ for every strategy
profile $\tau_{-i} \in \Sigma'_{\sigma,\sigma',I,-i}$.
\edfn
Intuitively, $\sigma \in \Sigma'_i$ is conditionally dominated on
$\Sigma'$ if there exists a mixed strategy 
$\sigma'$
that dominates $\sigma$
conditional on reaching $I$, when the other players are restricted to
using strategies 
in $\Sigma'_{-i}$ consistent with both $\sigma$ and $\sigma'$ reaching $I$. 

The following equivalent formulation will be convenient for our
purposes (and is arguably also a more natural definition of
conditional dominance). 
Given a player $i$, 
a set $\Sigma' = \Sigma'_i \times \Sigma'_{-i}$, a strategy $\sigma
\in \Sigma'_i$ and an information set $I$ for player $i$, 
let $\Sigma'_{\sigma,I,-i}$ consist of all the strategy profiles
$\tau_{-i} \in \Sigma'_{-i}$ such that $(\sigma,\tau_{-i})$ reaches
$I$.

\dfn\label{dfn.cdprime} Given a set $\Sigma' = \Sigma'_{i} \times
\Sigma'_{-i} \subseteq \Sigma(\Gamma)$, a strategy 
$\sigma \in \Sigma'_i$ is \emph{conditionally dominated$'$} on $\Sigma'$ 
if there exists an 
information set $I$ for player $i$ such that 
(a) $\Sigma'_{\sigma,I,-i} \neq \emptyset$, 
and (b) there exists a mixed strategy $\sigma'$ 
that differs from $\sigma$ only at information sets for $i$ at or below $I$
such that 
$u_i(\sigma',\tau_{-i}) > u_i(\sigma,\tau_{-i})$
for every strategy
profile $\tau_{-i} \in \Sigma'_{\sigma,I,-i}$.
\edfn

\pro
\label{conddomprim.prop} Given a set $\Sigma' = \Sigma'_{i} \times
\Sigma'_{-i} \subseteq \Sigma(\Gamma)$, a strategy 
$\sigma \in \Sigma'_i$ is conditionally dominated$'$ on $\Sigma'$
iff it is conditionally dominated on $\Sigma'$.
\epro
\prf
Assume $\sigma$ is conditionally dominated$'$ on $\Sigma$; that is,
there exists a
player $i$ and an information set $I$ for player $i$
such that
$\Sigma'_{\sigma,I,-i} \neq \emptyset$, and a mixed strategy
$\sigma'$ that differs from $\sigma$ only at or below $I$
such that for every strategy
profile $\tau_{-i} \in \Sigma'_{\sigma,I,-i}$, $u_i(\sigma',\tau_{-i})
> u_i(\sigma,\tau_{-i})$.
Since $\sigma'$ 
differs from $\sigma$ only at or below $I$, we have that
for every strategy $\sigma''$ in the support of $\sigma'$,
$\Sigma'_{\sigma,I,-i} = \Sigma'_{\sigma'',I,-i}$.
It follows that $\Sigma'_{\sigma,\sigma',I,-i} =
\Sigma'_{\sigma,I,-i}$;
thus, $\sigma$ is conditionally dominated.

For the converse, suppose that $\sigma$ is conditionally
dominated; that is, there exists a player $i$, an 
information set $I$ for player $i$, and a mixed strategy $\sigma'$ for $i$
such that $\Sigma'_{\sigma,\sigma',I,-i} \neq \emptyset$, and
for every $\tau_{-i} \in \Sigma'_{\sigma,\sigma',I,-i}$,
$u_i(\sigma',\tau_{-i}) > u_i(\sigma,\tau_{-i})$ . 
Choose a strategy $\tau_{-i} \in \Sigma'_{\sigma,\sigma',I,-i}$.  Since
$I$ is a game of perfect recall, $\sigma$ is a pure strategy,
$(\sigma,\tau_{-i})$ reaches $I$ and, for every strategy $\sigma''$ in
the support of $\sigma'$, $(\sigma'',\tau_{-i})$ reaches $I$, it must
be the case that $\sigma$ and $\sigma''$ perform exactly the same actions on
all information sets that precede $I$.  
Thus, we must have $\Sigma'_{\sigma,I,-i} = \Sigma'_{\sigma'',I,-i}$ for
all $\sigma''$ in the support of $\sigma'$, so $\Sigma'_{\sigma,I,-i}
= \Sigma'_{\sigma,\sigma',I,-i}$.

Now consider the strategy
$(\sigma,I,\sigma')$ that uses $\sigma$ up to $I$ and then switches
to $\sigma'$.  Clearly $(\sigma,I,\sigma')$ differs from 
$\sigma$ only at or below $I$.  Moreover, since $(\sigma,I,\sigma')$ and
$\sigma'$ 
agree
with each other (and agree with $\sigma$) on all information
sets preceding $I$, it is easy to see that
$((\sigma,I,\sigma'),\tau_{-i})$ and $(\sigma',\tau_{-i})$ generate the
same distribution over outcomes for all $\tau_{-i} \in
\Sigma'_{\sigma,\sigma',I,-i}$.  Thus,
$u_i((\sigma,I,\sigma'),\tau_{-i}) = u_i(\sigma',\tau_{-i}) >
u_i(\sigma,\tau_{-i})$ for all $\tau_{-i} \in
\Sigma'_{\sigma,I,-i} =
\Sigma'_{\sigma,\sigma',I,-i}$. It follows that $\sigma$ is
conditionally dominated$'$.
\eprf

The following lemma is 
a corollary of Propositions \ref{pro:Pearce} and \ref{conddomprim.prop}.

\lem\label{lem:Pearce} A strategy $\sigma \in \Sigma'_i$ is not
conditionally dominated on 
$\Sigma'$ iff, for each information set $I$ of player $i$, either (a)
$\Sigma'_{\sigma,I,-i} = \emptyset$ or 
(b) there are some beliefs 
$\mu_{\sigma,I}$ that player $i$ can have on strategies in
$\Sigma'_{\sigma,I,-i}$ 
such that 
$\sigma$ is a best response conditional on reaching $I$ given these
beliefs; more precisely, for every strategy $\sigma'$ for player $i$
that agrees with $\sigma$ on every information set for player $i$
that precedes $I$, we have that 
$$\sum_{\tau_{-i} \in\Sigma'_{\sigma,I,-i}} \mu_{\sigma,I}(\tau_{-i})
u_i(\sigma,\tau_{-i}) \ge 
\sum_{\tau_{-i} \in\Sigma'_{\sigma,I,-i}} \mu_{\sigma,I}(\tau_{-i})
u_i(\sigma',\tau_{-i}).$$
\elem

Note that in Lemma~\ref{lem:Pearce} there are no ``coherence''
requirements on the beliefs $\mu_{\sigma,I}$; that is, if $I$ and $I'$ are two
information sets for player $i$, there is no necessary connection
between $\mu_{\sigma,I}$ and $\mu_{\sigma,I'}$.  
As we now show, it is not hard to add such coherence
constraints. Specifically, we can require that a player's beliefs are
updated by conditioning.
However, to do so, we have to model a player's beliefs so as to
allow for conditioning on information sets that, intuitively, are not
reachable according to the agents beliefs about other players'
strategies.  
We thus assume that a player's beliefs are represented by an LPS.
\commentout{
Recall that an LPS is a tuple  $(\Omega,\F,(\mu_0,\mu_1,
\ldots, \mu_k))$, where $\F$ is a $\sigma$-algebra over $\Omega$ and $\mu_0,
\ldots, \mu_k$ are probability distributions on $(\Omega,\F)$.
Intuitively, the first measure in the sequence, $\mu_0$, is the most
important one, followed by $\mu_1$, $\mu_2$, and so on.   Conditioning
on an event $E$ in an LPS is allowed as long as $\vecmu(E) > \vec{0}$;
that is, as long as $\mu_j(E) > 0$ for some $j$ with $0 \le j \le k$.  
(We remark that, in the sequel, we write $\vecmu(E) = \vec{0}$ and
$\vecmu(E) = \vec{1}$ as well, with the obvious interpretation.)
In this case, $\vecmu \mid E =
(\mu_{k_0}(\cdot \mid E), \ldots, \mu_{k_m}(\cdot \mid E))$, where $(k_0,
\ldots, k_m)$ is the subsequence of all indices for which the
probability of $E$ is positive.
} 
We say that $\sigma$ is an \emph{almost-best response} to
beliefs characterized by an LPS $\vecmu = (\mu_0, \ldots)$
if $\sigma$ is a best response to $\mu_0$.  This means that 
$\sigma$ is an almost best response to $\vecmu$ even if there is a
strategy $\sigma'$ for player $i$ such that $\sigma$ and $\sigma'$ give
the same expected utility with respect to $\mu_0$ and $\sigma'$ has
higher expected utility than $\sigma$ with respect to $\mu_1$.  Thus,
an almost-best response is perhaps best understood as an $\epsilon$-best
response, for an infinitesimal $\epsilon$, since we are ignoring differences
that arise due to the ``less important'' distributions.  We use
almost-best response here because it corresponds most closely to
Pearce's definitions.

\lem\label{lem:Pearce1} A strategy $\sigma \in \Sigma'_i$ is not
conditionally dominated on 
$\Sigma'$ iff there exists an LPS $\vecmu_\sigma$ on the strategies in
$\Sigma'_{-i}$ such that 
the support of $\vecmu_{\sigma}$ is all of $\Sigma'_{-i}$ and,
for each information set $I$ of player $i$, either (a)
$\Sigma'_{\sigma,I,-i} = \emptyset$ or  (b)
$\sigma$ is an almost-best response conditional on reaching $I$ to the 
beliefs $\vecmu_\sigma \mid \Sigma'_{\sigma,I,-i}$.   
\elem

\prf It immediately follows from Lemma~\ref{lem:Pearce} that if a
distribution $\vecmu_\sigma$ exists, then $\sigma$ is not conditionally
dominated on $\Sigma'$.  For the converse, suppose that $\sigma$ is not
conditionally dominated on $\Sigma'$. 
We construct $\vecmu$ inductively.  Consider an information set $I_0$
for player $i$ that is not preceded by any other information set for
player $I$.  By Lemma~\ref{lem:Pearce}, there is a belief
$\mu_{\sigma,I_0}$ that player $i$ could have such that $\sigma$ is an
almost-best
response to $\mu_{\sigma,I_0}$ conditional on reaching $I_0$.  Let $\mu_0 =
\mu_{\sigma,I_0}$.  It is easy to check that if $I'$ is an information
set for player $i$ such that
$\mu_0(\Sigma'_{\sigma,I',-i}) >  
0$, then $\sigma$ is an almost-best response conditional on reaching $I'$ to the  
beliefs $\mu_0 \mid \Sigma'_{\sigma,I',-i}$.   If there are information
sets $I'$ for player $i$ such that $\mu_0(\Sigma'_{\sigma,I',-i}) = 0$, let
$I_1$ be an information for player $i$ such that
$\mu_0(\Sigma'_{\sigma,I_1,-i}) = 0$ and $I_1$ is 
not preceded by an information set $I'$ for player
$i$ such that 
$\mu_0(\Sigma'_{\sigma,I',-i}) = 0$.  Let $\mu_1$ be the
distribution $\mu_{\sigma,I_1}$ guaranteed to exist by
Lemma~\ref{lem:Pearce}.  We continue inductively.  Suppose that we have
defined $\mu_0, \ldots, \mu_k$ and 
information sets $I_0, \ldots, I_k$ such that $(\mu_0, \ldots,
\mu_{j-1})(\Sigma'_{\sigma,I_j,-i}) = \vec{0}$, $I_j$ is not preceded 
by an information set $I'$ for player $i$ such that
$(\mu_0,\ldots,\mu_{j-1})(\Sigma'_{\sigma,I',-i}) = \vec{0}$, and
$\mu_j = \mu_{\sigma,I_j}$.  If there is an information set $I'$ such that 
$(\mu_0,\ldots,\mu_{k})(\Sigma'_{\sigma,I',-i}) = \vec{0}$, then we
continue in this way to construct $I_{k+1}$ and $\mu_{k+1}$.  Clearly
this construction must terminate, since there are only finitely many
pure strategy profiles.
Finally, for all the strategy profiles $\tau_{-i} \in \Sigma'_{-i}$
that are not in the support of the LPS constructed so far, we simply add
an additional component in the LPS assigning the uniform
distribution to these remaining profiles. 
\eprf

Before defining EFR, it is worth comparing
conditional domination to weak domination: weak domination is
characterized by exactly the same conditions as in Lemma
\ref{lem:Pearce1}, except that instead of requiring that the LPS have
full support, we require that the first component of the LPS have full
support. More
formally, 
\lem\label{lem:Pearce2} A strategy $\sigma \in \Sigma'_i$ is not
weakly dominated on 
$\Sigma'$ iff there exists an LPS $\vecmu_{\sigma} = (\mu_{\sigma,0},
\ldots),$ on the strategies in 
$\Sigma'_{-i}$ such that
the support of $\sigma,0$ is all of $\Sigma'_{-i}$ and,
for each information set $I$ of player $i$, either (a)
$\Sigma'_{\sigma,I,-i} = \emptyset$ or  (b)
$\sigma$ is an almost-best response conditional on reaching $I$ to the 
beliefs $\vecmu_\sigma \mid \Sigma'_{\sigma,I,-i}$.   
\elem

\prf Suppose that there exists a distribution $\vecmu_{\sigma}$ as
specified by the lemma. Since $\cup_{I} \Sigma'_{\sigma,I,-i} =
\Sigma'_{-i}$, it follows that $\sigma$ is a best response to
the full-support belief $\mu_{\sigma,0}$ and thus, by
Proposition \ref{pro:Pearce}, is not weakly dominated.

For the converse, suppose that $\sigma$ is not
weakly dominated on $\Sigma'$. By Proposition \ref{pro:Pearce}, there
exists a distribution $\mu$ on $\Sigma'_{-i}$ whose support is all of
$\Sigma'_{-i}$ such that $\sigma$ is an almost-best response to $\mu$.
Define $\vecmu_{\sigma} = (\mu, \mu, \ldots)$. Clearly the first
component of $\vecmu_{\sigma}$ has full support; moreover, for every
$I$ such that $\Sigma'_{\sigma,I,-i}$ is non-empty, 
$\sigma$ is an almost-best response conditional on reaching $I$ to the beliefs
$\vecmu_{\sigma} \mid  \Sigma'_{\sigma,I,-i}$.
\eprf

We now define EFR.
Define a sequence of sets $\NCD^0_j, \NCD^1_j, \cdots$ of pure strategies for
each player $j$ as 
follows: $\NCD^0_j = \Sigma_j(\Gamma)$, and $\NCD^{k+1}_j$ consists of
the strategies $\sigma \in \NCD^k_j$ that are not conditionally dominated
on $\NCD^k_1 \times \cdots \times \NCD^k_n$.  Clearly we have $\NCD^0_j
\supseteq \NCD^1_j \supseteq \ldots$.  Let $\NCD^\infty_j = \inter_{k=1}^\infty
\NCD^k_j$.  It easily follows from Lemma~\ref{lem:Pearce} that $\NCD^k_j \ne
\emptyset$ for all $k$.  Since $\NCD^0_j$ is finite, we must have
$\NCD^\infty_j = \NCD^k_j$ for some $k$.  

\dfn A strategy $\sigma$ for player $j$ is \emph{extensive-form
rationalizable} if $\sigma \in \NCD^\infty_j$.
\edfn

\commentout{
Battigalli and Sinischalchi \citeyear{BS02} 
provide an epistemic characterization of EFR using a notion of
\emph{strong belief}; these are beliefs that are maintained unless
evidence shows that the beliefs are inconsistent.  For example, if player 1
has a strong belief in player 2's rationality, then whatever moves
player 2 makes, player 1 will revise her beliefs and, in particular, her
beliefs about player 2's beliefs, in such a way that she
continues to believe that player 2 is rational (so that she believes that
player 2 is making a best response to his beliefs), unless it is
inconsistent for her to believe that player 2 is rational.  Battigalli
and Sinischalchi characterize EFR in terms of common strong belief of
rationality.   like BFK, Battigalli and Sinischalchi assume complete
structures.  Here we show how EFR can be characterized using the
``all I know'' notion.  The analysis brings out the relationship between
EFR and iterated admissibility.
}

We now provide an epistemic characterization of EFR using the ``all I
know'' operator.
The notion of a structure appropriate for an extensive-form game $\Gamma$ is
a little different than that of a structure appropriate for a normal-form
game.  Now a structure has the form of a tuple
$(\Omega,\strat, \F, \PR_1, \ldots, \PR_n)$, where $\Omega$ and 
$\strat$ are as before, $\F$ is a $\sigma$-algebra over $\Omega \times \IN$,
and $\PR_i$ associates, with each state $\omega$ and natural number $m$
(thought of as a time step), player
$i$'s beliefs at $(\omega,m)$, but now the beliefs are LPSs, so
$\PR_i(\omega,m)$ has the form $(\mu_{i0}, \ldots, \mu_{ik})$ for some $k
\ge 0$, where $\mu_{ij}$ is a probability distribution on $(\Omega,\F)$.  
For convenience, we denote $\mu_{ij}$ as $\PR_{ij}(\omega,m)$.  
Let $\node(\omega,m)$ denote the $m$th node on 
the
path in the game tree
determined by $\strat(\omega)$, where we take $\node(\omega,0)$ to be
the root of the tree and take $\node(\omega,m)$ to be the leaf of the
path if the path has length less than $m$.  Intuitively,
$\PR_i(\omega,m)$ represents $i$'s beliefs at $\node(\omega,m)$.

A structure for an extensive-form game $\Gamma$ is \emph{appropriate} if
it satisfies the six conditions below.
The first four conditions are essentially the same as the
corresponding conditions for appropriateness in normal-form games.
The fifth condition says that a player must ascribe probability 1 to
being in his information set, and the sixth condition is a (significant)
weakening of the requirement that updating proceeds by conditioning.
\begin{enumerate}
\item For each strategy $\sigma_i$ for player $i$, 
$\intension{\sigma_i} = \{(\omega,m): \strat_i(\omega) = \sigma_i\} \in \F$.
\item $\PR_{i}(\omega,m)(\intension{\strat_i(\omega)}) = \vec{1}$.
\item For each LPS $\vec{\mu}$
on $(\Omega, \F)$ and player $i$, $\intension{\vec{\mu},i}
=\{(\omega,m) : \PR_i(\omega,m) = \vec{\mu}\} \in \F$;
\item $\PR_{i}(\omega,m)(\intension{\PR_i(\omega,m),i}) = \vec{1}$.
\item If $\node(\omega,m) = v$ and $v$ is in $i$'s information set $I$,
then $\PR_{i}(\omega,m)(\intension{I}) = \vec{1}$,  where $\intension{I} =
\{(\omega',m'): \node(\omega',m') \in I\}$. 
Moreover, even if player $i$ does not move at $\node(\omega,0)$,
$\PR_i(\omega,0)(\{(\omega',0): \omega'\in \Omega\}) = \vec{1}$. 
\item 
If $\PR_i(\omega,m+1)(\intension{\play_{-i}(\tau_{-i})}) \ne \vec{0}$, then
$\PR_i(\omega,m)(\intension{\play_{-i}(\tau_{-i})}) \ne \vec{0}$. 
\end{enumerate}
We can think of this definition as generalizing our earlier definition
in the following sense.  A normal-form game can be viewed as a special
case of an extensive-form game where the game tree has only one node and
all players move at that node.  A structure appropriate for a normal-form
game $\Gamma$ then satisfies the six conditions above (the last two are
satisfied vacuously), with the added constraint that $\PR_i(\omega,m)$ is
an LPS of length one.  
It is worth 
noting that  here we do \emph{not} require that players update their
beliefs by conditioning.  
Requirement 6 above is a (significant)
weakening of that requirement.  

We now make precise what it means to update by conditioning in our
setting.  There is an issue here because, in game theory, a player $i$'s 
beliefs are typically defined only at an information set $I$ where
player $i$ moves.  Here, the function $\PR_i$ is defined for all
$(\omega,m)$.  It turns out that it doesn't really matter how we define
player $i$'s beliefs at $(\omega,m)$ if $i$ player does not move at
$\node(\omega,m)$.  For definiteness, we assume that player $i$'s
beliefs do not change at nodes where $i$ does not move.

\dfn An appropriate structure $M = (\Omega,\strat,\F,\PR_1,\ldots,
\PR_n)$ for $\Gamma$ \emph{respects conditioning}
if it satisfies the following condition:
\begin{itemize}
\item if player $i$ does not move at $\node(\omega,m)$ and $m > 0$, then 
$\PR_i(\omega,m) = \PR_i(\omega,m-1)$; if $i$ does move at
$\node(\omega,m)$, $\node(\omega,m)$ is in information set $I$, 
and $\PR'_i(\omega,m)$ is the projection of $\PR_i(\omega,m)$ onto $\Omega$,
then $\PR'_i(\omega,m)(\omega') = \PR'_i(\omega,m-1)'(\omega' \mid
\intension{I})$.  
\end{itemize}
\edfn

We want to define the truth of a formula at a pair $(\omega,m)$ in an
appropriate structure $M$.
The definitions of the operators $\Diamond$ and $O_i^\L$ for a given
language $\L$ remain the same as before. However, we have to extend the
definitions of $B_i$%
, $\dbi$, 
and $\RAT_i$ to this setting.   We take $B_i \phi$
(resp., $\dbi \phi$)
to be true at $(\omega,m)$ if $\PR_i(\omega,m)(\intension{\phi}) =
\vec{1}$ 
(resp., $\PR_i(\omega,m)(\intension{\phi}) >
\vec{0}$). 
But for $\RAT_i$, we consider only $\PR_{i0}$, not $\PR_i$,
since this seems most appropriate when capturing EFR (see the
discussion just before Lemma~\ref{lem:Pearce1}).
Also, we 
take a player to be trivially rational at a
node if she does not move there.  
\begin{itemize}
\item $(M,\omega,m) \sat B_i \phi$ if there exists a set $F \in \F$ such
that $F \subseteq \intension{\phi}$ and $\PR_i(\omega,m)(F) = \vec{1}$.
\item $(M,\omega,m) \sat \dbi \phi$ if there exists a set $F \in \F$ such
that $F \subseteq \intension{\phi}$ and $\PR_{i}(\omega,m)(F) \ne \vec{0}$.
\item $(M,\omega,m) \sat \RAT_i$ if either (a) player $i$ does not move at
$\node(\omega,m)$ or (b) $i$ moves at $\node(\omega,m)$, $\node(\omega,m)$ is
in information set $I$, and $\strat_i(\omega)$ is a best response,
conditional on reaching information set $I$, 
given player $i$'s beliefs on the strategies of other players induced by
$\PR_{i0}(\omega,m)$.
\end{itemize}
We also use two more operators, $\init$ and $G$.
The formula $\init(\phi)$ is true at $(\omega,m)$ 
if $\phi$ is true at $(\omega,0)$;
$G \phi$ is true at $(\omega,m)$ if $\phi$ is true at all the nodes below 
$(\omega,m)$ (i.e., at $(\omega,m')$ for all $m' \ge m$).  Formally,
\begin{itemize}
\item 
$(M,\omega,m) \sat \init(\phi)$ if $(M,\omega,0) \sat \phi$;
\item $(M,\omega,m) \sat G \phi$ if  $(M,\omega,m') \sat \phi$ for all
times $m' \ge m$.
\end{itemize}

We now define the formula $\RATEFR^k_i$ by induction on $k$.
$\RATEFR^0_i$ is just $\true$, and 
$\RATEFR^{k+1}_i$ is an abbreviation of
$$\begin{array}{cc}
\init(B_i G (\RAT_i)) \land \PLAYCONefr^k_i \land
\init(O^{\L^0_{-i}}_i (\land_{j\ne i} \RATEFR^k_j)),
\end{array}$$
where $\PLAYCONefr^k_i$ is an abbreviation of
$$\land_{\sigma' \in \Sigma_i(\Gamma)} (\play_i(\sigma') \rimp
\Diamond(\play_i(\sigma') \land \RATEFR^k_i)).
$$ 
$\RATEFR^k_i$ is intended to be true at a world exactly if the strategy
used at that world has survived $k$ levels of iterated deletion of
conditionally dominated strategies.  
This is similar in spirit to the definition of $\RATzero_i^k$.  The
condition $\RAT_i$ (which is equivalent to $B_i \RAT_i$) is replaced by 
$\init(B_i G (\RAT_i))$; this effectively says that player $i$ is
rational at every node in the tree that is reachable by some strategy
profile in the structure.  Similarly, $\init(O^{\L^0_{-i}}_i (\land_{j\ne i}
\RATEFR^k_j))$ guarantees that all the players are $k$-level rational
at the beginning of play.  
Somewhat informally, $\RATEFR^{k+1}_i$
is true at  $(\omega,m)$ iff (a) $\strat_i(\omega)$ survives
$k$ rounds of iterated deletion of conditionally dominated 
strategies, (b) at the begining of play (i.e., at $(\omega,0)$), $i$
considers 
possible all strategies for other players that 
survive $k$ rounds of iterated deletion of conditionally dominated
strategies, and (c) conditional on every information set 
that could arise when he is using strategy $\sigma_i$ where $i$ plays,
$\sigma_i$ is a best response to the strategies played by the other
players.

At first sight, it might seem surprising that $\RATEFR^k_i$ captures
that player $i$'s strategy survives $k$ levels of iterated deletion of
conditionally dominated strategies: In the definition of $\RATEFR^{k+1}_i$
we require that \emph{all} player $i$ knows (about players $-i$) is
that they satisfy $\RATEFR^k_{-i}$; thus, roughly speaking, $i$ should assign
positive probability to all strategies consistent with
$k$-level rationality. Intuitively, this should lead to removing
strategies that are 
\emph{weakly} dominated. In contrast, the definition of conditional
dominance asks for removing strictly dominated strategies.
This ``paradox'' is resolved by noting that our current definition of ``all
$i$ knows'' requires only that $i$ assign positive probability to all
strategies consistent with $k$-level rationality at \emph{some} level in
the LPS 
describing his beliefs, whereas our definition of $\RAT_i$ requires $i$ to
be playing a best response with respect to only his \emph{top} level
belief; this is what prevents us from also removing
weakly dominated strategies.

On the other hand, $B_i \RAT_i$
requires 
$\RAT_i$ to hold at every world that $i$ assigns positive probability
to at \emph{some} level in its LPS; this is what makes us remove
strategies that are strictly dominated at any information set that can
be reached by strategies consistent with $(k-1)$-level rationality. In
other words, we remove all conditionally dominated strategies.

\thm\label{thm:charefr} The following are equivalent for a strategy
$\sigma \in \Sigma_i(\Gamma)$:
\begin{itemize}
\item[(a)] $\sigma \in \NCD_i^k$; 
\item[(b)] there exists an $\L^1$-measurable structure  $M$
that is appropriate for $\Gamma$ and respects conditioning, a state
$\omega$, and a time $m$ such that 
$\strat_i(\omega) = \sigma$ and $(M,\omega,m) \sat \RATEFR^{k}_i$;
\item[(c)] there exists a structure  $M$
that is appropriate for $\Gamma$ and respects conditioning, a state
$\omega$, and a time $m$ such that $\strat_i(\omega) = \sigma$ and
$(M,\omega,m) \sat \RATEFR^{k}_i$ for all $k \ge 0$;  
\item[(d)] there exists an $\L^1$-measurable structure  $M$
that is appropriate for $\Gamma$, a state $\omega$, and a time $m$ such that
$\strat_i(\omega) = \sigma$ and $(M,\omega,m) \sat \RATEFR^{k}_i$;
\item[(e)] there exists a structure  $M$
that is appropriate for $\Gamma$, a state $\omega$, and a time $m$ such that
$\strat_i(\omega) = \sigma$ and $(M,\omega,m) \sat \RATEFR^{k}_i$ for all
$k \ge 0$.  
\end{itemize}
In addition,  there is an appropriate structure $\bar{M}^k =
(\Omega^k,\strat, \F, \PR_1, \ldots, \PR_n)$ 
that respects conditioning
such that 
$\Omega^k = \{(k',i,\vec{\sigma}): k' \le k, 1 \le i \le n, \vec{\sigma}
\in \NCD^{k'}_1 \times \cdots \times \NCD^{k'}_n
\}$, 
$\strat(k',i,\vec{\sigma}) = 
\vec{\sigma}$, 
$\F = 2^{\Omega^k \times \IN}$, and
for all states 
$(k', i, \vec{\sigma}) \in \Omega^k$ and $m \ge 0$,
$(\bar{M}^k,(k',i, \vec{\sigma}),m) \sat \land_{j \ne i} \RATEFR_j^{k'}$.
\ethm

\prf As before, we proceed by induction on $k$, proving both the
equivalence of (a), (b), (c), (d), and (e)  
and the existence of a structure $\bar{M}^k$ with the required
properties.

The result clearly holds if $k=0$.
Suppose that the result holds for $k$; we prove  it for $k+1$.
We first show that (e) implies (a).  Suppose that 
there exists a structure  $M$
that is appropriate for $\Gamma$ and a state $\omega$ such that
$\strat_i(\omega) = \sigma$ and $(M,\omega,m) \sat \RATEFR^{k+1}_i$.
Since $(M,\omega,m) \sat \RATEFR^{k+1}_i$, by definition, we must have 
$(M,\omega,m) \sat \PLAYCONefr^k_i$, so there must exist some structure $M'$
appropriate for $\Gamma$, state $\omega'$, and time $m'$ such that
$(M',\omega',m') \sat 
\play_i(\sigma) \land \RATEFR^k_i$.  Thus, by the induction hypothesis, 
$\sigma \in NCD^k_i$.  To show that $\sigma \in \NCD^{k+1}_i$, by
Lemma~\ref{lem:Pearce}, it suffices to show that at each information set
$I$ for player $i$ such that $\NCD^k_{\sigma,I,-i} \ne \emptyset$, there
are beliefs $\mu_{\sigma,I}$ that $i$ can have on $\NCD^k_{\sigma,I,-i}$
such that 
$\sigma$ is a best response to these beliefs conditional on reaching
$I$.   So suppose that $\NCD^k_{\sigma,I,-i} \ne \emptyset$.  Choose some 
strategy $\tau_{-i} \in \NCD_{\sigma,I,-i}$ and let $v \in I$ be a node
reached by $(\sigma,\tau_{-i})$.  
Since $(M,\omega,m) \sat
\init(O^{\L^0_{-i}}_i (\land_{j\ne i} \RATEFR^k_j)) \land 
\init(B_i G (\RAT_i))$, it must be the case that  
$(M,\omega,0) \sat O^{\L^0_{-i}}_i (\land_{j\ne i} \RATEFR^k_j) \land 
B_i G (\RAT_i)$.  By the induction
hypothesis, 
$(\bar{M}^{k},(k,i,(\sigma,{\tau}_{-i})),0) \sat  
\play_{-i}(\tau_{-i}) \land (\land_{j\ne i} \RATEFR^k_j)$.
Thus, $(M,\omega,0) \sat \Diamond (\play_{-i}(\tau_{-i}) \land
(\land_{j\ne i}\RATEFR^k_j))$.  It follows that $\PR_i(\omega,0)(\omega',0) \ne
\vec{0}$ for some state $\omega'$ such that 
$\strat(\omega') = (\sigma,\tau_{-i})$.  
Moreover, we must have $(M,\omega',0) \sat G (\RAT_i)$.
Choose $m''$ such that $\node(\omega',m'') = v$.
We must have  $(M,\omega',m'') \sat \RAT_i$.  This
means that $\PR_{i}(\omega',m'')$ projected onto strategies defines a
probility $\mu_{\sigma,I}$ on $\Sigma_{-i}$ 
for which $\sigma$ is a best
response.  
Moreover, by condition 6, every strategy profile to which
$\mu_{\sigma,I}$ gives positive probability must 
get positive probability according to $\PR_i(\omega',0)$.  But
$\PR_i(\omega',0) = \PR_i(\omega,0)$, so,
since $(M,\omega,0) \sat O^{\L^0_{-i}}_i (\land_{j\ne i}
\RATEFR^k_j)$, by the induction hypothesis, the support of
$\mu_{\sigma,I}$ must be in $\NCD^k_{-i}$.  
This completes the argument.

We next construct the structure $\bar{M}^{k+1} =
(\Omega^{k+1},\strat, \F, \PR_1, \ldots, \PR_n)$. 
The construction is essentially identical to the one used in our
characterization of iterated admissibility (i.e., in the proof of
Theorem \ref{thm:charwd}); we just need to adapt the construction to take
into account the fact that we are considering extensive-form games, and,
in particular, 
to ensure that 
beliefs are updated by conditioning. 
We define $\Omega^{k+1}$, $\strat$, and $\F$ as required in the
theorem.  It remains to define $\PR_1, \ldots, \PR_n$.  
For each strategy $\sigma \in \Sigma_j(\Gamma)$, we define 
$\PR_j(\omega,m)$ at all states $\omega = (k', i,\vec{\sigma})$ such that
$\sigma_j = \sigma$ 
by induction on $k'$.
If $\omega$ is a state of the form $(0, i,\vec{\sigma})$, given $m$,
let $\PR_j(\omega,m)$ be an arbitrary distribution over states.
Consider states $\omega = (k', i,\vec{\sigma})$
where $k' \geq 1$.
By Lemma~\ref{lem:Pearce1}, there is 
a full-support 
LPS $\vecmu_\sigma$ 
on the strategies in $\NCD^{k'}_{-j}$ such that, 
for each information set $I$ of player $j$, either (a)
$\NCD^{k'}_{\sigma,I,-i} = \emptyset$ or  (b)
$\sigma$ is a best response conditional on reaching $I$ to the 
beliefs $\vecmu_\sigma \mid \NCD^k_{\sigma,I,-i}$.   
Let $m' \leq m$ be the largest integer such that
$\node(\vec{\sigma},m)$ is in an information set $I$ where $j$ moves
(i.e., $m'$ is the latest time that $j$ moves);
if no such information set exists, let $m'=0$.
If $m' > 0$ then let $(\mu'_0,\ldots, \mu'_k) = \vecmu_\sigma \mid
\Sigma'_{\sigma,I}$; otherwise, let $(\mu'_0,\ldots, \mu'_k) =
\vecmu_\sigma$. 
\begin{itemize}
\item If $i\ne j$, then $\PR_j(\omega,m) = (\mu''_0, \ldots,
\mu''_k)$, where  
$\mu''_j((k'',i',\vec{\tau}),m'') = \mu_j'(\vec{\tau})$ if
$k'' = k'-1$, $i' = j$, $\tau_j = \sigma_j$, and $m'' = m$, and 0 otherwise.
\item If $i= j$, then $\PR_j(\omega,m) = (\mu''_0, \ldots, \mu''_k)$, where 
$\mu''_j((k'',i',\vec{\tau}),m'') = \mu_j'(\vec{\tau})$ if
$k'' = k'$, $i' = j$, $\tau_j = \sigma_j$, and $m'' = m$, and 0 otherwise.
\end{itemize}

We leave it to the reader to check that
$\bar{M}^k$ is appropriate
and respects conditioning%
, and that $(\bar{M}^k,(k',i,\vec{\sigma}),m)
\sat \land_{j \ne i} \RATEFR_j^{k'}$.

To see that (a) implies (b), suppose that $\sigma_j \in \NCD^{k+1}_j$.
Choose a state $\omega$ in 
$\bar{M}^{k+1}$ of the form $(k+1,i,\vec{\sigma})$, where $i \neq
j$ and $v$ is the root of the tree.
As we just showed, 
 $(\bar{M}^{k+1}, \omega,m) \sat \RATEFR^{k+1}_j$ and
$\strat_j(\omega) = \sigma_j$.  Moreover, $\bar{M}^{k+1}$ is
$\L^1$-measurable (since $\F$ consists of all subsets of $\Omega^{k+1}$).

Clearly (b) implies both (c) and (d), and both (c) and (d) imply (e).
This completes the proof.
\eprf

We end this section by formally comparing EFR to IA.
Recall our current definition of $\dbi$: $(M,\omega,m)
\sat \dbi \phi$ if there exists a  set $F \in \F$ such
that $F \subseteq \intension{\phi}$ and $\PR_{i}(\omega,m)(F) \ne
\vec{0}$; in other words, $\dbi \phi$ holds at $\omega$ if \emph{some}
component of $i$'s LPS assigns $\phi$ positive probability. We now
show that if we instead require that the \emph{first} component of $i$'s LPS
assigns $\phi$ positive probability, then $\RATEFR_i^{k}$ instead
characterizes strategies in $\NWD_i^k$. As the only place where $\dbi$
is used in $\RATEFR_i^{k}$ is in the definition of the ``all I know''
operator, the difference between EFR and iterated admissibility can be
understood as a difference in how strongly players believe that all
they know about other players is that they are consistent with
appropriate levels of rationality.

More precisely, we define $(M,\omega) \sat \dbip \phi$ 
if there exists a set $F \in \F$ such
that $F \subseteq \intension{\phi}$ and $\PR_{i0}(\omega,m)(F) \ne
0$.  Define $(O')^{\L}_i$ just as $O^{\L}_i$, except
that $\dbi$ is replaced by $\dbip$.  Finally, define
$\RATEFRp_i^k$ just as $\RATEFR_i^k$, except that 
$O^{\L}_i$ is replaced by $(O')^{\L}_i$.

\thm\label{thm:charefrwd} 
The following are equivalent: 
\begin{itemize}
\item[(a)] The strategy $\sigma \in \Sigma_i(\Gamma)$ is in $\NWD_i^k$; 
\item[(b)] there exists an $\L^1$-measurable structure  $M$
that is appropriate for $\Gamma$ and respects conditioning, a state
$\omega$, and a time $m$ such that 
$\strat_i(\omega) = \sigma$ and $(M,\omega,m) \sat \RATEFRp^{k}_i$;
\item[(c)] there exists a structure  $M$
that is appropriate for $\Gamma$ and respects conditioning, a state
$\omega$, and a time $m$ such that $\strat_i(\omega) = \sigma$ and
$(M,\omega,m) \sat \RATEFRp^{k}_i$ for all $k \ge 0$;  
\item[(d)] there exists an $\L^1$-measurable structure  $M$
that is appropriate for $\Gamma$, a state $\omega$, and a time $m$ such that
$\strat_i(\omega) = \sigma$ and $(M,\omega,m) \sat \RATEFRp^{k}_i$;
\item[(e)] there exists a structure  $M$
that is appropriate for $\Gamma$, a state $\omega$, and a time $m$ such that
$\strat_i(\omega) = \sigma$ and $(M,\omega,m) \sat \RATEFRp^{k}_i$ for all
$k \ge 0$.  
\end{itemize}
In addition,  there is an appropriate structure $\bar{M}^k =
(\Omega^k,\strat, \F, \PR_1, \ldots, \PR_n)$ 
that respects conditioning
such that 
$\Omega^k = \{(k',i,\vec{\sigma}): k' \le k, 1 \le i \le n, \vec{\sigma}
\in \NCD^{k'}_1 \times \cdots \times \NCD^{k'}_n \}$, 
$\strat(k',i,\vec{\sigma}) = \vec{\sigma}$, 
$\F = 2^{\Omega^k \times \IN}$, and for all states 
$(k', i, \vec{\sigma}) \in \Omega^k$ and $m \ge 0$,
$(\bar{M}^k,(k',i, \vec{\sigma}),m) \sat \land_{j \ne i} \RATEFRp_j^{k'}$.
\ethm

\prf As before, we proceed by induction on $k$, proving both the
equivalence of (a), (b), (c), (d), and (e)  
and the existence of a structure $\bar{M}^k$ with the required
properties.  The result clearly holds if $k=0$.
Suppose that the result holds for $k$; we prove it for $k+1$.
We first show that (e) implies (a). This follow as in the
proof that (c) implies (a) in Theorem \ref{thm:charwd}, since
$\init(B_i G (\RAT_i))$ implies $\RAT_i$, $\PLAYCONefr^k_i$ implies
$\PLAYCONzero^k_i$, and $\dbip$ is equivalent to $\dbi$ in the setting
of Theorem~\ref{thm:charwd} (which uses probability measures---length 1 LPSs).
The remainder of the proof proceeds exactly as the proof of Theorem
\ref{thm:charefr},
except that instead of using Lemma \ref{lem:Pearce1}, we
appeal to Lemma \ref{lem:Pearce2}.
\eprf
}

\section{Discussion}
\label{sec:discussion}
\fullv{
We have provided a logical formula
that captures
the intuition that ``all a player knows''
is that players satisfy appropriate rationality
assumptions.
Our formalization of ``all player $i$ knows is $\phi$'' is in terms of a
language $\L$: roughly speaking, we require that $i$ assigns positive
probability to all formulas $\psi \in \L$ that are consistent with
$\phi$. 
We provided a formula that is intended to capture the intuition that all
player $i$ knows (in language $\L$) is that all the other players are
$k$-level rational.
We showed that when $\L$ expresses statements about the strategies
played by the other players, our logical formula characterizes
strategies that are iterated admissible (i.e., survive iterated deletion
of weakly dominated strategies). 
On the other hand, when $\L$ is less expressive (e.g., empty) the same
logical formula instead characterizes strategies surviving iterated
deletion of strictly dominated strategies (and thus also characterizes
rationalizable strategies). Thus, the expressiveness of the language
used by the players to describe their beliefs about other 
players can be viewed as affecting how the game is played. 
In addition, we showed that if $\L$ expresses statements about the
strategies used by the other players, then a variant of this formula,
more appropriate for reasoning about extensive-form games, characterizes
Pearce's notion of extensive-form rationalizability.

These results suggest that the notion of ``all I know'' is really at the
heart of many of the intuitions underlying solution concepts.
We plan to consider the effect of the players using other languages to
describe their beliefs.  For example, we are interested in the solution
concept that arises if the language includes
$\RAT_i$ for each player $i$ but does not include $\play_i(\sigma)$, so
that players can talk about the rationality of other players without
talking about the strategies they use.  
We would also like to consider other ways of incorporating restrictions
on how players form beliefs about other players. For example, we could
restrict players' beliefs to 
be consistent with a theory $\TT$; namely, we might require $i$ to assign
positive probability to all and only formulas $\psi$ consistent with both
rationality and $\TT$
(or, essentially equivalently, to all and only formulas that can be
satisfied in some class of Kripke structures).
Such restrictions provide a straighforward way of capturing
players' prior beliefs about other players. 
They may also be used to capture the way
``boundedly rational'' players reason about each other. But what are
``natural'' restrictions? And how do such restriction affect how the
game will be played?  
We leave an exploration of these questions for future reseach.
}
\shortv{
We have used the ``all I know'' operator introduced in our earlier
paper to provide an epistemic characterization of IA that deals with
Samuelson's conceptual concerns.  We actually provided two
characterizations, one in LPS structures and one in probability
structures.  The former uses a generalized belief operator, while the
latter uses a generalized approximate belief operator.  These operators
may be of independent interest.  For example, a logic with a
generalized belief operator may be an appropriate logic in which to
describe belief revision \cite{agm:85,katsuno_mendelzon:91} and
iterated belief revision \cite{Darwiche94}, since it allows us to
describe how beliefs would be revised.   It clearly has deep
connections with counterfactual reasoning as well.
For example, in a logic of counterfactuals, a formula such as
$B_i(\phi_1,\phi_2)$ can be viewed as an abbreviation of $B_i(\phi_1)
\land (\neg B_i(\phi_1) \Cond_i B_i(\phi_2))$, where $\Cond_i$ is a
counterfactual operator (see \cite{Hal31} for a discussion of and
semantics for this standard operator); the formula $B_i(\phi_1, \ldots,
\phi_k)$ can be expressed using counterfactuals in a similar way.
It would of interest to axiomatize the logic of generalized belief.

The more quantitative operator $B_i^\delta$ may also be of independent
interest.  Interestingly, in \emph{cognitive hierarchy theory} (CHT)
\cite{CHC04}, there are assumed to be different types of
players: roughly speaking, \emph{level-$k$} players are assumed to be
$k$-level rational, and players assign probabilities to a player being 
of level-$k$.  Whereas in our characterization of IA, level-$k$
players are assigned the highest probability, followed by level-$k-1$,
and so on, in CHT it is the other way around.  In any case, having an
operator like $B_i^\delta$ may allow a more realistic characterization
of players beliefs than a purely qualitative generalized belief
operator.

Finally, while we have focused here only on IA, in other work
\cite{HP11a}, we have also used the notion of all I
know to characterize Pearce's notion of
\emph{extensive-form rationalizability} \cite{Pearce84}, a
well-studied solution concept in extensive-form games that also
involves iterated deletion.  That
characterization too used a variant of the $\PLAYCONzero_i$ formula, and
thus does not address Samuelson's concerns.  Although we have not
yet checked details, it seems that we should also be able to get a
characterization of extensive-form rationalizability using the
techniques of this paper.  All this suggests that thinking in terms of
an ``all I know'' operator and generalized belief may provide further
insights into solution concepts.
}

\fullv{
\appendix
\section{Proofs}
In this appendix, we prove the results stated earlier.
We repeat the statements for the reader's convenience.
We start with Proposition~\ref{lem:GRAT}, which is needed for the proof of
Theorem~\ref{thm:charwdg}. 

\medskip

\opro{lem:GRAT}
  Suppose that $M$ is an appropriate model for game $\Gamma$,
  $(M,\omega) \sat \GRATzero_i^{k+1}$, and $\PR_i(\omega) = \vec{\mu}
  = (\mu_0, \ldots, \mu_m)$.
\begin{itemize}
  \item[(a)] For all strategy profiles $\vec{\tau}_{-i}
    \in\Sigma_{-i}$, we have 
    $\vec{\mu}(\intension{\play(\tau_{-i})}) \ne \vec{0}$. 
  \item[(b)] If  $\NWD^1_j \ne \NWD^0_j$ for at least two players $j$,
    then $(M,\omega) \sat \neg \GRATzero^1_i \land
      \ldots \land \neg  \GRATzero^{k}_i$.
    \item[(c)] If $h < h' \le k$, then $\intension{H^{h'}_{-i}}
      \gg_{\mu} \intension{H^{h}_{-i}}$, 
      where $H^{k'}_{-i}$ is an abbreviation of the formula
$\GRAT_{-i}^{k'} \land \neg
      \GRAT_{-i}^{k'+1} \land \ldots \land \neg \GRAT_{-i}^{k}$ (so
      $H^k_{-i}$ is $\GRAT_{-i}^k$).
      Moreover, 
      for all $h \le k$ and strategy profiles
      $\vec{\tau}_{-i}$ and $\vec{\tau}'_{-i}$, if
      $\intension{\GRAT_{-i}^h \land \play(\vec{\tau}_{-i})} \ne
  \emptyset$, and     $\intension{(\GRAT_{-i}^h \lor \ldots \lor
    \GRAT_{-i}^k) \land 
  \play(\vec{\tau}_{-i}')} =  \emptyset$, then
$\intension{\play(\vec{\tau}_{-i})} 
      \gg_{\vec{\mu}} \intension{\play(\vec{\tau}'_{-i})}$. 
\end{itemize}
\eopro
\prf
Assume that the hypotheses of the proposition holds.  Since
$(M,\omega) \sat \GRATzero_i^{k+1}$, we have that
$(M,\omega) \sat O^{\L^0_{-i}}_i (\GRATzero^k_{-i},
\ldots, \GRATzero^0_{-i})$.
For part (a), note that the formulas $H_{-i}^0, \ldots, H_{-i}^k$ are mutually
exclusive and exhaustive (so that $H_{-i}^0 \lor \ldots \lor
  H_{-i}^k$ is equivalent to $\true$).  Thus, there must be some $h$
  with $0 \le h \le k$  
such that $(M,\omega) \sat \Diamond(H_{-i}^h \land \play(\vec{\tau}_{-i}))$.
It follows from the 
definition of $O^{\L^0_{-i}}_i (\GRATzero^k_{-i},
\ldots, \GRATzero^0_{-i})$ that
$\vec{\mu}(\intension{\play(\vec{\tau}_{-i})} \mid 
\intension{\neg \GRAT_{-i}^k \land \ldots \land \neg
  \GRAT_{-i}^{k-h}}) \ne \vec{0}$, so
$\mu(\intension{\play(\vec{\tau}_{-i})}) \ne \vec{0}$.

For part (b), we proceed by induction on $h$ to show that if $\NWD^1_{j} \ne
\NWD^0_{j}$ for at least two players $j$, then for all $h > 1$,
all states $\omega'$ in $M$, and all players $j$, we 
have $(M,\omega') \sat\GRATzero_{j}^{h} \rimp (\neg \GRATzero_{j}^1
\land \ldots \land \neg \GRATzero_j^{h-1})$.
For the base case, note that
if $(M,\omega') \sat \GRATzero_j^1$ and
$\PR_j(\omega') = (\mu_0',\ldots)$, then $\mu_0'$ must give positive
probability to all strategies in $\Sigma_{-j}$,
since $\GRATzero_j^1$ is an abbreviation of $RAT_j \land O^{\L^0_{-i}}_j (\true)$.
On the other hand, if 
$(M,\omega') \sat
\GRATzero_j^2$, then $\mu_0'$ must give positive probability to all
and only strategy profiles in $\Sigma_{-j}$ that are compatible with
$\GRATzero_{-j}^1$, which means that they are compatible with
$\RAT_{-j}$, and hence in $\NWD_{-j}^1$.  
Since $\NWD^0_{-j} \ne \NWD^1_{-j}$ by assumption, it follows that
$(M,\omega') \sat \GRAT_j^2 \rimp \neg \GRAT_1^1$. 
For the inductive step, note that if $h > 2$ and $(M,\omega') \sat
\GRATzero_j^h$, then
$\mu_0(\intension{\GRATzero_{-j}^{h-1}}) = 1$.
On the other hand, if 
$2 \le h' \le h-1$ and $(M,\omega') \sat \GRATzero_j^{h'}$,
then $\mu_0(\intension{\GRATzero_{-j}^{h'-1}}) = 1$.  Since, by the induction
hypothesis, $\intension{\GRATzero_{-j}^{h-1}} \inter
\intension{\GRATzero_{-j}^{h'-1}} = 
\emptyset$, we must have $(M,\omega') \sat \GRATzero_j^h \rimp \neg
\GRATzero_j^{h'}$.   Moreover, $(M,\omega') \sat \GRATzero_j^h \rimp \neg
\GRATzero_j^{1}$, since if $(M,\omega') \sat \GRATzero_j^1$, as we have
observed before, $\mu_0'$ it must give positive probability to all
strategies, while $\GRATzero_{-j}^{h-1} \rimp \RAT_{-j}$, so if 
$(M,\omega') \sat \GRATzero_j^h$, then $\mu_0$ gives positive
probability only to strategies in $\Sigma_{-j}$ compatible with rationality.

For part (c), we first show that $\intension{H^{h'}_{-i}} 
\gg_{\vec{\mu}} \intension{H^{h}_{-i}}$ if $0 \le h < h' \le k$.
Since $O^{\L^0_{-i}}_i (\GRATzero^k_{-i},
\ldots, \GRATzero^0_{-i})$ implies
$B_i(\GRATzero^k_{i})$, it follows that $\min_\ell\{\ell:
\mu_\ell(H^k_{-i}) > 0\} = 0$ and that, for all formulas $\phi$,
$\min_\ell\{\ell: \mu_\ell(\intension{\neg \GRAT^k_{-i} \land \phi}\} > 0$.
It follows that $\intension{H^k_{-i}} \gg_{\mu} \intension{H^h_{-i}}$
for all $h$ with $0 \le 
h < k$.  It also follows from the definition of $O^{\L^0_{-i}}_i
(\GRATzero^k_{-i}, \ldots, 
\GRATzero^0_{-i})$ that $\vec{\mu}(\intension{\GRAT_{-i}^{h'}} \mid
\intension{\neg \GRAT_{-i}^{h'+1} \land \ldots \land \neg
  \GRAT_{-i}^k}) = 1$.  Thus, for all formulas $\phi$, we have that
$$\min_\ell\{\ell: \mu_\ell(\intension{H^{h'}_{-i}}) > 0\} < \min\{\ell:
\mu_\ell(\intension{\neg \GRAT^h 
  \land \ldots \land \neg \GRAT^k_{-i} \land \phi}\} > 0.$$ It
  follows that $\min_\ell\{\ell: \mu_\ell\intension{(H^{h'}_{-i}}) > 0\} <  \min_\ell\{\ell: \mu_\ell(\intension{H^{h}_{-i}}) > 0\}$ for all $h' < h$, as
    desired.

    Now if $\intension{\GRAT_{-i}^h \land
      \play(\vec{\tau}_{-i})} \ne \emptyset$, then there exists $h'
    \ge h$ such that
    $(M,\omega) \sat \Diamond (\play(\vec{\tau}_{-i})
    \land H_{-i}^{h'})$.  It easily follows from the definition of 
$O^{\L^0_{-i}}_i (\GRATzero^k_{-i}, \ldots,
    \GRATzero^0_{-i})$ that   $\min_\ell\{\ell:
    \mu_\ell(\intension{H^{h'}_{-i} \land \play(\tau_{-i})} > 0\} =
        \min_\ell\{\ell: \mu_\ell(\intension{H^{h'}_{-i}}) > 0\}$.
 As we have already observed, $\intension{H_{-i}^k \lor \ldots
   \lor H_{-i}^0} = \Omega$.  Since
$\intension{(\GRAT_{-i}^h \lor \ldots \lor \GRAT_{-i}^k) \land 
   \play(\vec{\tau}_{-i}')} =  \emptyset$,
 and $\intension{\GRAT_{-i}^h \lor \ldots \lor \GRAT_{-i}^k} =
 \intension{H_{-i}^h \lor \ldots \lor H_{-i}^k}$, 
 it follows that
 $\intension{\play(\vec{\tau}_{-i}')} \subseteq
  \intension{H_{-i}^0 \lor \ldots \lor H_{-i}^{h-1}}$.  Thus,
$\min_\ell\{\ell: \mu_\ell(\intension{\play(\tau_{-i})}) > 0\}
\le  \min_\ell\{\ell: \mu_\ell(\intension{H_{-i}^{h-1}}) > 0\}$.  
Since $h' \ge h-1$, it follows from the first hlaf of part (c) that
$\intension{\play(\vec{\tau}_{-i})} 
   \gg_{\vec{\mu}} \intension{\play(\vec{\tau}'_{-i})}$, as desired.
  \eprf

  We can now prove Theorem~\ref{thm:charwdg}.

\medskip
  
\othm{thm:charwdg} The following are equivalent:
\begin{itemize}
\item[(a)] the strategy $\sigma$ for player $i$ survives $k$ rounds of
iterated deletion of weakly dominated strategies in $\Gamma$;
\item[(b)] there exists
  a fully measurable
LPS structure $M^{k}$
appropriate for $\Gamma$ and a state $\omega^{k}$ in $M^{k}$ such that
$\strat_i(\omega^{k}) = \sigma$ and $(M^{k},\omega^{k}) \sat
\GRATzero^{k}_i$.%
\footnote{It follows from the proof that we can take all the LPSs in $M^k$
  to have length $k+1$.}
\end{itemize}
In addition,
if $\vec{\sigma} \in \NWD^k$, then
there is a 
fully measurable
LPS structure $\bar{M}^k =
(\Omega^k,\strat,\F, \PR_1^k, \ldots, \PR_n^k)$ such that 
$\Omega^k = \{(k',i,\vec{\sigma}): 0 \le k' \le k, 1 \le i \le n, \vec{\sigma}
\in \NWD^{k'}\}$,
$\strat^k(k',i,\vec{\sigma}) = 
\vec{\sigma}$, $\F^k = 2^{\Omega^k}$, and
for all states 
$(k', i, \vec{\sigma}) \in \Omega^k$,  
$(\bar{M}^k,(k',i, \vec{\sigma})) \sat \GRATzero_{-i}^{k'}$.
\eothm

\prf The proof is similar in spirit to that of
Theorem~\ref{thm:charwd}, although we must make modifications to deal
with LPSs.
\fullv{Again, we proceed by induction on $k$,}
\shortv{We proceed by induction on $k$,}
proving both the equivalence of (a) and (b)
and the existence of an LPS  structure $\bar{M}^k$ with the required properties.

\fullv{Again, the result clearly holds if $k=0$.}
\shortv{The result clearly holds if $k=0$.}
Suppose that the result
holds for $k$; we prove it for $k+1$.  To show that (b)
implies (a), suppose that (b) holds,  
$\strat_i(\omega) = \sigma$, and $(M,\omega) \sat \GRATzero^{k+1}_i$.  
Let $\PR_i(\omega) = \vecmu = \<\mu_0,\ldots, \mu_m\>$.  Since 
$(M,\omega) \sat \RAT_i$, $\sigma$ must be a best response with respect to 
$\vecmu$.  Suppose, by way of contradiction, that
$\sigma$  is not in $\NWD_i^{k+1}$.
Let $h$ be the smallest index such that $\sigma \notin \NWD_i^{h+1}$.
Then $\sigma$ is weakly dominated 
by some mixed strategy $\sigma'$ such that all the strategies in the support of
$\sigma'$ are in $\NWD^h_i$, for some $h \le k$.   Thus, for all
strategy profiles $\tau_{-i} \in  \NWD^h_{-i}$, we have
$u_i(\sigma',\tau_{-i}) \ge u_i(\sigma,\tau_{-i})$, and for at least
one strategy profile $\tau'_{-i} \in \NWD^h_{-i}$, we have  
$u_i(\sigma',\tau'_{-i}) > u_i(\sigma,\tau'_{-i})$.
Note that since $h \le k$ and $\tau_{-i}' \in \NWD_{-i}^h$, it follows
from the induction hypothesis that there exists some state $\omega'$ in
$M^k$ such that $\strat_{-i}(\omega') = \tau_{-i}$ and $(M^k,\omega')
\sat \GRATzero^h_{-i}$.  Let $\ell$ be the least index such that
$\mu_{\ell}(\play(\vec{\tau}'_{-i})) > 0$ (by
Proposition~\ref{lem:GRAT}(a), there is such an index).  
By Proposition~\ref{lem:GRAT}(c), if
$\mu_{\ell'}(\play(\vec{\tau}''_{-i})) > 0$ for some $\ell' \le \ell$
and some strategy profile $\vec{\tau}''_{-i}$, then it must be the
case that $\intension{\play(\vec{\tau}''_{-i}) \land \GRAT_{-i}^{h'}} \ne
\emptyset $ for $h'$ with $h \le h' \le k$.  By the induction
hypothesis, it follows that $\vec{\tau}''_{-i} \in \NWD_{-i}^{h'}
\subseteq \NWD^h_{-i}$.  Thus, $u_i(\sigma',\vec{\tau}_{-i}'')  \ge
u_i(\sigma,\vec{\tau}_{-i}'')$.  It follows that $\sigma$ is
\emph{not} a best response with respect to $\vec{\mu}$, since $i$'s
expected utility is higher with $\sigma'$.

\fullv{The construction of $\bar{M}^{k+1}$ is similar in spirit to
  that in the Theorem~\ref{thm:charwd}.
  Again, as required, we define}
\shortv{We next construct the structure $\bar{M}^{k+1} =
    (\Omega^{k+1},\strat^{k+1},\F^{k+1}, \PR_1^{k+1}, \ldots, \PR_n^{k+1})$.  
As required, we define}
$\Omega^{k+1} = \{(k',i,\vec{\sigma}): k' \le k+1, 1 \le i \le n,
\vec{\sigma} \in
\NWD^{k'}\}$, $\strat^{k+1}(k', i, \vec{\sigma}) = \vec{\sigma}$, and $\F^{k+1} =
2^{\Omega^{k+1}}$.  For a state $\omega$ of the form $(0,
i,\vec{\sigma})$, let $\PR_j^{k+1}(\omega)$ be
the LPS $(\mu_0, \ldots, \mu_0)$, where $\mu_0$ is 
the uniform distribution over states (we could
actually use an arbitrary distribution here);
for a state $\omega$ of the form $(k', i,\vec{\sigma})$, 
where $k'\geq 1$, since $\sigma_j \in \NWD^{k'}_j$,
and hence also in $\NWD^{k''}_j$ for $0 \le k'' < k$, 
by Proposition \ref{pro:Pearce}, there exist distributions
$\mu_{k'',\sigma_j}$, $1 \le k'' \le k$, on strategies such that the support of
$\mu_{k'',\sigma_j}$ is all of $NWD_{-j}^{k''-1}$ and 
$\sigma_j$ is a best response to $\mu_{\sigma_{k'',\sigma_j}}$.  We can
extend $\mu_{k'',\sigma_j}$ to a distribution $\mu_{k'',i,\sigma_j}'$
on $\Omega^{k+1}$ 
as follows:
\begin{itemize}
\item if $i\neq j$, then $\mu_{k'',i,\sigma_j}'(m,i',\vec{\tau}) = 
\mu_{k'',\sigma_j}(\vec{\tau}_{-j})$ if $i'=j, m = k''-1$, and $\tau_j =
  \sigma_j$, and 0 otherwise;   
\item $\mu_{k',j,\sigma_j}'(m,i',\vec{\tau}) = 
\mu_{k',\sigma_j}(\vec{\tau}_{-j})$ if $i'=j$, $m = k''$, and $\tau_j = \sigma_j$, and 0 otherwise.  
\end{itemize}
Let $\PR_j(k',i,\vec{\sigma}) = (\mu_{k',i,\sigma_j},
\mu_{k'-1,i,\sigma_j}, \ldots, \mu_{1,i,\sigma_j})$.
We leave it to the reader to check that this
structure is appropriate,  
and that 
$(\bar{M}^{k+1},(k',i,\vec{\sigma})) \sat \GRATzero^{k'}_{=i}$.
One key point is that since $\sigma_j$ is a best response with to each
of $\mu_{k',i,\sigma_j}, \ldots, \mu_{1,i,\sigma_j}$, it is not hard
to show that it is a best response with respect to the LPS.  

\fullv{The argument that (a) implies (b) now follows just as in the proof of
Theorem~\ref{thm:charwd} (except that we replace $\RATzero^{k'}_{j}$
by $\GRATzero^{k'}_{j}$).}
\shortv{
To see that (a) implies (b), suppose that $\sigma_i \in \NWD^{k+1}_i$.
Choose a state $\omega$ in 
$\bar{M}^{k+1}$ of the form $(k+1,j,\vec{\sigma})$, where $i \neq j$.  
It follows from the properties of $\bar{M}^{k+1}$ that
$(\bar{M}^{k+1}, \omega) \sat \GRATzero^{k+1}_i$ and
$\strat_i(\omega) = \sigma_i$.  
Moreover, $\bar{M}^{k+1}$ is
fully measurable
(since $\F^{k+1}$ consists of all subsets of $\Omega^{k+1}$).}
\eprf

\medskip

\othm{thm:charwdg1} For all finite games $\Gamma$ and all
sufficiently small $\epsilon > 0$, there 
   exists some $\delta > 0$ such that 
the following are equivalent: 
\begin{itemize}
\item[(a)] the strategy $\sigma$ for player $i$ survives $k$ rounds of
iterated deletion of weakly dominated strategies in $\Gamma$;
\item[(b)] there exists a 
fully measurable structure  $M^{k}$
appropriate for $\Gamma$ and a state $\omega^{k}$ in $M^{k}$ such that
$\strat_i(\omega^{k}) = \sigma$ and $(M^{k},\omega^{k}) \sat
\GRAT^{k,\delta,\epsilon}_i$.
\end{itemize}
In addition,
if $\vec{\sigma} \in \NWD^k$, then
for all sufficiently small $\epsilon > 0$, there exists some $\delta >
0$ and 
a fully measurable structure $\bar{M}^{k,\delta} =
(\Omega^{k,\delta},\strat^{k,\delta},\F^{k,\delta}, \PR_1^{k,\delta},
\ldots, \PR_n^{k,\delta})$ such that  
$\Omega^{k,\delta} = \{(k',i,\vec{\sigma}): 0 \le k' \le k, 1 \le i \le n, \vec{\sigma}
\in \NWD^{k'}\}$,
$\strat^{k,\delta}(k',i,\vec{\sigma}) = 
\vec{\sigma}$, $\F^{k,\delta} = 2^{\Omega^k}$, 
and
for all states 
$(k', i, \vec{\sigma}) \in \Omega^{k,\delta}$,   we have 
$(\bar{M}^{k,\delta},(k',i, \vec{\sigma})) \sat
\GRAT_{-i}^{k',\delta,\epsilon}$.%
\footnote{How small $\epsilon$ has to be depends only on the game.  
Although the choice of
  $\delta$ depends on the choice of $\epsilon$, $\epsilon$ plays no role in the
construction of $\bar{M}^{k,\delta}$ (which is why we did not write
$\bar{M}^{k,\delta,\epsilon}$).}
\eothm

\prf The proof has the same high-level structure as the proof of 
Theorem~\ref{thm:charwdg}, but we now need to be a bit more careful in
analyzing the probabilities.
%
As before, we proceed by induction on $k$,
proving both the equivalence of (a) and (b)
and the existence of a structure $\bar{M}^k$. 
%
The (b) implies (a) direction works for any choice of $\epsilon>0$
as long as $\delta$ is significantly smaller than $\epsilon$.
The (a) implies (b) direction, on the other hand, will
additionally require $\epsilon$ to be sufficiently small (and
$\delta$ to be no greater).
It follows that for all sufficiently small
$\epsilon>0$ there exists a $\delta>0$ such that the equivalence
holds.

The result clearly holds if $k=0$.
Suppose that the result
holds for $k$; we prove it for $k+1$.  To show that (b)
implies (a), suppose that (b) holds,  
$\strat_i(\omega) = \sigma$, and $(M,\omega) \sat
\GRAT^{k+1,\delta,\epsilon}_i$.
Since $\Gamma$ is a finite game, there is some number $\gap > 0$,
which is  
the largest value such that, for all players $i$ and 
strategies $\sigma_i$ for player $i$, if $\sigma_i$ is deleted in
round $h$ in the deletion process, 
then there exists a mixed strategy $\sigma_i'$ and a strategy profile
$\tau_{-i}' \in \NWD^{h-1}_{-i}$ such that 
$u_i(\sigma', \tau_{-i}) \ge u_i(\sigma, \tau_{-i})$ for all $\tau_{-i} \in
\NWD^{h-1}_{-i}$ and   $u_i(\sigma', \tau'_{-i}) > u_i(\sigma,
\tau'_{-i}) + \gap$.
Since $\Gamma$ has only finitely many strategies, we must have $\gap >
0$; since all utilities are in $[0,1]$, we must have $\gap \le 1$.  
Set
$\delta = \frac{\gap \cdot \epsilon}{4}$.
Note for future reference that 
$\delta < \frac{1}{2}$ and that $\delta < \frac{\epsilon}{2}$.

Let $\mu= \PR_i(\omega)$.
Since 
$(M,\omega) \sat \RAT_i$, $\sigma$ must be a best response with respect to 
$\mu$.  Suppose, by way of contradiction, that $\sigma$  
is not in $\NWD^{k+1}_i$.
Then, for some $h \le k$, $\sigma \in \NWD^h_i - \NWD^{h+1}_i$.  Thus,
$\sigma$ is weakly dominated 
by some mixed strategy $\sigma'$ such that all the strategies in the support of
$\sigma'$ are in $\NWD^h_i$ and there exists a 
strategy profile $\tau'_{-i} \in \NWD^h_{-i}$ such that
$u_i(\sigma',\tau'_{-i}) \geq u_i(\sigma,\tau'_{-i})) + \gap$.

Define the random variable $\Adv$ that measures the utility advantage
of using $\sigma'$ as opposed to $\sigma$:
$$\Adv(\omega') = u_i(\sigma', \strat_{-i}(\omega')) - u_i(\sigma,
\strat_{-i}(\omega')).$$
We now show that $\E_{\mu}(\Adv) > 0$, which contradicts our
assumption that
$\sigma$ is a best response, since if the expected utility of using
(the mixed strategy) $\sigma'$ is greater than that of using $\sigma$,
there also exists some pure strategy $\sigma''$ in the support of
$\sigma'$ that has the same expected utility as that of $\sigma'$
(with respect to $\mu)$, and thus $\sigma$ cannot be a best response with
respect to $\mu$.

Let $\GRAT_{i}^{\geq h+1,\delta,\epsilon} =
\GRAT_{-i}^{h+1,\delta,\epsilon} \lor \GRAT_{-i}^{h+2,\delta,\epsilon}
\lor \ldots \lor \GRAT_{-i}^{k,\delta,\epsilon}$. 
By the law of total expectation, 
$$\begin{array}{ll}
  &\E_{\mu}(\Adv)\\
  = &\E_{\mu}(\Adv \mid \intension{\GRAT_{-i}^{\geq
      h+1,\delta,\epsilon}}) \mu(
\intension{\GRAT_{-i}^{\geq h+1,\delta,\epsilon}}) + \E_{\mu}(\Adv
\mid \intension{ \neg
\GRAT_{-i}^{\geq h+1,\delta,\epsilon}}) \mu( \intension{\neg \GRAT_{-i}^{\geq
    h+1,\delta,\epsilon}}).
\end{array}
$$ 
By the induction hypothesis, if $\strat_{-i}(\omega') \in
\intension{\GRAT_{-i}^{\ge h+1,\delta,\epsilon}}$, then
$\strat_{-i}(\omega') \in \union_{h' = h+1}^k
\NWD^{h'}_{-i} = \NWD^{h+1}_{-i}$ (since $\NWD^j_{-i} \subseteq
\NWD_{-i}^{h+1}$ for all $j\geq h+1$).  Since $\sigma'$ weakly
dominates $\sigma \in NWD_i^h - \NWD^{h+1}$, we have that
$\mu_i(\sigma',\tau_{-i}) \ge \mu_i(\sigma,\tau_{-i})$ for all
$\tau_{-i} \in  \NWD^h_{-i} \supseteq \NWD^{h+1}_{-i}$. Thus, the first term in the expression
above is at least 0.
Since $\mu( \intension{\neg \GRAT_{-i}^{\geq h+1,\delta,\epsilon}})
\le 1$, it follows that
$\E_{\mu}(\Adv) \ge \E_{\mu}(\Adv \mid 
\intension{\neg \GRAT_{-i}^{\geq h+1,\delta,\epsilon}}).$
Now applying the law of total expectation 
to $\E_{\mu}(\Adv \mid \intension{\neg \GRAT_{-i}^{\geq h+1,\delta,\epsilon}})$,
we have
that
$$\begin{array}{lll}
&\E_{\mu}(\Adv)\\ \geq &\E_{\mu}(\Adv \mid \intension{\neg
  \GRAT_{-i}^{\geq h+1,\delta,\epsilon}}) 
\\  =    
& \E_{\mu}(\Adv \mid \intension{\GRAT_{-i}^{h,\delta,\epsilon} \land \neg
  \GRAT_{-i}^{\geq h+1,\delta,\epsilon}}) \mu(\intension{\GRAT_{-i}^{h,\delta,\epsilon}} \mid \intension{\neg
 \GRAT_{-i}^{\geq h+1,\delta,\epsilon}})\mu(\intension{\neg \GRAT_{-i}^{\geq
      h+1,\delta,\epsilon}}) \\ &+
\E_{\mu}(\Adv \mid \intension{\neg \GRAT_{-i}^{h,\delta,\epsilon} \land \neg
  \GRAT_{-i}^{\geq h+1,\delta,\epsilon}}) \mu(\intension{\neg \GRAT_{-i}^{h,\delta,\epsilon}} \mid \intension{\neg
  \GRAT_{-i}^{\geq h+1,\delta,\epsilon}})\mu(\intension{\neg \GRAT_{-i}^{\geq
      h+1,\delta,\epsilon}}). 
\end{array}
$$
Since $(M,\omega) \sat  O^{\L_{0}, \delta, \epsilon} (\GRAT_{-i}^{k,\delta,\epsilon})$,
we have that
\begin{itemize}
\item[(a)] $\mu(\intension{\GRAT_{-i}^{\geq h+1,\delta,\epsilon}}) > 0$,
\item[(b)] $\mu(\intension{\GRAT_{-i}^{h,\delta,\epsilon}} \mid \intension{\neg
  \GRAT_{-}^{\geq h+1,\delta,\epsilon}}) 
    \geq 1- \delta$, and
 \item[(c)] 
$\mu (\intension{\play_{-i}(\tau_{-i})} \mid \intension{\neg
      \GRAT_{-i}^{\geq h+1,\delta,\epsilon}}) \geq \epsilon$. 
\end{itemize}
%
From (b), it follows that $\mu (\intension{\neg \GRAT_{-i}^{h,\delta,\epsilon}} \mid \intension{\neg
\GRAT_{-i}^{\geq h+1,\delta,\epsilon}}) \leq \delta$. 
Now using (c) and the law of total probability,\footnote{By the law of total
  probability, 
$\mu(X \mid Z) = \mu(X \mid Y \inter Z) \mu(Y\mid Z) + \mu(X \mid
\bar{Y} \inter Z) \mu(\bar{Y}\mid Z) \le
  \mu(X \mid Y \inter Z) + \mu(\bar{Y} \mid Z)$, so   $\mu(X \mid Y
  \inter Z)   \ge  \mu(X \mid Z) - 
    \mu(\bar{Y} \mid Z)$.} 
we get
\begin{equation}\label{eq0}\begin{array}{lll}
  &\mu (\intension{\play_{-i}(\tau_{-i})} \mid
\intension{\GRAT_{i}^{h,\delta,\epsilon} \land\neg \GRAT_{-i}^{\geq
    h+1,\delta,\epsilon}})\\ \geq
&\mu (\intension{\play_{-i}(\tau_{-i})} \mid \intension{\neg
      \GRAT_{-i}^{\geq h+1,\delta,\epsilon}}) 
- \mu(\intension{\neg \GRAT_{-i}^{h,\delta,\epsilon}} \mid \intension{\neg
  \GRAT_{-}^{\geq h+1,\delta,\epsilon}}) \\
\ge &\epsilon-\delta\\ \geq &\frac{\epsilon}{2}.\end{array}\end{equation}

Since the range of the utility
function $u_i$ is $[0,1]$, we must have
$\Delta^{\sigma',\sigma}(\omega) \ge -1$ for all worlds $\omega$.
Again using the fact that $\mu (\intension{\neg
  \GRAT_{-i}^{h,\delta,\epsilon}} \mid \intension{\neg 
\GRAT_{-i}^{\geq h+1,\delta,\epsilon}}) \leq \delta$, it follows that
the second term of the expression above is bounded below
by $-\delta \cdot \mu( \intension{\neg \GRAT_{-i}^{\geq h+1,\delta,\epsilon}})$.
From (b), (\ref{eq0}), and our assumptions on $\sigma'$, it follows that
first term in the expression above is bounded below by  
$$\gap \cdot \frac{\epsilon}{2} \cdot (1-\delta) \mu (\GRAT_{-i}^{\geq
  h+1,\delta,\epsilon}) > \frac{\gap \cdot 
    \epsilon}{4} \cdot \mu (\GRAT_{-i}^{\geq h+1,\delta,\epsilon}),$$
where the strict inequality follows from (a) and the fact that
$\delta < \frac{1}{2}$.
We conclude that 
$$\E_{\mu}(\Adv) > \mu(\intension{\GRAT_{-i}^{\geq h+1,\delta,\epsilon}}) \left
(\frac{\gap \cdot
    \epsilon}{2} - \delta \right) \geq 0,$$ 
which gives the desired contradiction.

We next construct the structure $\bar{M}^{k+1,\delta} =
(\Omega^{k+1,\delta},\strat^{k+1,\delta},\F^{k,\delta},
\PR_1^{k+1,\delta}, \ldots, \PR_n^{k+1,\delta})$. 
The construction is very similar to that of Theorem~\ref{thm:charwdg}.
As required, we define
$\Omega^{k+1,\delta} = \{(k',i,\vec{\sigma}): k' \le k+1, 1 \le i \le n,
\vec{\sigma} \in
\NWD^{k'}\}$, $\strat^{k+1,\delta}(k', i, \vec{\sigma}) =
\vec{\sigma}$, and $\F^{k+1,\delta} = 
2^{\Omega^{k+1}}$.  For a state $\omega$ of the form $(0,
i,\vec{\sigma})$, let $\PR_j^{k+1,\delta}(\omega)$ be
the uniform distribution over states; for 
a state $\omega$ of the form $(k', i,\vec{\sigma})$, 
where $k'\geq 1$, since $\sigma_j \in \NWD^{k'}_j$,
and hence also in $\NWD^{k''}_j$ for $0 \le k'' < k$, 
by Proposition \ref{pro:Pearce}, there exist distributions
$\mu_{k'',\sigma_j}$, $1 \le k'' \le k$, on strategies such that the support of
$\mu_{k'',\sigma_j}$ is all of $NWD_{-j}^{k''-1}$ and 
$\sigma_j$ is a best response to $\mu_{\sigma_{k'',\sigma_j}}$.
We can extend $\mu_{k'',\sigma_j}$ to a distribution $\mu_{k'',i,\sigma_j}'$
on $\Omega^{k+1}$ 
\commentout{
as follows:
\begin{itemize}
\item if $i\neq j$, then $\mu_{k'',i,\sigma_j}'(m,i',\vec{\tau}) = 
\mu_{k'',\sigma_i}(\vec{\tau}_{-j})$ if $i'=j, m = k'-1$, and $\tau_j =
\sigma_j$, and 0 otherwise;   
\item $\mu_{k',j,\sigma_j}'(m,i',\vec{\tau}) = 
\mu_{k',\sigma_j}(\vec{\tau}_{-j})$ if $i'=j$, $m = k''$, and $\tau_j = \sigma_j$, and 0 otherwise.  
\end{itemize}
}
just as we did in the construction of Theorem~\ref{thm:charwdg}.

Let $\epsilon_0$ be the smallest \emph{non-zero} probability assigned to some
strategy by any of the probability distributions $\mu_{k',i,\sigma_j}$,
for $k' \in \{1,\ldots, k\}$, players $i$ and $j$, and strategies
$\sigma_j$.
The construction works for all $\epsilon \leq \epsilon_0$.
(Thus, the ``sufficiently small'' in the theorem statement means ``at most
$\epsilon_0$''; note that $\epsilon_0$ is determined by the game $\Gamma$.) 
Choose $\delta$ so the $\epsilon \le \delta < 1/2$. 
We now define $\PR_j^{k+1,\delta}(k',i,\vec{\sigma})$ as
$$\delta^0 (1-\delta) \mu_{k',i,\sigma_j} +
\delta^1  (1-\delta) \mu_{k'-1,i,\sigma_j}, +
\delta^{k'-1} (1-\delta) \mu_{1,i,\sigma_j} + \delta^{k'}
(1-\delta)\mu_0 + \delta^{k'+1}\mu_0,$$
where $\delta' = 1/(1-\delta^{k'})$ is a normalizing factor
(since 
$\delta^0 + \cdots + \delta^{k'-1} = 1/(1-\delta^{k'})$).
$\PR^{k_1,\delta}_j$ is a convex combination of the distributions 
$\mu_{k',i,\sigma_j}$, \ldots, $\mu_{1,i,\sigma_j}$, and 
hence is a well-defined probability distribution.
We leave it to the reader to check that 
this
structure is appropriate
and that 
$(\bar{M}^{k+1,\delta},(k',i,\vec{\sigma})) \sat
\GRAT^{k',\delta,\epsilon}_{-i}$.
The latter argument depends on the following observations:
(1) Since $\sigma_j$ is a best response with respect to each of
$\mu_{k',i,\sigma_j}, \ldots, \mu_{1,i,\sigma_j}$, it is not hard to
show that it is a best response with respect to the convex
combination (despite the fact that these distributions have different
supports); (2)  $\mu_{k'',i,\sigma_j}$ ascribes probability 1 to 
$\intension{\GRAT_{-i}^{k'',\delta,\epsilon}}$ and, since $\delta < 1/2$, the same
argument as that in Proposition~\ref{lem:GRAT}(b) shows that these sets are
disjoint; and (3) for all strategy profiles $\tau_{-i} \in
\NWD^{k''}_{-i}$, $\mu_{k'',i,\sigma_k}$ ascribe probability at least
$\epsilon$ to  $\tau_{-i}$.

\commentout{
To see that (a) implies (b), suppose that $\sigma_j \in \NWD^{k+1}_j$.
Choose a state $\omega$ in 
$\bar{M}^{k+1}$ of the form $(k+1,i,\vec{\sigma})$, where $i \neq j$.  
As we just showed,
$(\bar{M}^{k+1}, \omega) \sat \GRATzero^{k+1}_j$ and
$\strat_j(\omega) = \sigma_j$.  Moreover, $\bar{M}^{k+1}$ is
$\L^0$-measurable (since $\F$ consists of all subsets of
$\Omega^{k+1}$).}
The argument that (a) implies (b) is essentially identical to that in
the proof of Theorem~\ref{thm:charwdg}, so we omit it here.
Since it depends on $\bar{M}^{k,\delta}$, the constraints on
$\epsilon$ needed to ensure the existence of $\bar{M}^{k,\delta}$
apply here as well (that is, $\epsilon$ must be sufficiently small).
\eprf
}

\subsubsection*{Acknowledgements}
The first author is supported in part by NSF grants 
IIS-178108 and IIS-1703846, a grant from the Open
Philanthropy Foundation, and ARO grant W911NF-17-1-0592.
The second author is supported in part by 
NSF grant IIS-1703846.

\fullv{\bibliographystyle{chicagor}
\bibliographystyle{chicago}}
\bibliographystyle{eptcs}
\bibliography{z,joe,refs}
\end{document}

%% file: defn.tex
\newtheorem{THEOREM}{Theorem}[section]
\newenvironment{theorem}{\begin{THEOREM} \hspace{-.85em} {\bf :} }%
                        {\end{THEOREM}}
\newtheorem{LEMMA}[THEOREM]{Lemma}
\newenvironment{lemma}{\begin{LEMMA} \hspace{-.85em} {\bf :} }%
                      {\end{LEMMA}}
\newtheorem{COROLLARY}[THEOREM]{Corollary}
\newenvironment{corollary}{\begin{COROLLARY} \hspace{-.85em} {\bf :} }%
                          {\end{COROLLARY}}
\newtheorem{PROPOSITION}[THEOREM]{Proposition}
\newenvironment{proposition}{\begin{PROPOSITION} \hspace{-.85em} {\bf :} }%
                            {\end{PROPOSITION}}
\newtheorem{DEFINITION}[THEOREM]{Definition}
\newenvironment{definition}{\begin{DEFINITION} \hspace{-.85em} {\bf :} \rm}%
                            {\end{DEFINITION}}
\newtheorem{CLAIM}[THEOREM]{Claim}
\newenvironment{claim}{\begin{CLAIM} \hspace{-.85em} {\bf :} \rm}%
                            {\end{CLAIM}}
\newtheorem{EXAMPLE}[THEOREM]{Example}
\newenvironment{example}{\begin{EXAMPLE} \hspace{-.85em} {\bf :} \rm}%
                            {\end{EXAMPLE}}
\newtheorem{REMARK}[THEOREM]{Remark}
\newenvironment{remark}{\begin{REMARK} \hspace{-.85em} {\bf :} \rm}%
                            {\end{REMARK}}

\newcommand{\thm}{\begin{theorem}}
\newcommand{\lem}{\begin{lemma}}
\newcommand{\pro}{\begin{proposition}}
\newcommand{\dfn}{\begin{definition}}
\newcommand{\rem}{\begin{remark}}
\newcommand{\xam}{\begin{example}}
\newcommand{\cor}{\begin{corollary}}
\newcommand{\prf}{\noindent{\bf Proof:} }
\newcommand{\ethm}{\end{theorem}}
\newcommand{\elem}{\end{lemma}}
\newcommand{\epro}{\end{proposition}}
\newcommand{\edfn}{\bbox\end{definition}}
\newcommand{\erem}{\bbox\end{remark}}
\newcommand{\exam}{\bbox\end{example}}
\newcommand{\ecor}{\end{corollary}}
\newcommand{\eprf}{\bbox\vspace{0.1in}}
\newcommand{\beqn}{\begin{equation}}
\newcommand{\eeqn}{\end{equation}}

\newcommand{\bbox}{\vrule height7pt width4pt depth1pt}

\newcommand{\clm}{\begin{claim}}
\newcommand{\eclm}{\end{claim}}

\newcommand{\sat}{\models}

\newcommand{\rimp}{\Rightarrow}

\newcommand{\union}{\cup}
\newcommand{\inter}{\cap}

\newcommand{\IN}{\mbox{$I\!\!N$}}

\renewcommand{\phi}{\varphi}

\newcommand{\F}{{\cal F}}
\newcommand{\E}{{\cal E}}

\newcommand{\Z}{{\cal Z}}

\newcommand{\<}{\langle}
\renewcommand{\>}{\rangle}

\newcommand{\ol}{\setlength{\itemsep}{0pt}\begin{enumerate}}
\newcommand{\eol}{\end{enumerate}\setlength{\itemsep}{-\parsep}}
\newcommand{\ul}{\setlength{\itemsep}{0pt}\begin{itemize}}
\newcommand{\dl}{\setlength{\itemsep}{0pt}\begin{description}}
\newcommand{\edl}{\end{description}\setlength{\itemsep}{-\parsep}}
\newcommand{\eul}{\end{itemize}\setlength{\itemsep}{-\parsep}}

\newcommand{\true}{{\it true}}

\newcommand{\RAT}{\mathit{RAT}}
\newcommand{\play}{\mathit{play}}
\newcommand{\dbi}{\langle B_i \rangle}
\newcommand{\dbj}{\langle B_j \rangle}
\newcommand{\commentout}[1]{}

\newcommand{\bi}{\begin{itemize}}
\newcommand{\ei}{\end{itemize}}
\newcommand{\be}{\begin{enumerate}}
\newcommand{\ee}{\end{enumerate}}